\pdfoutput=1  % Instructs arXiv to run this with pdflatex rather than latex.

\documentclass[aps,prd,superscriptaddress,floatfix,nofootinbib,preprintnumbers,eqsecnum]{revtex4-1}

\usepackage{amsmath,float,graphicx,placeins,amssymb,multirow,pgfplots,pgfplotstable}
\usepackage{empheq,enumitem}
\usepackage{tikz}
\usepackage{comment}
\usepackage{enumerate,slashed,cancel,comment,color,bbm,graphicx,rotating,placeins,mathtools,multirow,hhline}
\usepackage[normalem]{ulem}
\usepackage{array}

\usepackage{hyperref}
\usepackage{color}
 \hypersetup
 {
   colorlinks,
   linkcolor={blue!80!black},
   citecolor={blue!70!black},
   urlcolor={blue!70!black}
 }

\usepackage{lipsum}
\usepackage{diagbox}
\usetikzlibrary{shapes.geometric,arrows}
\graphicspath{}

\def\beq{\begin{equation}}
\def\eeq{\end{equation}}
\def\beqn{\begin{eqnarray}}
\def\eeqn{\end{eqnarray}}
\def\half{\mbox{\small ${\frac{1}{2}}$}}

\def\quarter{\mbox{\small ${\frac{1}{4}}$}}

\newcommand{\newc}{\newcommand}
\def\calZ{{\cal Z}}
\def\calM{{\cal M}}
\def\calV{{\cal V}}
\def\calF{{\cal F}}

\def\bQ{{\bf Q}}
\def\bT{{\bf T}}

\def\Qs{{\bf q}}

\def\barOmega{{\overline{\Omega}}}

\def\half{{\textstyle{1\over 2}}}
\def\quarter{{\textstyle{1\over 4}}}
\def\ie{{\it i.e.}\/}
\def\eg{{\it e.g.}\/}
\def\etc{{\it etc}.\/}
\def\inbar{\,\vrule height1.5ex width.4pt depth0pt}
\def\IR{\relax{\rm I\kern-.18em R}}
 \font\cmss=cmss10 \font\cmsss=cmss10 at 7pt
\def\IQ{\relax{\rm I\kern-.18em Q}}
\def\IZ{\relax\ifmmode\mathchoice
 {\hbox{\cmss Z\kern-.4em Z}}{\hbox{\cmss Z\kern-.4em Z}}
 {\lower.9pt\hbox{\cmsss Z\kern-.4em Z}}
 {\lower1.2pt\hbox{\cmsss Z\kern-.4em Z}}\else{\cmss Z\kern-.4em Z}\fi}
\begin{document}

\setcounter{page}{0}

  \title{
   \rule[-0.2in]{\hsize}{0.01in}\\\rule{\hsize}{0.01in}\\
   \vskip 0.05 in {%\normalsize
           \rm Submitted to the  Proceedings of the US Community Study\\
    \vskip -0.035 in
       on the Future of Particle Physics (Snowmass 2021)}\\
   \vskip -0.035 in
       \rule{\hsize}{0.01in}\\\rule[+0.2in]{\hsize}{0.01in} 
   ~\\ ~\\ ~\\ ~\\
  {\LARGE
      More is Different:\\
   Non-Minimal Dark Sectors and their Implications\\
    for Particle Physics, Astrophysics, and Cosmology\\
      \vskip 0.08 truein
      \textbf{\textit{ {--- 13 Take-Away Lessons for Snowmass 2021 --- } }}}}

%   \title{
%   \begin{center}
%   \rule[-0.2in]{\hsize}{0.01in}\\\rule{\hsize}{0.01in}\\
%   \vskip 0.1in Submitted to the  Proceedings of the US Community Study\\
%   on the Future of Particle Physics (Snowmass 2021)\\
%   \rule{\hsize}{0.01in}\\\rule[+0.2in]{\hsize}{0.01in} 
%   \end{center}
%   More is Different:\\
     %   Non-Minimal Dark Sectors and their Implications\\
    %   for Particle Physics, Astrophysics, and Cosmology\\
    %   %\vskip 0.01 truein
    %   {\it --- 13 Take-Away Lessons for Snowmass 2021 --- }}
%   

  %  \title{More is Different:\\
       %  Non-Minimal Dark Sectors and their Implications\\
      %  for Particle Physics, Astrophysics, and Cosmology\\
      %  %\vskip 0.01 truein
      %  {\it --- 13 Take-Away Lessons for Snowmass 2021 --- }}

\def\andname{\hspace*{-0.5em}} % gets rid of "and" in author list

\author{\Large ~\\   Keith R. Dienes,}
\email[\large Email address: ]{dienes@arizona.edu}
\affiliation{\large Department of Physics, University of Arizona, Tucson, AZ 85721 USA}
\affiliation{\large Department of Physics, University of Maryland, College Park, MD 20742 USA}
\author{\Large Brooks Thomas}
\email[\large Email address: ]{thomasbd@lafayette.edu}
\affiliation{\large Department of Physics, Lafayette College, Easton, PA  18042 USA}

\begin{abstract}
 {\large \rm ~\\ ~\\ ~\\ ~\\  ~\\ 
   \begin{center}{\large\bf Abstract\\}\end{center}  
    The phrase ``more is different'' is often used to refer
  to the new, unexpected collective phenomena that can arise
  when the number of states in a given system is large.
  In this contribution to the Snowmass 2021 Study, we
  describe 13 unexpected collective 
  phenomena that can arise when the dark sector
  contains a large number of states, contrary to the usual assumptions.
  These 13 take-away lessons stretch across all of the domains of
  relevance for dark-matter physics, including 
  collider signatures, direct-detection signatures, indirect-detection signatures, 
   new perspectives on dark-matter complementarity,
  and even unexpected astrophysical and cosmological phenomena that transcend
     those normally associated with single-component dark-matter scenarios.
    These lessons --- and the phenomena on which they are based --- thereby illustrate the need to maintain a broad perspective
  when contemplating the possible signatures and theoretical possibilities
  associated with non-minimal dark sectors.
   }
\end{abstract}
\maketitle

\tableofcontents

\def\ie{{\it i.e.}\/}
\def\eg{{\it e.g.}\/}
\def\etc{{\it etc}.\/}
\def\taubar{{\overline{\tau}}}
\def\qbar{{\overline{q}}}
\def\kbar{{\overline{k}}}
\def\bQ{{\bf Q}}
\def\calT{{\cal T}}
\def\calN{{\cal N}}
\def\calF{{\cal F}}
\def\calM{{\cal M}}
\def\calZ{{\cal Z}}

\def\beq{\begin{equation}}
\def\eeq{\end{equation}}
\def\beqn{\begin{eqnarray}}
\def\eeqn{\end{eqnarray}}
\def\apo{\mbox{\small ${\frac{\alpha'}{2}}$}}
\def\half{\mbox{\small ${\frac{1}{2}}$}}
\def\sqapo{\mbox{\tiny $\sqrt{\frac{\alpha'}{2}}$}}
\def\sqap{\mbox{\tiny $\sqrt{{\alpha'}}$}}
\def\sqapxtwo{\mbox{\tiny $\sqrt{2{\alpha'}}$}}
\def\aptwo{\mbox{\tiny ${\frac{\alpha'}{2}}$}}
\def\apofour{\mbox{\tiny ${\frac{\alpha'}{4}}$}}
\def\bosqtwo{\mbox{\tiny ${\frac{\beta}{\sqrt{2}}}$}}
\def\btosqtwo{\mbox{\tiny ${\frac{\tilde{\beta}}{\sqrt{2}}}$}}
\def\apofour{\mbox{\tiny ${\frac{\alpha'}{4}}$}}
\def\sqaptwo{\mbox{\tiny $\sqrt{\frac{\alpha'}{2}}$}  }
\def\apoeight{\mbox{\tiny ${\frac{\alpha'}{8}}$}}
\def\sapoeight{\mbox{\tiny ${\frac{\sqrt{\alpha'}}{8}}$}}

\newc{\gsim}{\lower.7ex\hbox{{\mbox{$\;\stackrel{\textstyle>}{\sim}\;$}}}}
\newc{\lsim}{\lower.7ex\hbox{{\mbox{$\;\stackrel{\textstyle<}{\sim}\;$}}}}
\def\calM{{\cal M}}
\def\calV{{\cal V}}
\def\calF{{\cal F}}
\def\bQ{{\bf Q}}
\def\bT{{\bf T}}
\def\Qs{{\bf q}}

\def\half{{\textstyle{1\over 2}}}
\def\quarter{{\textstyle{1\over 4}}}
\def\ie{{\it i.e.}\/}
\def\eg{{\it e.g.}\/}
\def\etc{{\it etc}.\/}
\def\inbar{\,\vrule height1.5ex width.4pt depth0pt}
\def\IR{\relax{\rm I\kern-.18em R}}
 \font\cmss=cmss10 \font\cmsss=cmss10 at 7pt
\def\IQ{\relax{\rm I\kern-.18em Q}}
\def\IZ{\relax\ifmmode\mathchoice
 {\hbox{\cmss Z\kern-.4em Z}}{\hbox{\cmss Z\kern-.4em Z}}
 {\lower.9pt\hbox{\cmsss Z\kern-.4em Z}}
 {\lower1.2pt\hbox{\cmsss Z\kern-.4em Z}}\else{\cmss Z\kern-.4em Z}\fi}

\newcommand{\Dsle}[1]{\hskip 0.09 cm \slash\hskip -0.26 cm #1}
\newcommand{\Dirsl}[1]{\hskip 0.09 cm \slash\hskip -0.20 cm #1}
\newcommand{\met}{{\Dsle E_T}}

\def\lesson#1#2{{
\vskip 0.10 truein
\setlength{\fboxsep}{3pt}
\noindent \fbox{
\setlength{\fboxsep}{5pt}
\noindent
\hskip -8pt 
\fbox{
\parbox{6.70 truein}{
\begin{itemize}[leftmargin=*]
\item  \vskip -0.08 truein
       \underbar{\bf Lesson \#{{#1}}}:
      {{#2}}
\end{itemize}
\vskip -0.08 truein
}}}
\vskip 0.1 truein}}

\def\summarylesson#1{{
\vskip 0.10 truein
\setlength{\fboxsep}{3pt}
\noindent \fbox{
\setlength{\fboxsep}{5pt}
\noindent
\hskip -8pt 
\fbox{
\parbox{6.70 truein}{
\begin{itemize}[leftmargin=*]
\item  \vskip -0.08 truein
       \underbar{\bf Take-Away Summary Lesson}\/:
      {{#1}}
\end{itemize}
\vskip -0.08 truein
}}}
\vskip 0.1 truein}}

%=====================================================================================================

\large

\FloatBarrier
\section{\large Introduction:  ~More is Different \label{intro}}

In 1972, P.~W.~Anderson published an influential article~\cite{anderson}
 with the title ``More Is Different''.
In this article, he emphasized the fact that systems with many degrees of freedom
can often give rise to surprising collective phenomena that transcend the sorts of expectations
that arise from studies based on systems with only a few degrees of freedom.
While Anderson's primary interest was in condensed-matter physics (and indeed
his examples focused on phenomena such as superconductivity), we believe
a similar set of lessons may apply to the dark sector.
Indeed, the common paradigm for the dark sector presupposes that the dark matter
within the universe is comprised of only one, or a few, dark-matter states.
In this article, by contrast, we highlight several of the surprising ``collective'' phenomena
that can emerge if the dark sector is composed of a large
number 
of dark states, each with its
own mass, lifetime, and cosmological abundance.

Of course, many models of decaying dark matter transcend the canonical WIMP or axion frameworks and populate new regions
of the dark-matter parameter space or
give rise to modified signatures.  However perhaps none do so as dramatically as those that arise within the Dynamical Dark Matter (DDM) framework~\cite{Dienes:2011ja}.  Indeed, as we shall see, the DDM framework intrinsically relies on the supposition that the number of states in the dark sector is large and perhaps even infinite, all while satisfying all known phenomenological constraints.   In this article, we shall therefore adopt the DDM framework
as our exemplar and focus on the surprising ``collective'' features that can arise within such a framework.
Given these phenomena, we shall then extract 13 different ``take-away lessons''
that we believe should be borne in mind when contemplating future theoretical possibilities
for dark-matter physics. 

We stress that our purpose in this article is not to provide a review of theories with non-minimal dark sectors and/or multiple-component dark matter.  Indeed, there is already a huge literature on these topics to which we refer the reader for further information.
{\it Rather, our goal here is to highlight the extreme limit in which large numbers of dark-sector states conspire together to produce new and unexpected collective phenomena. }
We therefore adopt  DDM as the ideal theoretical framework within which to discuss these possibilities.

%=========================================================

\FloatBarrier
\section{\large The Dynamical Dark Matter (DDM) Framework as Exemplar:\\  Motivation and General Overview \label{sec:dynDM}}

The traditional view of dark matter asserts that one (or several) dark-matter particles $\chi$ carry the entire dark-matter abundance $\Omega_\chi=\Omega_{\rm CDM} \approx 0.26$.  Such particles must be hyperstable, with lifetimes exceeding the age of the universe by many orders of magnitude, usually $\tau_\chi\gtrsim {\cal O}(10^{28})\,{\rm s}$.  This bound arises because any such particle which decays too rapidly into Standard-Model states is likely to upset BBN and and light-element abundances, and also leave undesirable imprints in the CMB and diffuse photon/X-ray backgrounds.  Stability is thus critical for traditional dark matter.  Indeed, the resulting theory is essentially ``frozen in time'', with quantities such as $\Omega_{\rm CDM}$ remaining fixed.

Let us however imagine that the dark sector of the universe consists of $N$ states, with $N\gg 1$ --- {\it i.e.}\/, an entire {\it ensemble}\/ of states.  In this case, no individual state needs to carry the full abundance $\Omega_{\rm CDM}$ as long as the sum of their abundances continues to match $\Omega_{\rm CDM}$.  In particular, these individual components  can carry a wide variety of abundances, some large but some very small.  {\it However, a given dark-matter component need not be stable if its abundance at the time of its decay is sufficiently small.}\/ Indeed, a sufficiently small abundance assures that the disruptive effects of the decay of such a particle will be minimal, and that all constraints from BBN, CMB, {\it etc.}\/, will continue to be satisfied.
For this reason, a multi-component dark sector need not require stability for each component.  Instead, {\it we need simply require that states with larger abundances have smaller decay widths, but states with smaller abundances can have larger decay widths.  As long as decay widths are generally balanced against abundances in this way across our entire dark-sector ensemble, all phenomenological constraints can be satisfied.}

%=============BEGIN FIGURE=================%
\begin{figure}[h!]
    \begin{center}
    \mbox{\hskip -0.15 truein\includegraphics[width=0.54\textwidth, keepaspectratio]{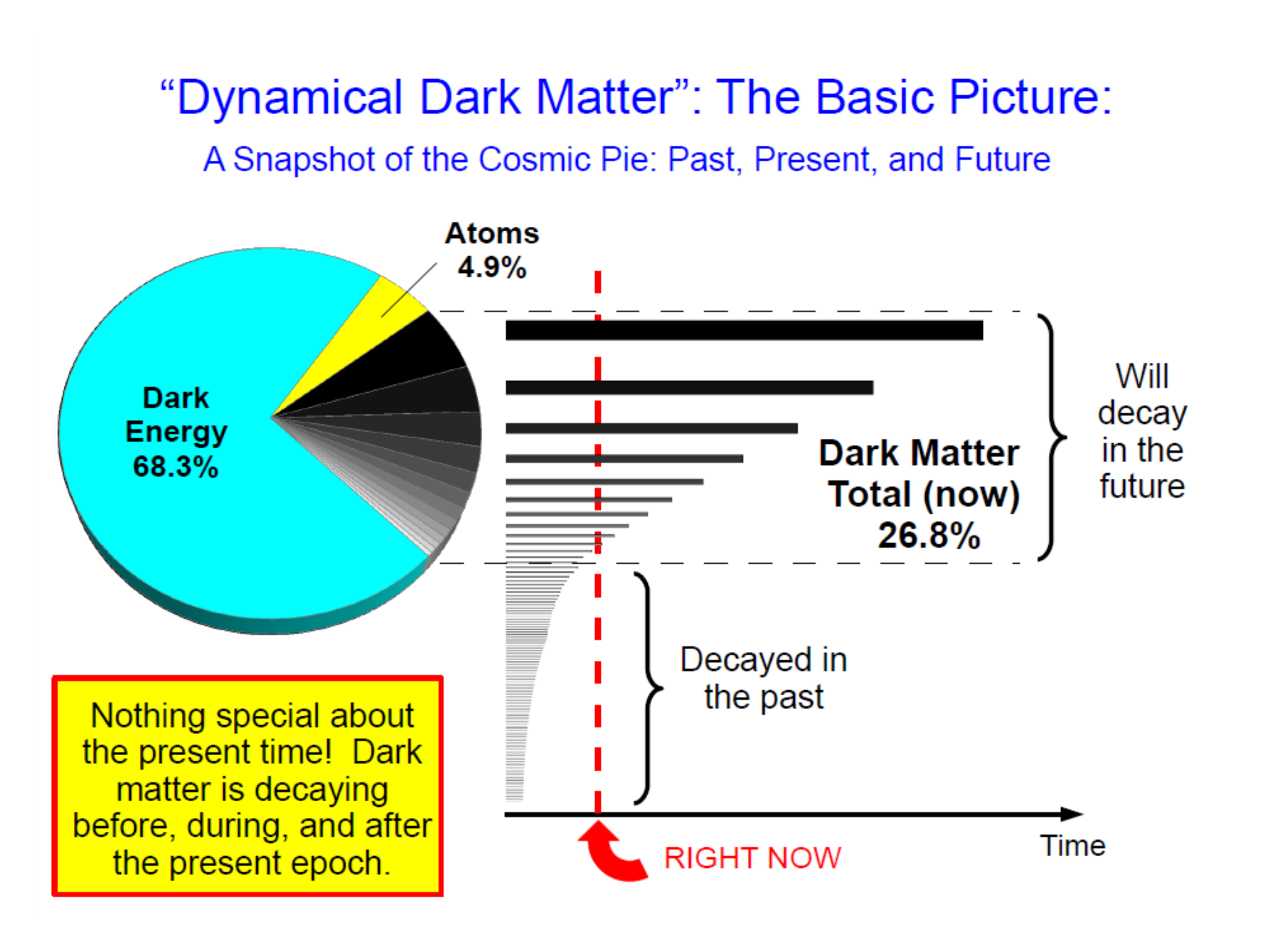}
     \hskip -0.28truein
    \includegraphics[width=0.54\textwidth, keepaspectratio]{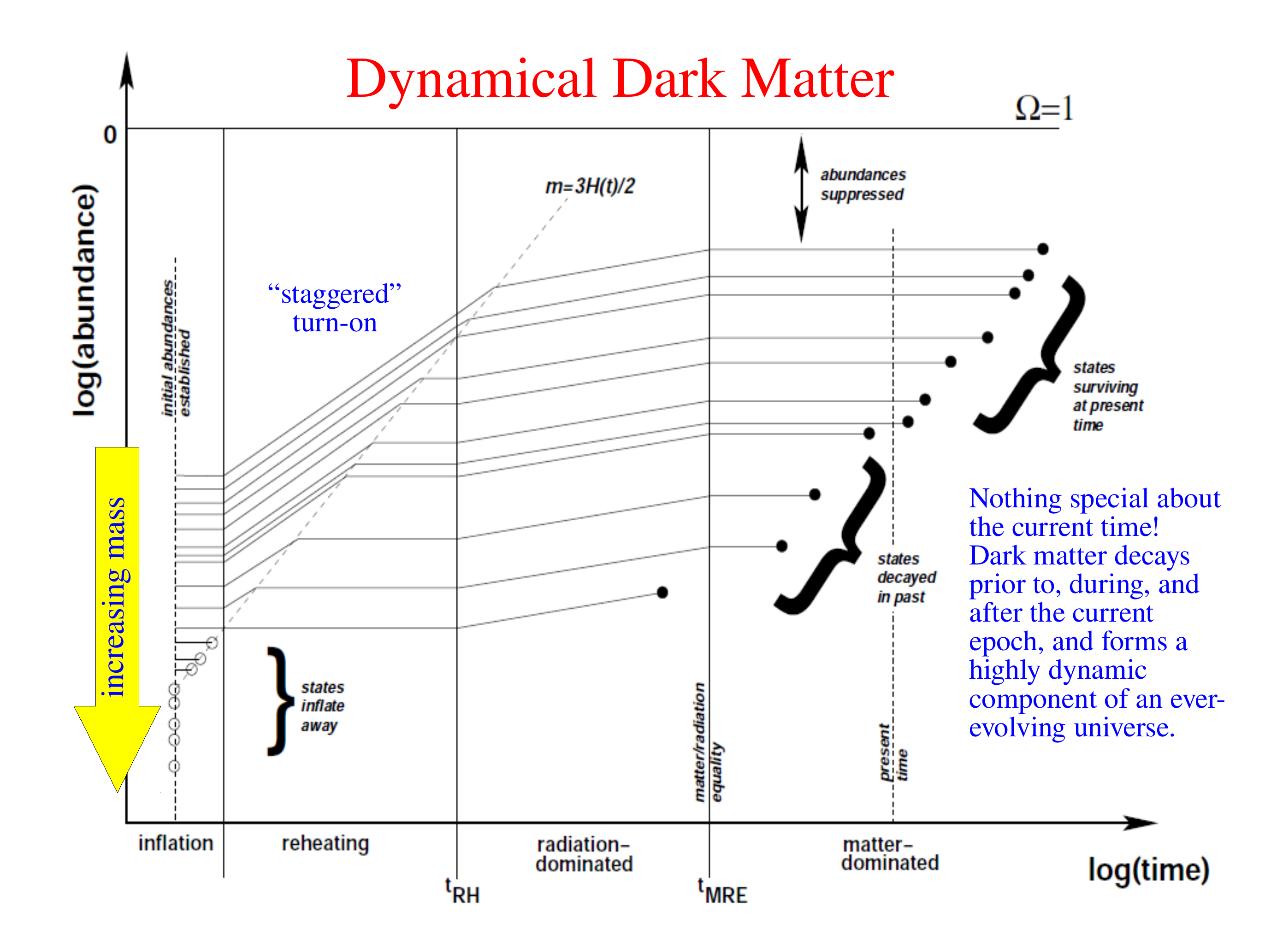}}
        \caption{
        The basic features of the 
        DDM framework.  
        {\it Left panel}\/:  The dark sector of the cosmic pie is composed of a number of different components which each propagate forward in time until eventually decaying.  In each case, states with larger abundances are presumed to have longer lifetimes, and thus decay later than 
        those with smaller abundances. 
        {\it Right panel}\/:  The individual dark-sector abundances are sketched
        as functions of time through different epochs of the early universe. This panel is taken from Ref.~\cite{Dienes:2011ja}, where further details are provided.
        We thus see that dark-matter decays are occurring throughout the history of the universe, leading to a very ``dynamical'' dark sector.}
          \label{introfigs}
    \end{center}
\end{figure}
%===============END FIGURE=================%

This, then, is the Dynamical Dark Matter (DDM) framework~\cite{Dienes:2011ja}.
In a nutshell, the DDM framework  posits that the dark matter in the Universe comprises a vast ensemble of interacting fields with a variety of different masses, lifetimes, and cosmological abundances. Moreover, rather than imposing stability for each field individually, the DDM framework rests upon a balancing of
lifetimes against cosmological abundances across the entire ensemble.
For this reason, individual constituents of the DDM ensemble are decaying {\it throughout}\/ the evolution of the Universe, from early times until late times and even today and beyond. In general, the decay products can involve SM states as well as other, lighter ensemble components.  DDM is thus a highly dynamical scenario in which cosmological quantities such as the total dark-matter abundance $\Omega_{\rm CDM}$ experience a non-trivial time dependence beyond those normally associated with cosmological expansion.    Moreover, because the DDM ensemble cannot be characterized in terms of a single well-defined mass, decay width, or interaction cross section, the DDM framework gives rise to many striking experimental and observational signatures which transcend those usually associated with dark matter and which ultimately reflect the collective behavior of the entire DDM ensemble.

These basic features of the DDM framework are illustrated in Fig.~\ref{introfigs}.  The left panel illustrates the idea that the dark sector of the cosmic pie is composed of a number of different components which each propagate forward in time until eventually decaying.  The right panel illustrates this same behavior as the abundances of these different components pass through the different epochs of the universe, from the time of their production (here sketched as having taken place during inflation) through the present day and beyond.   In each case, states with larger abundances are presumed to have longer lifetimes and thus decay later.  These figures are drawn under the implicit assumption that each dark-matter component of the DDM ensemble decays directly to states in the visible sector.  In such cases, the dark-matter pie slice is continually shrinking while the visible-sector pie slice is growing.  However, it is also possible to have {\it intra-ensemble}\/ decays in which heavier dark-matter states within the ensemble decay to lighter ensemble states (with or without the associated production of visible-sector states).  All of these properties are specified once a particular {\it model}\/ is constructed within the overall DDM framework.

The DDM framework was originally introduced in Ref.~\cite{Dienes:2011ja}, while in Refs.~\cite{Dienes:2011sa, Dienes:2012jb} explicit models within this framework were constructed which were shown to satisfy all known collider, astrophysical, and cosmological constraints.  
Since then, there has been considerable work in fleshing out this framework and exploring its consequences.  One major direction of research consists of analyzing the various signatures by which the DDM framework might be experimentally tested and constrained. These include unique DDM signatures at direct-detection experiments~\cite{Dienes:2012cf}, at indirect-detection experiments~\cite{Dienes:2013lxa,Boddy:2016fds,Boddy:2016hbp}, and at colliders~\cite{Dienes:2012yz,Dienes:2014bka,Curtin:2018ees,Dienes:2019krh,Dienes:2021cxr}.     DDM scenarios can also leave observable imprints across the cosmological timeline, stretching from structure formation~\cite{Dienes:2020bmn,Dienes:2021itb} all the way to late-time supernova recession data~\cite{Desai:2019pvs,Anchordoqui:2022gmw} 
and unexpected implications for evaluating Ly-$\alpha$ constraints~\cite{Dienes:2021cxp}.  Such imprints can also potentially include a non-traditional time-evolution for the Hubble parameter~\cite{Desai:2019pvs,Anchordoqui:2022gmw}.  DDM scenarios also give rise to enhanced complementarity relations~\cite{Dienes:2014via,Dienes:2017ylr} between different types of experimental probes.

A second direction of DDM research over the past decade
involves DDM {\it model-building} --- {\it i.e.}\/, examining the various
ways in which suitably-balanced DDM ensembles can emerge naturally from different models of BSM physics.  It turns out that DDM models often share certain characteristics which allow their consequences to be studied together in a model-independent way.  
For example, the masses $m_\ell$,  corresponding lifetimes $\tau_\ell \equiv 1/\Gamma_\ell$, and cosmological abundances $\Omega_\ell$
of the individual states within the DDM ensemble at the time of their initial production often turn out to be tied together through general (approximate or exact) scaling relations of the form
\beq
       \Gamma_\ell = \Gamma_0\left(\frac{m_\ell}{m_0}\right)^\gamma~,~~~~~
       \Omega_\ell = \Omega_0 \left(\frac{m_\ell}{m_0}\right)^\alpha~~~~~~ {\rm where}~~~
        m_\ell = m_0 + (\Delta m) \ell^\delta~.
\label{scalings}
\eeq
Here $\lbrace \alpha, \gamma, \delta\rbrace$ are scaling exponents and
$\lbrace m_0, \Delta m, \Gamma_0,  \Omega_0\rbrace$ are additional
free parameters.  Such scaling relations usually hold either across the entire DDM ensemble or within those portions of the ensemble that are relevant for various phenomenological questions.
Specific DDM models that one can construct within the DDM framework then correspond to  different sets of parameter values.  Indeed, specific DDM models that have been constructed
include theories involving large extra dimensions, both flat~\cite{Dienes:2011ja,Dienes:2011sa} and warped~\cite{Buyukdag:2019lhh}; theories involving strongly-coupled hidden sectors~\cite{Dienes:2016vei,Dienes:2018tux, Buyukdag:2019lhh}; theories involving large spontaneously-broken symmetry groups~\cite{Dienes:2016kgc}; and even string theories~\cite{Dienes:2016vei}.
Indeed, it is even possible to consider a limit with $m_0>0$ and $\Delta m\to 0$, thereby giving rise to a ``continuum dark matter'' model~\cite{Csaki:2021gfm,Csaki:2021xpy}
in which the DDM masses populate a gapped continuum which can be self-consistently realized in a soft-wall background~\cite{Cabrer:2009we}.
Moreover, the dark states within different DDM models can accrue suitable cosmological abundances in a variety of ways.  These include not only non-thermal generation mechanisms such as misalignment production~\cite{Dienes:2011ja, Dienes:2011sa}
and mass-generating phase transitions in the early Universe~\cite{Dienes:2015bka,Dienes:2016zfr,Dienes:2019chq},
but also thermal mechanisms such as freeze-out~\cite{Dienes:2017zjq}.
In fact, DDM ensembles satisfying the scaling relations in Eq.~(\ref{scalings}) can even give rise to new theoretical possibilities for stable mixed-component cosmological eras~\cite{Dienes:2021woi}.

Each of these different possibilities for model-building corresponds to a specific set of values for the parameters in Eq.~(\ref{scalings}).  For example, realizing the DDM ensemble as
  the Kaluza-Klein (KK) excitations of a dark five-dimensional scalar
  field compactified on a circle of radius $R$ (or a $\mathbb{Z}_2$ orbifold thereof) then results in either  $\lbrace m_0,\Delta m,\delta\rbrace = \lbrace m, 1/R, 1\rbrace$
  or $\lbrace m_0,\Delta m,\delta\rbrace =\lbrace m, 1/(2 m R^2), 2\rbrace$,
  depending on whether $m R \ll1$ or $mR\gg 1$,
  respectively, where $m$ denotes the four-dimensional scalar mass~\cite{Dienes:2011ja, Dienes:2011sa}.   
  Alternatively, taking the DDM ensemble constituents to be  the bound states of a dark strongly-coupled
  gauge theory yields $\delta = 1/2$, where $\Delta m$ and $m_0$ are determined by
  the Regge slope and intercept of the strongly-coupled theory, respectively~\cite{Dienes:2016vei}.
  Thus $\delta=\lbrace 1/2, 1,2\rbrace$ serve as compelling ``benchmark'' values.
  Likewise, $\gamma$ is generally
  governed by the particular
  decay modes associated with the decay rates $\Gamma_\ell$.  For example, if the DDM constituents $\phi_\ell$ decay purely to photons
  through a dimension-$d$ contact operator of the form ${\cal O}_\ell \sim c_\ell \phi_\ell {\cal F}/\Lambda^{d-4}$
  where $\Lambda$ is an appropriate mass scale and where ${\cal F}$ is an operator built from photon fields, we have $\gamma= 2d - 7$.
  Thus values such as $\gamma=\lbrace 3,5,7\rbrace$ can serve as relevant benchmarks.
  Finally, $\alpha$ is governed
  by the original production mechanism for the DDM constituents.
  For example, one typically finds that $\alpha<0$ for misalignment production~\cite{Dienes:2011ja, Dienes:2011sa}, while
  $\alpha$ can generally be of either sign for thermal freeze-out~\cite{Dienes:2017zjq}.

While there are many motivations for considering the DDM framework to be a realistic description of the dark sector, perhaps the most compelling emerges within string theory.  In many string scenarios, the Standard Model is presumed to live on a four-dimensional brane while many other gauge-neutral fields (such as 
gravitons, gravitini, axions, other axion-like particles, string-theory moduli, right-handed neutrinos, and a plethora of dark-matter candidates) 
are presumed to live in a higher-dimensional bulk.   However, because these fields carry no SM quantum numbers, they are effectively ``dark'';   likewise, because they live in a higher-dimensional bulk, they appear as infinite towers of Kaluza-Klein states.   Indeed, within such string theories, nothing stabilizes these KK towers.   As a result, string theories naturally give rise to DDM-like dark sectors of the sort we have outlined here~\cite{Dienes:2011ja,Dienes:2016vei}.  

The DDM framework is thus a rich arena in which to explore new theoretical and phenomenological possibilities for non-minimal dark sectors when the number of dark states grows large and new collective phenomena emerge.  Many of these new possibilities will be outlined below.
However, given the above discussion of the DDM framework, 
we believe that there are two general lessons which we can already draw concerning such non-minimal dark sectors.

\lesson{1}{
      The dark sector need not have a unique mass, decay cross-section, or cosmological abundance.
       Accordingly, it may be necessary to rethink the manner in which  experimental or observational bounds on the dark sector are typically quoted --- including those which are generally plotted
      in the mass/cross-section $(m_\chi,\sigma_\chi)$-plane.  Indeed, it may be necessary to shift
       to more general parameters --- such as the scaling exponents and associated variables outlined in Eq.~(\ref{scalings}) --- which allow for more general dark sectors.
}

\lesson{2}{
   Stability need no longer be the governing principle for the dark sector.   Indeed, dark matter may
     be decaying {\it throughout}\/ the history of the universe, so long as the cosmological abundances of the
    individual dark-matter components are properly balanced against their decay widths.
         This balancing between lifetimes and abundances is actually the fundamental principle governing the dark sector,
       and reduces to the traditional notion of dark-matter stability 
        only in the limit in which the number of states in the dark sector is taken to one.
}

%=====================================================================================
\FloatBarrier

\section{\large Lessons Regarding the Experimental and Observational Signatures of the Dark Sector  \label{sec:signatures}}

We now turn to lessons concerning the potential experimental and/or observational signatures of such non-minimal dark sectors.
As we shall see, these span the space from signatures accessible to
collider experiments to those accessible via direct-detection experiments as well as indirect-detection observations.
As we shall see, these signatures are often extremely different from what might be expected from single-component dark sectors.
Non-minimal dark sectors also lead to significant enhancements in our traditional view of the {\it complementarities}\/
between these different dark-matter discovery channels.

%=========================================

\FloatBarrier
\subsection{\large Unexpected collider signatures:  ~New shapes for kinematic distributions \hfill \label{sec:colliderkinematics}}

Models within the DDM framework can give rise to a variety of
distinctive signatures at colliders.  One such signature involves
a modification of the distributions of kinematic variables for collider
processes involving substantial missing transverse energy $\met$.
The choice of which such variables are the most auspicious for distinguishing DDM ensembles
from traditional, single-particle dark-matter candidates
ultimately depends on the properties of the DDM ensemble constituents $\chi_\ell$ and the
manner in which they couple to the fields of the visible sector.  

One possibility is that the dark and visible sectors are coupled by a mediator
$\psi$ which is charged under the Standard-Model $SU(3)$ color gauge group.  This possibility
is of particular interest both because the mediator can be produced copiously
through strong interaction at a hadron collider and because the decay of
each mediator particle necessarily yields both dark- and visible-sector particles.
For example, if the mediator $\psi$ is an $SU(3)$-octet fermion and the DDM ensemble constituents $\chi_\ell$
are SM-singlet fermions, $\psi$ will decay primarily via processes of the form
$\psi \rightarrow \overline{q}q\chi_\ell$.  Since all of the ensemble
constituents have identical quantum numbers, any $\chi_\ell$ with a mass
$m_\ell < m_\psi$ can appear in the final state of this decay.  However, since
they have different masses, the distribution of the invariant mass $m_{jj}$
for the resulting pair of hadronic jets is different for each
final-state $\chi_\ell$.  The overall $m_{jj}$ distribution, which receives
contributions from all of these decay processes, can therefore differ
dramatically from the $m_{jj}$ distribution characteristic of a
single-particle dark-matter scenario with the same coupling
structure~\cite{Dienes:2012yz}.

Of course, in cases in which only a small number of the $\chi_\ell$ are
kinematically accessible through $\psi$ decays, these invariant-mass distributions will be distinguished by the presence of a multiple
kinematic edges. 
By contrast, in cases in which the number of kinematically-accessible $\chi_\ell$  is large, the decay phenomenology of the $\psi$
particles will depend more sensitively on the full structure of the DDM ensemble and qualitatively different features can emerge.
{\it Indeed, in such scenarios, individual kinematic edges will no longer be manifest.   Instead, the invariant-mass distributions can exhibit
distinctive smooth shapes not realized in single-particle dark-matter scenarios.}\/

%=============BEGIN FIGURE=================%
\begin{figure}[b!]
    \begin{center}
\mbox{
    \includegraphics[width=0.48\textwidth]{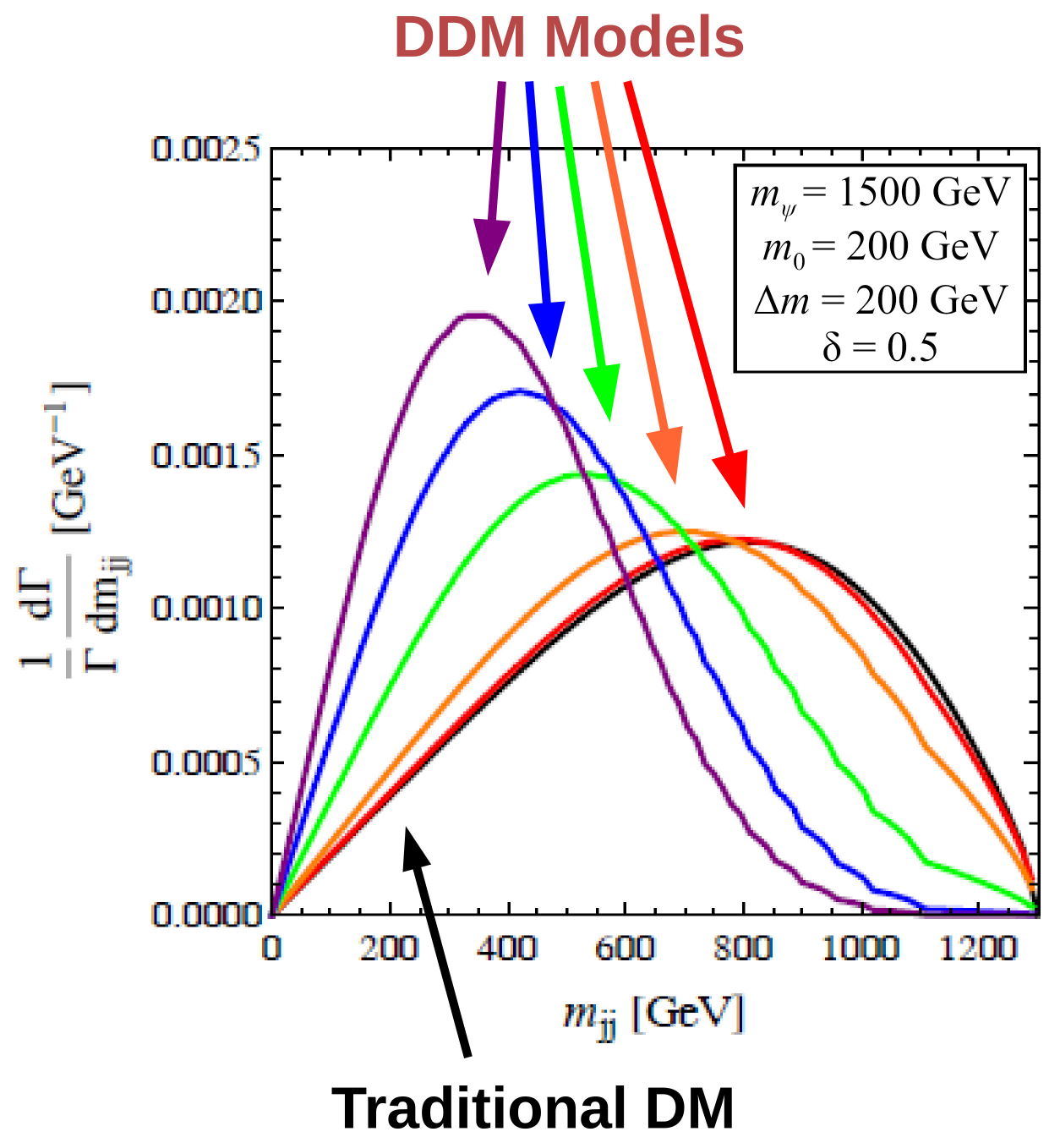}
    ~~~
    \includegraphics[width=0.48\textwidth]{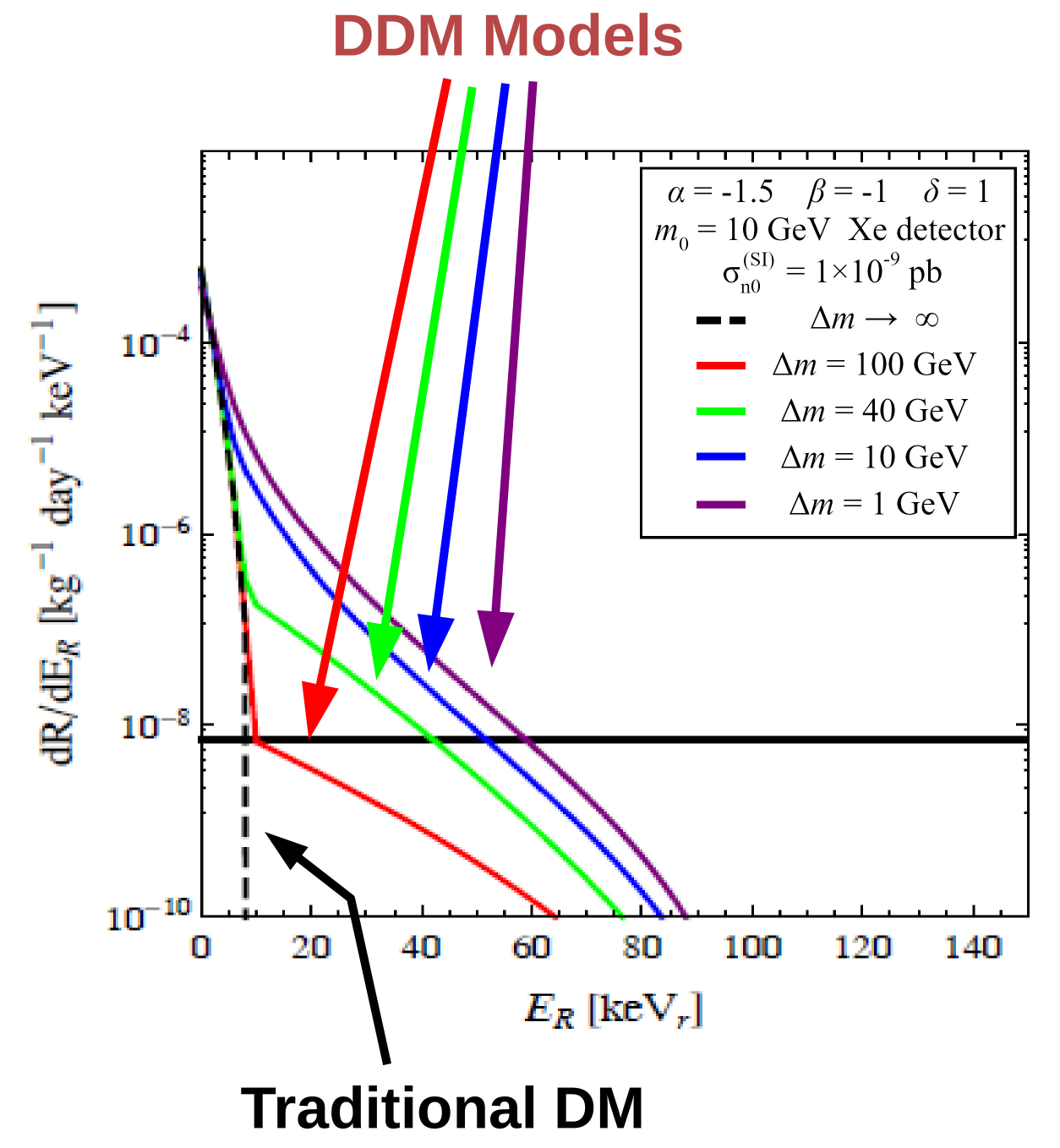}
 }
        \caption{
        In a non-minimal dark sector 
         with many dark-matter states, the distributions associated with kinematic process can be deformed to take 
         smooth, wholly new characteristic shapes that which significantly differ from those associated with single-component
           dark matter. 
          {\it Left panel}\/:   An example from collider physics.  Here we plot the invariant-mass distributions for a
          pair of hadronic jets produced by the decay
          $\psi \rightarrow \overline{q}q\chi_\ell$ of
          a fermionic mediator particle $\psi$ to the constituents
          $\chi_\ell$ of a DDM
          ensemble at a hadron collider.  This figure is taken
          from Ref.~\cite{Dienes:2012yz}.
          {\it Right panel}\/:  An example from a direct-detection experiment, to be discussed in Sect.~\ref{sec:directdetection}.~
          Here we plot the nuclear recoil-energy spectrum that arises when a flux of dark-sector constituents from the galactic halo undergoes elastic spin-independent scattering off nuclei within the detector volume.
          This figure is taken from Ref.~\cite{Dienes:2012cf}.
 }
\label{fig:newcurves}
\end{center}
\end{figure}
%===============END FIGURE=================%

In order to demonstrate this, let us consider decays of the form
$\psi\to \chi_\ell jj$ where each $j$ signifies a jet.  Let us also assume a mass spectrum for the $\chi_\ell$ of the form in Eq.~(\ref{scalings}), and as an example let us also choose  
benchmark values $m_\psi=1.5~{\rm TeV}$, $m_0=\Delta m = 200~{\rm GeV}$, and 
$\delta=0.5$.  We find that $N\approx 43$ ensemble states will be kinematically accessible.  In the left panel of Fig.~\ref{fig:newcurves}, we show the resulting smooth kinematic distributions that emerge if we assume that the $\psi\to \chi_\ell jj$ decays have couplings $c_\ell$ which scale across the DDM ensemble as $c_\ell\sim c_0 (m_\ell/m_0)^\eta$ where the curves shown in red, orange, green, blue, and purple correspond to
$\eta = \lbrace -2,-1,0,1,2\rbrace$, respectively.   
By contrast, in black we show the kinematic distribution that emerges for a traditional dark-matter particle of mass $m_0$.
While the invariant-mass distributions for our multi-component dark sectors closely resemble the traditional single-component distribution for $\eta \ll 0$ (in which case the decay $\psi\to \chi_0 jj$ dominates), we see that this distribution begins to take an entirely different shape as $\eta$ increases.
Indeed, as $\eta$ increases, the decaying $\psi$ parent is effectively allowed to decay into increasing numbers of the DDM constituents.  More of the ensemble thus becomes kinematically accessible, and the invariant-mass distribution assumes a new shape entirely~\cite{Dienes:2012yz}.

we see, then, that these distributions for non-minimal dark sectors can differ significantly from those associated with
single-particle dark-matter scenarios.  For this reason, these distributions can provide a
powerful experimental discriminant between DDM ensembles and more traditional dark-matter candidates.  Indeed, we see from Fig.~\ref{fig:newcurves}
that one feature that may entirely disappear is the sharp kinematic edge
at $m_\psi - m_\chi$, where $m_\chi$ is the mass of the single dark-matter particle.
Then then potentially robs us of a means of 
determining the dark-matter mass --- all of which is consistent with Lesson~\#1, in which we pointed out that no single mass can be associated with a multi-component dark sector.
Interpreting the properties of the underlying DDM ensemble from these new
distributions is thus an important challenge.

Distinctive distributions for other kinds of kinematic variables can also arise within the context of non-minimal dark sectors.  For example, in DDM scenarios in which
the mediator is an $SU(3)$ color triplet scalar $\phi$ and the $\chi_\ell$ are SM-singlet
fermions, $\phi$ decays primarily via processes of the form
$\phi \rightarrow q\chi_\ell$.  Thus, the pair production of these mediators gives
rise to a populations of dijet events with substantial $\met$.  The distributions
of the kinematic variables $\met$ and $M_{T_2}$ obtained for such events in
DDM scenarios can likewise differ significantly from the corresponding distributions
obtained for single-particle dark-matter scenarios~\cite{Dienes:2014bka}.

Although our discussion has focused on only a few examples,
we believe that the general lesson is clear.

\lesson{3}{
     In a collider experiment, the distributions of kinematic variables associated with a given dark sector can exhibit a variety of shapes that are {\it qualitatively different}\/ from those that are typically assumed for a single dark-matter particle.   Indeed, these new shapes can be entirely smooth and even lack expected features such as  kinematic endpoints.
     In a similar way, different kinematic cuts may become appropriate when analyzing such collider data, and these cuts will generally be highly correlated with these kinematic distributions.
     These issues are discussed further in  Refs.~\cite{Dienes:2012yz,Dienes:2014bka}.
}

%===========================================================
\FloatBarrier
\subsection{\large Unexpected collider signatures:  ~Extended mediator-induced decay chains \hfill \label{decaychains}}  

Non-minimal dark sectors can also give rise to rather exotic collider processes, and these in turn can also produce unexpected collider signatures.

As an example, let us consider a process which is very natural from the perspective of a non-minimal dark sector.
In general, in order to suppress the couplings between the dark
and visible sectors, a standard assumption is that these two sectors communicate only through a
mediator.  
{\it However, if the dark sector contains
multiple components with similar quantum numbers, then this mediator also generically gives rise
to intra-ensemble dark-sector decays, with heavier dark components decaying to lighter dark components.}\/  
Indeed, the heavier states within a DDM ensemble can often have significantly shorter
lifetimes than the lighter ensemble constituents which dominate the dark-matter
abundance at the present time, which in turn implies that that a significant number of these heavier $\chi_\ell$ can
be unstable on collider timescales.  As a result, extended decay chains can
arise involving multiple intra-ensemble decay steps, with each step of the decay chain also
producing visible matter.  The visible by-products of such mediator-induced decay chains can
therefore serve as a unique signature of such scenarios.  

%=============BEGIN FIGURE=================%
\begin{figure}[h!]
    \begin{center}
    \includegraphics[width=0.90\textwidth, keepaspectratio]{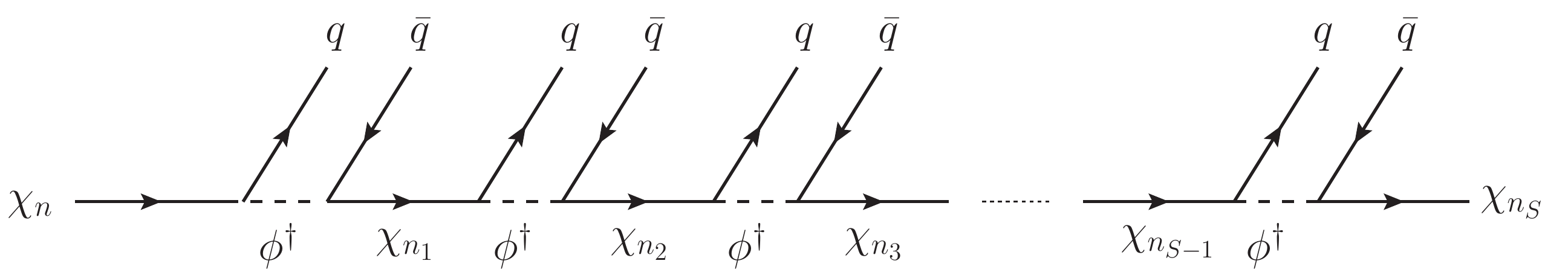}~
        \caption{
            An extended decay chain that can arise due to the existence of a non-mimimal dark sector.  Such decay chains involve multiple successive intra-ensemble decays of the form
            $\chi_n \rightarrow \overline{q}q \chi_{n'} $, where $\chi_{n'}$ is lighter than $\chi_n$.  Each step of the decay chain shown here is facilitated
            by an off-shell $t$-channel mediator particle $\phi^\dagger$ and yields a pair of hadronic jets.  Such a decay chain ultimately 
            terminates in a
            light collider-stable particle $\chi_{n_S}$ for some $n_S$
            which manifests itself as $\met$.
            This diagram is taken from Ref.~\protect\cite{Dienes:2019krh}.
            }
            \label{fig:SwathChain}
    \end{center}
\end{figure}
%===============END FIGURE=================%

In Fig.~\ref{fig:SwathChain}, we show an example of such a decay chain in which each decay step proceeds through an off-shell $t$-channel mediator
particle $\phi$ and yields a SM quark/antiquark pair as well as a single lighter ensemble
constituent.
Extended decay chains of
this sort can even give rise to events involving large jet multiplicities, both at the parton and detector levels~\cite{Dienes:2019krh}.

We emphasize that such extended decay chains
can arise in a variety of different theoretical contexts that need not necessarily involve dark-matter physics~\cite{
Strassler:2006im,
Strassler:2006qa,
Martin:2007gf,
Giromini:2008xh,
Strassler:2008jq,
Juknevich:2009ji,
Juknevich:2009gg,
Craig:2015pha, 
Schwaller:2015gea,
Cohen:2015toa,
Knapen:2016hky,
Park:2017rfb,
Cohen:2018cnq,
DAgnolo:2019cio}.
Indeed, 
  one would not normally interpret such decay chains as having any relation to dark matter, given that the dark matter that exists today must be comprised of one or more particles whose lifetimes are sufficiently long that such particles are necessarily collider-stable.
However, within the context of a non-minimal DDM-like dark sector, these dark-matter states are merely the lightest components of a unified tower of dark states which also include heavier states with extremely short lifetimes.  Thus such decay chains can actually arise from (and thereby provide evidence of) a non-minimal dark sector --- even if the individual steps of the decay chain involve particles that are too short-lived to comprise the dark matter today.

The above discussion assumes that the decay chain involves only prompt decays --- \ie, that the $\chi_\ell$ states involved in the chain have extremely short lifetimes --- while the final state produced is collider-stable.
{\it However, it is possible for additional experimental signatures to arise
from the decays of $\chi_\ell$ ensemble constituents whose  lifetimes $\tau_\ell$ lie between these extremes.}\/
Such ensemble constituents
can then be considered long-lived particles (LLPs), and the the decays of such particles with proper decay lengths
$c \tau_\ell \sim \mathcal{O}(10\,\mathrm{mm} - 100\,\mathrm{m})$ can give rise to
macroscopically displaced vertices within a collider detector.
The successive intra-ensemble decays of
such states as in Fig.~\ref{fig:SwathChain} can then give rise to 
``tumblers''~\cite{Dienes:2021cxr} --- \ie, sequences of displaced vertices
arising from sequential decays within the same decay chain.  Kinematic
correlations between the momenta of the visible particles in tumbler events
can then be used in order to distinguish tumblers from other events involving multiple
displaced vertices.

\lesson{4}{If the dark sector contains
multiple components with similar quantum numbers, then  mediators which connect this dark sector to the visible sector can also give rise to mediator-induced decay chains in which heavier dark components decay to lighter dark components and also simultaneously produce large numbers of states in the visible sector. Such decays may either be prompt, as in Ref.~\cite{Dienes:2019krh}, or correspond to displaced vertices, as in Ref.~\cite{Dienes:2021cxr}, the latter giving rise to so-called ``tumbler'' events.  Thus, contrary to expectations, the appearance  of such decay chains may well be associated with a dark sector --- even though such decay chains involve particles that are too short-lived to comprise the dark matter today.   For more information, see Refs.~\cite{Dienes:2019krh, Dienes:2021cxr}.
}

%===========================================================
\FloatBarrier
\subsection{\large Unexpected collider signatures:   ~Non-minimal dark sectors at dedicated LLP detectors \hfill \label{LLPs}} 

There is yet another unique possibility for collider signatures associated with non-minimal dark sectors.
Of course, ensemble constituents with lifetimes even longer than those considered above will appear only as $\met$
within the main collider detector wherein they were initially produced.   However, dark-sector constituents with lifetimes 
$c \tau_\ell \sim \mathcal{O}(10\,\mathrm{m} - 10^7\,\mathrm{m})$ could nevertheless lead to signals at dedicated LLP detectors such as
MATHUSLA~\cite{Chou:2016lxi,Curtin:2018mvb}, FASER~\cite{Feng:2017uoz},
and Codex-b~\cite{Gligorov:2017nwh,Aielli:2019ivi}.   MATHUSLA, for example,
is capable of probing regions of DDM parameter space inaccessible to
the ATLAS and CMS detectors themselves~\cite{Curtin:2018mvb,Curtin:2018ees}.
Moreover, correlating information obtained from LLP searches at MATHUSLA
with information obtained from a variety of searches at the main CMS detector
can yield further insight into the structure of a DDM ensemble and the
properties of its constituents.

It would not normally be expected that a dedicated LLP detector such as MATHUSLA or FASER should be relevant for dark-matter physics. Indeed, traditional dark-matter candidates are presupposed to have such long lifetimes that their ``decay lengths'' $c\tau$ greatly exceed the range to which such LLP detectors are sensitive.  However, within a non-minimal DDM-like dark sector, dark states of many different lifetimes are all locked together within a common ensemble whose properties are governed collectively, such as through the scaling relations in Eq.~(\ref{scalings}).   By observing the extra-ensemble decays of those dark-sector states whose lifetimes may lie within the range to which these LLP detectors are sensitive, it thereby becomes possible for such LLP detectors to indirectly constrain the properties of the dark matter today.   Indeed, LLP detectors such as MATHUSLA and FASER have extremely small backgrounds, thereby providing ideal environments for such studies.

\lesson{5}{Contrary to conventional expectations, dedicated LLP detectors such as MATHUSLA and FASER can provide important constraints on dark-sector physics.  Such detectors could therefore potentially serve as 
discovery and diagnosis tools for non-minimal dark sectors. Moreover,  because of their extremely small backgrounds, 
such LLP detectors may well be the
first or only discovery opportunities for non-minimal dark sectors --- especially if a majority of the 
dark-sector states have decay lengths exceeding
the sizes of the traditional main-line detectors.  For more 
details, see Refs.~\cite{Curtin:2018ees, Feng:2022inv}. }

%===========================================================
\FloatBarrier
\subsection{\large Unexpected signatures at direct-detection experiments \hfill \label{sec:directdetection}}

Many of the lessons learned for the collider signatures of non-minimal dark sectors also apply for the signatures of these sectors in direct-detection experiments.   This is especially true of Lesson~\#3.  

In a direct-detection experiment, the incoming dark-matter particle(s) scatters off a nucleus associated with the substance ({\it e.g.}\/, argon or xenon) contained within the detector volume.    One then measures the recoil-energy spectrum of this nucleus and thereby aims to deduce the mass of the dark-matter particle involved in the scattering.  If only one WIMP-like particle species comprises the dark matter, then a characteristic recoil-energy spectrum can be expected.   This recoil-energy spectrum falls extremely rapidly as a function of the recoil energy.   However, if the galactic halo contains significant abundances of many dark-sector ensemble constituents, our detector may be subjected to a flux of dark-sector particles whose relative flux fractions depend on the physics of the associated halo.

In the right panel of Fig.~\ref{fig:newcurves}, we illustrate the nuclear recoil-energy spectra that arise when the dark sector comprises a DDM ensemble.   We assume that the masses and abundances of the different dark-matter components in the galactic halo scale as in Eq.~(\ref{scalings}), and likewise we assume that each such state experiences an effective coupling to nuclei which scales similarly with a scaling exponent $\beta$.  We also assume that 
these couplings are dominated by elastic, spin-independent $\chi_\ell$/nucleon interactions.  Holding the $\chi_0$/nucleus elastic scattering cross-section $\sigma^{(0)}_{\rm SI}$  fixed
(where $\chi_0$ is the lightest ensemble constituent), we find the nuclear recoil spectra
shown, where the different colors correspond to different values of the mass-splitting parameter $\Delta m$ in Eq.~(\ref{scalings}).

As $\Delta m\to \infty$, only the lightest ensemble state contributes to scattering within the detector, thereby reproducing the traditional nuclear recoil spectrum.   However, as $\Delta m$ decreases, increasingly many ensemble states can scatter with the nuclei and thereby contribute to the nuclear recoil spectra.  For relatively large values of $\Delta m$, this gives rise to resolvable kinks in the recoil-energy spectra.  {\it However, once $\Delta m$ becomes sufficiently small, an entirely new shape for the nuclear recoil spectrum begins to emerge --- one which is completely smooth and which no longer resembles that for a single dark-matter particle.}\/  Measuring such spectra therefore becomes a powerful experimental way of identifying the existence of a non-minimal dark sector, and the precise shape that emerges can be exploited to learn
about the collective properties of the entire dark sector.

\lesson{6}{Within a direct-detection experiment, the individual dark-sector constituents which comprise a non-minimal dark sector can conspire to provide unexpected shapes for the corresponding nuclear recoil-energy spectra which cannot be obtained for single dark-matter components.   Indeed, as evident from the right panel of Fig.~\ref{fig:newcurves}, these recoil spectra can generally be wider, signalling enhancements in the recoil fluxes per unit energy bin, and pass through inflection points as functions of the recoil energy, all while remaining relatively smooth.  The shapes of these spectra can therefore be used in order to deduce the existence of a non-minimal dark sector and its properties within the galactic halo.
For more information, see Ref.~\cite{Dienes:2012cf}. 
}

%================================================================
\FloatBarrier

%===========================================================
\FloatBarrier
\subsection{\large Unexpected signatures at indirect-detection experiments\hfill \label{indirectdetection}}

Indirect-detection experiments are also highly sensitive to the potential
non-minimality of the dark sector.  Indirect-detection
experiments operate by measuring the
fluxes of SM particles which presumably originate as the products of the
annihilation or decay of dark-sector particles within the galactic halo and then
propagate across interstellar distances before being observed.
These SM particles include not only photons
but also a variety of
cosmic-ray particles such as electrons, positrons, nuclei and anti-nuclei,
and even neutrinos.  The corresponding flux spectra can thus provide
an ``indirect'' window into the properties of the dark sector from which
they originated.

Several data anomalies within different photon and cosmic-ray flux
spectra have received significant attention over the past decade.
These include an excess in gamma-rays emanating from the
direction of the galactic center, as observed by FERMI-LAT; an excess
in the flux of cosmic-ray antiprotons observed by AMS-02;  and
a similar excess in cosmic-ray positrons observed by a variety of indirect-detection
experiments including HEAT, PAMELA, AMS-01, FERMI-LAT, and AMS-02.
Possible explanations involving dark matter have been proposed for
each of these anomalies.  However, whether or not the flux spectrum
which results from any model of annihilating or decay dark matter
agrees with the observed spectrum depends on the masses, couplings, \etc.,
of the particle species which constitute the dark matter.
As a result, flux spectra with a particular shape or profile may be
difficult to realize within the context of scenarios
in which a single particle or a small number of particles
constitute the dark matter.  By contrast, scenarios involving large
numbers of dark-matter states can give rise to a far broader array of
flux spectra with qualitative features which are extremely
difficult to produce within the context of single-particle models.

The cosmic-ray positron excess furnishes one motivation for considering
the more general flux spectra associated with large numbers of dark-matter
states.  This flux is higher than anticipated based on estimates of standard
astrophysical backgrounds, suggesting that this flux may indeed be receiving
additional contributions from annihilating or decaying dark-matter particles.
The simplest models of annihilating or decaying dark-matter
particles predict that such a positron flux excess should exhibit a
distinctive peak which then rapidly falls as a function of energy.
Such an abrupt downturn was not observed in early AMS-02 results;
rather, the data suggested that the flux spectrum exhibited a plateau
at high energies.  While more recent AMS-02 data suggest that the cosmic-ray
positron flux may indeed experience a downturn at high energies, exactly
how abrupt this downturn might be remains uncertain.  

Within traditional minimal dark sectors, there are a number of possible
explanations which could account for the absence of an abrupt downturn in the
positron flux spectrum at high energies.  However, these explanations tend to
invoke either the existence of complex dark-matter annihilation or decay processes
involving exotic intermediate states which only subsequently decay into the
observed particles, or dark-matter particles which decay primarily via
three-body processes involving additional fields.  These mechanisms also
require that the dark-matter state(s) have rather heavy TeV-scale masses.
These features are required in order to ensure that the resulting electron
and positron flux spectra are significantly ``softer'' (i.e., broader and
more gently sloped) than those produced by a single dark-matter particle
undergoing a two-body decay directly to SM states.

%=============BEGIN FIGURE=================%
\begin{figure}[t!]
    \begin{center}
    \hskip -0.4 truein \includegraphics[width=0.9 \textwidth, keepaspectratio]{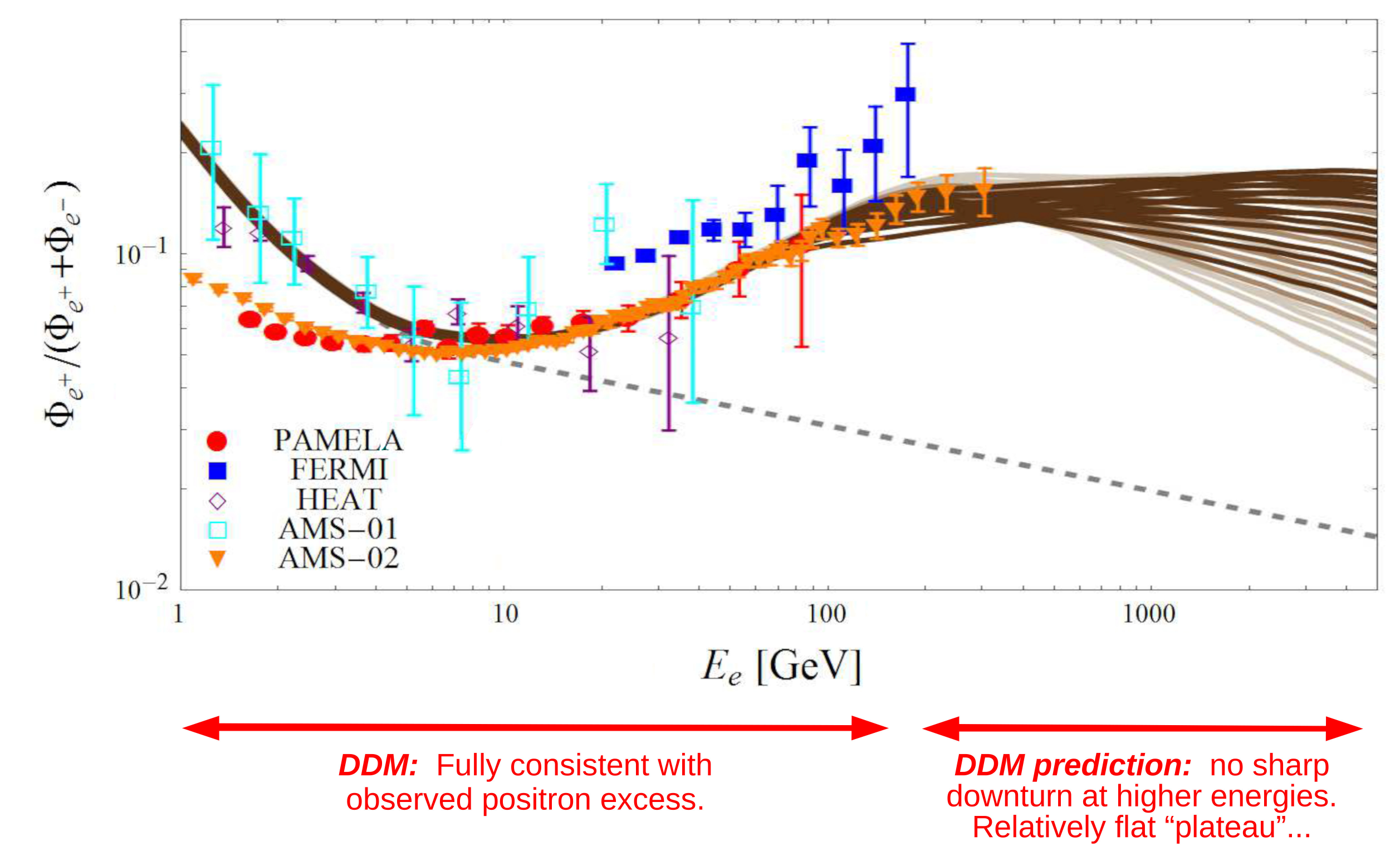}
        \caption{
         Non-minimal dark sectors can change the predicted cosmic-ray fluxes
         that are measured via indirect-detection experiments in ways that are
         difficult to realize with only a single species of dark-matter particle.  
         Here we show the positron fraction of the total $e^+$ and $e^-$ flux
         that would emerge from a DDM-like ensemble in which the dark-sector
         constituents $\phi_\ell$ have masses in the range
         $200\,{\rm GeV}\lsim  m_\ell \lsim 800\, {\rm GeV}$
         and primarily decay into $\mu^+ \mu^-$ pairs.
         Each of the brown curves indicates the flux predicted within a
         different possible DDM ensemble, where we have considered only those
         ensembles which are consistent (to within 3$\sigma$) with existing data.  
         While a single dark-matter candidate would generally predict a
         significant downturn in the positron flux at higher energies,
         the accumulation of positrons from the sequential decays of the
         different components of a non-minimal dark-sector ensemble continues
         to support the positron flux out to much higher energies --- precisely
         as was suggested by early AMS-02 results released in 2013.
       This figure is taken from Ref.~\cite{Dienes:2013lxa}.
       }  
          \label{fig:positron_excess}
    \end{center}
\end{figure}
%===============END FIGURE=================%

By contrast, non-minimal dark sectors can accomplish this feat with
significantly lighter dark-matter constituents undergoing simple
two-body decays to leptons~\cite{Dienes:2013lxa}.
Moreover, this can be done without running afoul of numerous other competing
constraints on decaying dark matter. Even more dramatically, DDM-like
frameworks make the fairly robust prediction that the positron fraction should,
after rising, ultimately level off and then remain roughly constant out to
approximately 1~TeV, without experiencing any sharp downturns.   This behavior
is shown in Fig.~\ref{fig:positron_excess} for DDM ensembles of states with
masses in the range $200\,{\rm GeV}\lsim  m_\ell \lsim 800\, {\rm GeV}$ which
primarily decay into $\mu^+ \mu^-$ pairs.  Indeed, it is ultimately the
successive decays of the different dark-matter constituents of the DDM-like
ensemble which are responsible for this success, with
the rigid structure of dark-sector ensemble --- as encapsulated within
the scaling relations in Eq.~(\ref{scalings}) --- requiring that any excess
positron flux which is already consistent with existing data at low energies
subsequently exhibit a ``plateau'' as a function of energy, declining at best
only slowly at higher energies.  Thus, in this way, the sequential decays
of the ensemble constituents have the net effect of ``supporting'' the
positron flux out to higher energies than could otherwise be achieved with
only one or several dark-matter states.  

Within the context of the galactic-center gamma-ray excess, the profile of the
corresponding flux spectrum can also serve to constrain models of
annihilating or decaying dark matter.  For example, it has been
shown~\cite{Boddy:2016hbp} that the flux spectra which arise in scenarios with
large numbers of dark-matter states are capable of modeling the observed
flux spectrum quite accurately, while at the same time manifesting other
characteristic features which provide clues as to their origin.

\lesson{7}{Non-minimal dark sectors can lead to photon and cosmic-ray flux
spectra which exhibit unorthodox energy-dependences whose features tend to
be broader than those associated with a single dark-matter particle.  These
flux spectra need not exhibit abrupt cutoffs at high energies, but
can actually experience a nearly flat plateau across a wide range of energies
as a result of the sequential decays of the different constituents within the
dark sector.  In this way, non-minimal dark sectors can model observed flux
spectra at indirect-detection experiments in ways that minimal dark sectors
cannot easily emulate.  For further discussion, see Ref.~\cite{Dienes:2013lxa}.}

%================================================================
\FloatBarrier
\subsection{\large New directions in dark-matter complementarity \hfill \label{complementarity}}

As we have discussed, there exist many search strategies for discovering dark matter.   These include the possibility of dark-matter production
at colliders; direct detection of cosmological dark matter through its elastic scattering off ordinary matter at underground experiments; and indirect detection of dark matter through observation of the remnants of the annihilation of cosmological dark matter into ordinary matter at terrestrial or satellite-based experiments. 

At first
glance, these different techniques may seem to be entirely independent, relying on
three independent properties of dark matter, specifically its
amplitudes for production, scattering, and annihilation.
However, these three amplitudes are often related to each
other through crossing symmetries. As a result,
these different search techniques are actually
correlated with each other through their dependence on a
single underlying interaction that couples dark matter
to ordinary matter.  It then follows that the results achieved through any
one of these search techniques will have immediate implications for the others. 
Indeed, this is the origin of the celebrated ``complementarity'' which connects the different existing dark-matter
search techniques.

\begin{figure}[h!]
    \begin{center}
    \mbox{\hskip 0.1 truein \includegraphics[width=0.55\textwidth, keepaspectratio]{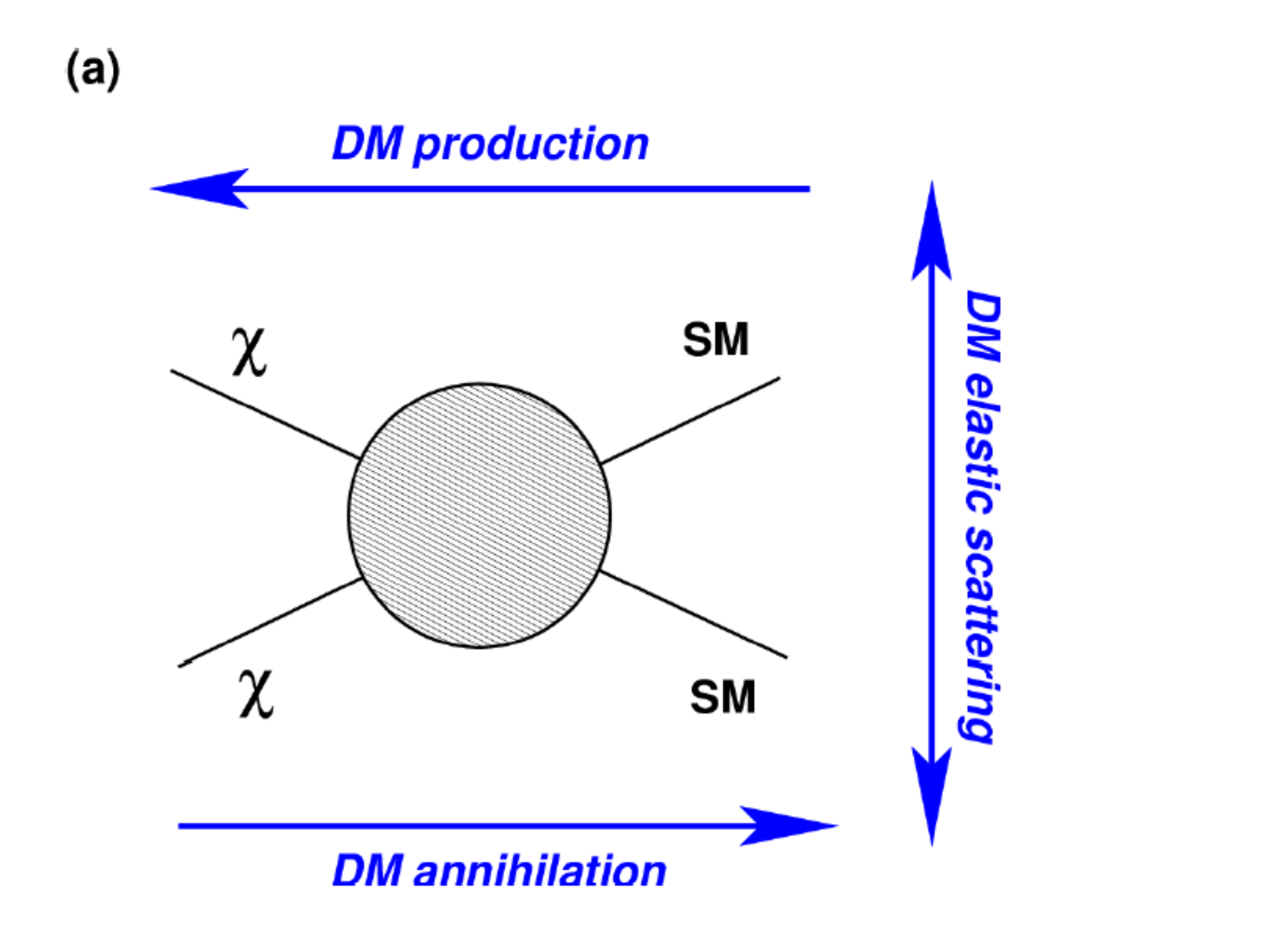}
   \hskip -0.7 truein
    \includegraphics[width=0.55\textwidth, keepaspectratio]{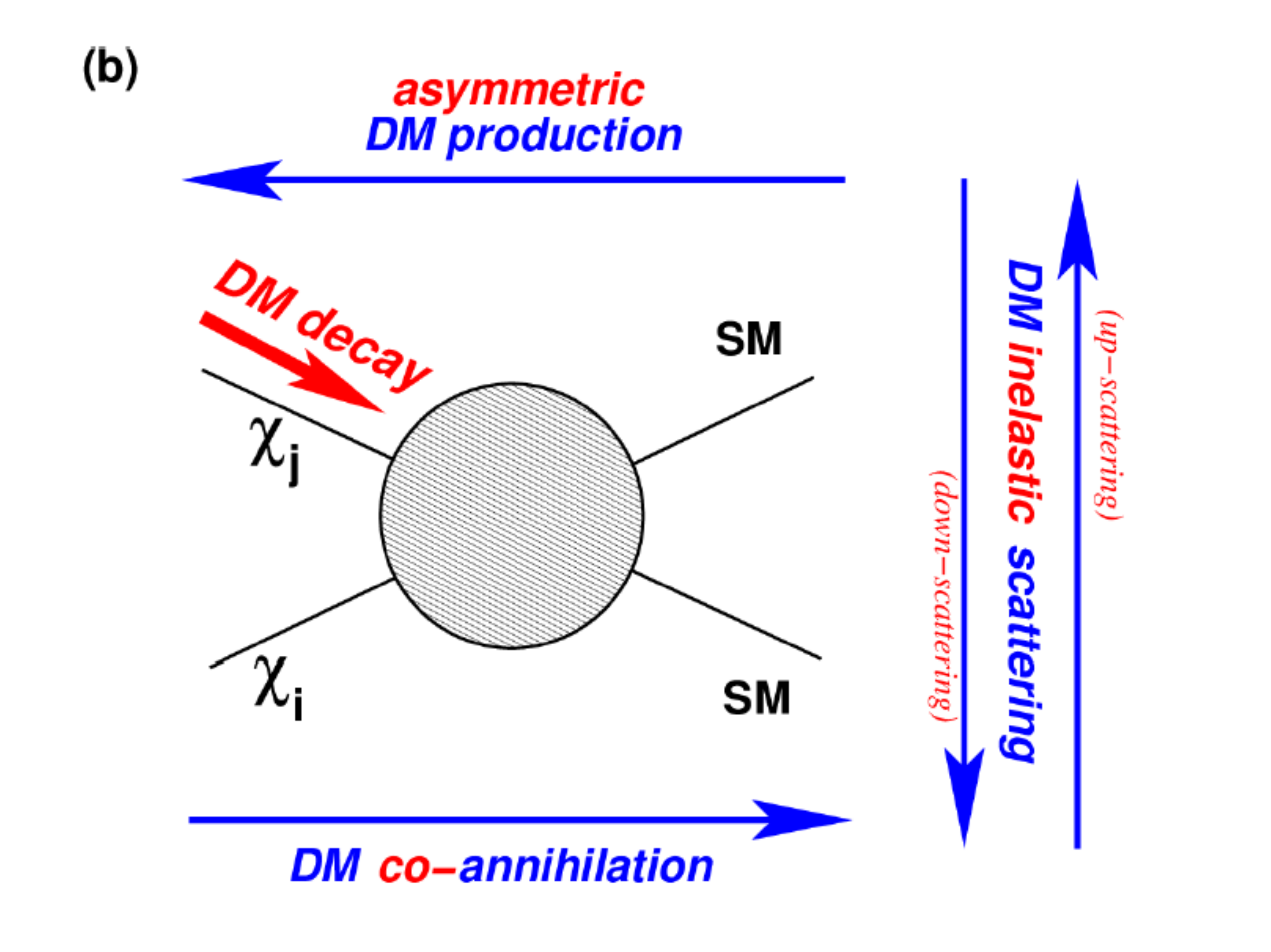}}
        \caption{
         A new direction for dark-matter complementarity. {\it Left panel}\/:  Considering the four-point interaction shown With only a single dark-matter state $\chi$, we see that dark-matter production, annihilation, and elastic scattering correspond to the three different directions for the flow of time indicated by blue arrows.  {\it Right panel}\/:  For a non-minimal dark sector, by contrast, the two $\chi$ legs of this diagram may correspond to dark-matter particles $\chi_i$ and $\chi_j$ with differing masses $m_j> m_i$.   In such cases, dark-matter annihilation becomes dark-matter {\it co}\/-annihilation, the dark-matter production becomes asymmetric, and the elastic scattering becomes inelastic and corresponds to either up-scattering or down-scattering.  However, we also see that a new ``diagonal'' direction for the flow of time becomes possible, as shown.  This corresponds to dark-matter {\it decay}\/ $\chi_j\to \chi_i\cdot {\rm SM}\cdot {\rm SM}$.
     }   
          \label{fig:complementarity}
    \end{center}\end{figure}

For a traditional dark-matter particle $\chi$,
this complementarity is often explained through a diagram such as that shown in the left panel of Fig.~\ref{fig:complementarity} in which an effective interaction involving two $\chi$ particles and two Standard-Model (SM) particles is shown.   For such an interaction, each of the different processes and search strategies outlined above simply corresponds to a different direction for the flow of time (as indicated via the blue arrows):  time flowing horizontally to the right corresponds to dark-matter annihilation, while time flowing horizontally to the left corresponds to dark-matter production (as in a collider) while time flowing vertically (either upward or downward) corresponds to dark-matter elastic scattering.  Indeed, viewed from this perspective, each of the different search strategies discussed above potentially provides a different set of constraints on the properties of this same fundamental interaction, such as its characteristic mass scale $\Lambda$.   Note that such a diagram is symmetric under inversion of the vertical direction.

%=============BEGIN FIGURE=================%
%===============END FIGURE=================%

This situation changes dramatically (and indeed becomes far richer) for a non-minimal dark sector.
In particular, if we assume that the dark sector consists of
at least two different dark-matter components $\chi_i$ and $\chi_j$ with different masses $m_j> m_i$, then our previous 
$(\chi\cdot \chi\cdot {\rm SM}\cdot {\rm SM})$ interaction can now become a
$(\chi_i \cdot \chi_j \cdot {\rm SM}\cdot {\rm SM})$
interaction.
In such cases, the resulting complementarity relations will differ from the single-particle case in two fundamental ways.   First, the kinematics associated with each
of the traditional complementarity directions is altered:  the dark-matter annihilation of one dark particle
against itself or its antiparticle becomes {\it co-annihilation}\/
between two different dark species; 
the dark-matter production becomes asymmetric rather than
symmetric; and the dark-matter scattering --- previously exclusively elastic --- now becomes
inelastic, taking the form of either ``up-scattering''
or ``down-scattering'' depending on whether it is the
incoming or outgoing dark-matter particle which has
greater mass (thereby reflecting the breaking of the vertical inversion symmetry of this diagram).   These kinematic changes are illustrated in red text in the right panel of Fig.~\ref{fig:complementarity}, 
and can significantly affect the phenomenology
of the corresponding processes. 

But perhaps even more
importantly, an entirely new direction for dark-matter
complementarity also opens up.   This is the possibility of
dark-matter (intra-ensemble) {\it decays}\/ from heavier to lighter dark-matter
components.   {\it Indeed, this process corresponds to a ``diagonal'' direction for the imagined flow of time, as shown
in Fig.~\ref{fig:complementarity},
and thus represents an entirely new direction
for dark-matter complementarity}\/! Indeed, such a diagonal direction was
not available in single-component theories of dark matter
due to phase-space constraints, and is ultimately driven
by the non-zero mass difference $m_{ij}\equiv m_j-m_i$ between the associated
parent and daughter dark-matter particles.  

At a practical level, this enhanced complementarity extends our parameter space from 
one direction parametrized by $\Lambda$ to a {\it plane}\/ parametrized by $(\Lambda, m_{ij})$.  Here $\Lambda$ might generally represent the scale of new physics associated with the four-point diagram under discussion, or other parameters associated with this process.  From this vantage point, traditional complementarity analyses can now be seen as the 
$m_{ij}\to 0$ limit within a larger structure.
Indeed, the constraints from the diagonal ``decay'' processes can be truly complementary to those from the other directions within Fig.~\ref{fig:complementarity} in the sense that the constraints from such intra-ensemble decays might overlap the previous constraints within  certain regions of the larger $(\Lambda, m_{ij})$-plane while also providing constraints at entirely new locations
within this plane which are beyond the reach of the other search strategies.  Explicit examples of this phenomenon within such enlarged parameter spaces are given in Refs.~\cite{Dienes:2014via, Dienes:2017ylr}.

Of course, we have already considered some of the constraints that might emerge from intra-ensemble dark-matter decays (see, \eg, Lesson~\#4 above).  
However, our point here is that 
these decays are actually part of a larger complementarity, probing the same effective dark/visible interaction, and that this enhanced complementarity can generically be an important ingredient in probing and constraining the parameter spaces of theories with non-minimal dark sectors. 

\lesson{8}{Non-minimal dark sectors generically lead to enhanced complementarities between different search strategies, effectively introducing a new, fourth ``diagonal'' direction corresponding to intra-ensemble dark-matter decay.  This feature is illustrated in Fig.~\ref{fig:complementarity}, and can provide additional constraints on the dark-matter parameter apace which are indeed complementary to the others, often inhabiting regions of dark-matter parameter space that are beyond those reached by the traditional search strategies based on dark-matter production, scattering, and annihilation. For further details and examples, see Refs.~\cite{Dienes:2014via, Dienes:2017ylr}.
}

%===============================================================================
\FloatBarrier
\section{\large Lessons Regarding Potential New Phenomenologies for the Dark Sector}

In this section we provide illustrative lessons concerning potential new collective phenomena that can emerge in non-minimal dark sectors and thereby have important ramifications for their eventual phenomenologies.
Our goal, as it has been throughout this work, is to emphasize features that transcend our usual expectations and which could thereby point the way to new research directions.

%======================================================
\FloatBarrier
\subsection{\large Decoherence:  ~A new way for the dark sector to stay dark \hfill \label{decoherence}}

Despite the existence of many different dark-matter components $\phi_\ell$ within a non-minimal dark sector,
it may often happen that only a particular linear combination
\beq 
          \phi' ~\equiv~ \sum_{\ell=0}^N
          \,c_\ell \phi_\ell
\label{linear_combination}
\eeq
couples to the visible sector, where $N\gg 1$.
While there are many scenarios under which such a coupling structure might arise, particularly natural examples include higher-dimensional setups in which the SM lives on a four-dimensional brane embedded within a higher-dimensional bulk.  Indeed, such scenarios are generic within string theory.  In such scenarios, the fields living in the bulk are necessarily neutral under SM gauge symmetries and therefore function as dark matter.   Such fields might include members of the gravity (super-)multiplet,
axions and other axion-like particles,
string-theory moduli, right-handed neutrinos, {\it etc}\/.  From a four-dimensional perspective, each of these higher-dimensional bulk fields appears as an infinite tower of Kaluza-Klein states and thus constitutes a DDM-like dark-matter ensemble.   For any such higher-dimensional field $\Phi$, the most natural coupling to the visible sector living on the brane takes the form 
\beq 
    {\cal L}_{\rm int} ~\ni  ~ \int d^4 x \,d^{n}y  ~\psi_{\rm SM}(x) \,
     \Phi(x,y) \,\delta^n(y-y_\ast)
\label{SMcoupling}
\eeq
where $y$ are the coordinates of $n$ compactified extra spacetime dimensions, where $y_\ast$ is the location of the brane within this higher-dimensional space, and where $\psi_{\rm SM}(x)$ schematically represents a four-dimensional visible-sector field.  
Decomposing $\Phi(x,y)$ into KK modes $\phi_\ell(x)$ and corresponding KK wavefunctions $f_\ell(y)$ appropriate for the compactification geometry via
\beq 
          \Phi(x,y) ~=~ \sum_{\ell=0}^\infty \,\phi_\ell(x) 
           \,f_\ell(y)
\eeq 
and subsequently performing the $y$-integral in Eq.~(\ref{SMcoupling}) then yields a four-dimensional coupling between $\psi_{\rm SM}(x)$ and the particular KK linear combination $\phi'(x)\equiv \sum_{\ell=0}^\infty c_\ell \phi_\ell(x)$ where $c_\ell = f_\ell(y_\ast)$. 

In such situations,
dark-matter production processes that involve the visible sector can  produce 
dark matter only in the  linear combination $\phi'$.   
However, once produced, $\phi'$  will rapidly decohere 
if $N\gg 1$.
This state will therefore no longer be detectable at later times through processes involving visible-sector interactions, since these processes continue to be sensitive to only the linear combination $\phi'$.
The dark-matter state $\phi'$ will thus be essentially invisible
as far as subsequent visible-sector detection is concerned. 

This decoherence phenomenon was first discussed within the context of large extra spacetime dimensions in Ref.~\cite{Dienes:1999gw}, and was subsequently extended in Refs.~\cite{Dienes:2011ja, Dienes:2011sa, Dienes:2012jb}.   It
can easily be understood through elementary quantum mechanics.  
Because each component state $\phi_\ell$ 
has its own mass $m_\ell$ and energy $E_\ell$, 
the different $\phi_\ell$ components within $\phi'$ fall out of phase with each other under time-evolution.  
For $N\gg 1$, the resulting destructive interference then prevents any meaningful
 reconstitution of the original linear combination $\phi'$ at any later time. 
Indeed, the physics behind this decoherence phenomenon is the same as that underlying neutrino oscillations, except that in this case the value of $N$ in Eq.~(\ref{linear_combination}) is presumed large.   In general, letting $P_{\phi'\to\phi'}(t)$ denote the $\phi'$ survival probability, we find that the time-averaged probability $\overline{P}$ generically scales as $1/N$.   We thus find that $\overline{P}\to 0$ as $N\to \infty$.  This occurs even if the masses/energies of the different components happen to be related to each other through rational multiplicative factors. 

%=============BEGIN FIGURE=================%
\begin{figure}[h!]
    \begin{center}
    \mbox{
  \hskip -0.2 truein
\includegraphics[width=0.8 \textwidth, keepaspectratio]{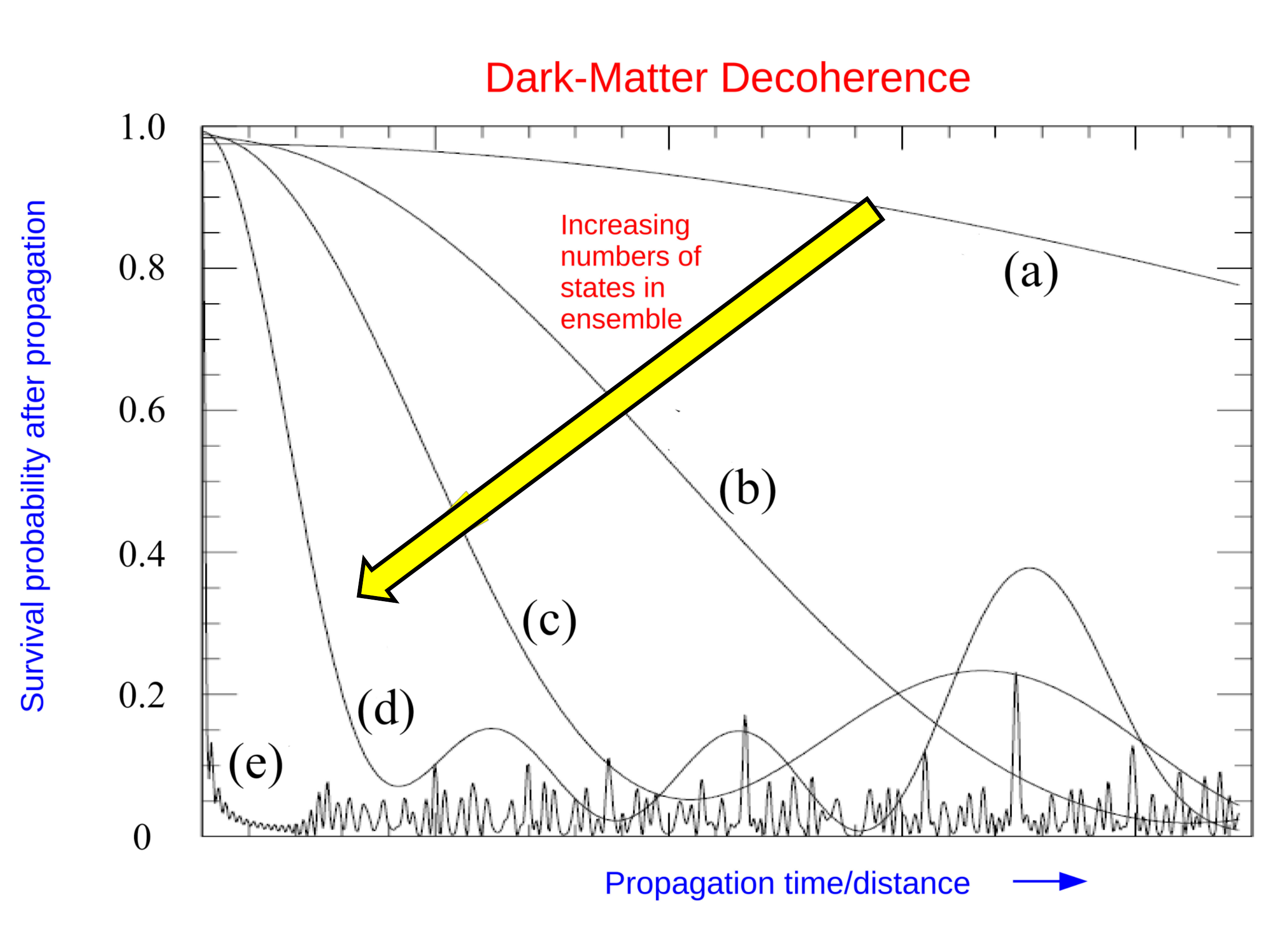}~
 }
        \caption{
        Dark-matter decoherence in a non-minimal dark sector.  If the visible sector couples to only a particular linear combination $\phi'$ of individual mass-eigenstate dark-matter components $\phi_\ell$, then only this linear combination can be produced or detected through processes involving the visible sector.   Here, within a certain model~\cite{Dienes:1999gw}, we show the probability $P_{\phi'\to\phi'}(t)$ that $\phi'$ survives after a certain time has elapsed (or after a certain propagation distance has been traversed) for linear combinations involving $N=1,2,3,5,30$ dark-sector states [curves (a) through (e) respectively].  We see that $\phi'$ decoheres rapidly, even for relatively small values of $N$, and thus becomes essentially invisible as far as subsequent couplings to the visible sector are concerned.  This figure is taken from Ref.~\cite{Dienes:1999gw}.}
    \label{fig:decoh}
    \end{center}
\end{figure}
%===============END FIGURE=================%
   
This decoherence phenomenon is illustrated in Fig.~\ref{fig:decoh}, where we plot the survival probability $P_{\phi'\to\phi'}(t)$ for a  $\phi'$ linear combination drawn from an example in Ref.~\cite{Dienes:1999gw} involving $N+1$ different components
for $N=1,2,3,5,30$.    We see that $\overline{P}\to 0$
as $N\to \infty$.  We also see that $\overline{P}$ is already very small (and thus below expected detector capabilities) even for relatively small values of $N$.    
Decoherence thus provides a novel ``collective'' mechanism through which the dark matter arising within a non-miminal dark sector may remain dark.

\lesson{9}{Within non-minimal dark sectors, dark matter can quickly ``decohere'' and thereby become invisible to subsequent detection within the visible sector, as illustrated in Fig.~\ref{fig:decoh}.  This decoherence phenomenon is fairly generic, emerging as a collective phenomenon involving multiple dark-sector components and requiring nothing more than a setup in which the visible sector couples to only a particular linear combination of these components.   Decoherence thus provides a new mechanism that may help dark matter remain dark.  For further details and phenomenological implications, see Refs.~\cite{Dienes:1999gw, Dienes:2011ja,Dienes:2011sa, Dienes:2012jb,decohtoappear}.}

%======================================================
\FloatBarrier
\subsection{\large Surprises in thermal freeze-out:  ~Rising, falling, and non-monotonic abundance distributions \hfill \label{thermalfreezeout}}

Thermal freeze-out is one of the most widely discussed and widely
exploited methods of abundance generation in the dark-matter literature, providing
a natural and versatile mechanism through which a neutral, weakly-interacting massive particle species
$\chi$ which is initially in thermal equilibrium can acquire
a present-day abundance $\Omega_{\rm CDM} \approx 0.26$.
This mechanism not only underpins the so-called
``WIMP miracle'' but can also yield similar abundances for dark-matter particles which do not
participate in SM weak interactions but which nevertheless have annihilation cross-sections similar to that
of a traditional WIMP.~  The range of dark-matter
masses $m_\chi$ for which thermal freeze-out is typically relevant
spans the range ${\cal O}(1\,{\rm keV}) \lsim m_\chi\lsim {\cal O}(100\,{\rm TeV})$.  The lower limit to this
range reflects the requirement that the dark-matter
candidate be ``cold'' --- {\it i.e.}\/, non-relativistic --- during the
freeze-out epoch, while the upper
limit stems from considerations related to perturbative
unitarity. However, there are ways of circumventing
this upper bound and thereby broadening the window of applicability. For example, this bound is considerably relaxed in
theories in which the dark and visible sectors thermally
decouple well before the freeze-out epoch.

This picture changes drastically for DDM-like non-minimal dark sectors~\cite{Dienes:2017zjq}. First, because each constituent state $\chi_\ell$ has its own mass $m_\ell$, the different states can freeze-out at different times and temperatures.  But even more importantly, the freezing-out of the dark sector ultimately endows each of the individual components with its own abundance $\Omega_\ell$, leading us to think of $\Omega_\ell$ as a {\it function}\/ of the corresponding mass $m_\ell$.   Of course, both of these features ultimately depend on the specific interactions that dominate the couplings between the dark and visible sectors and perhaps even couple the dark sector to itself.  

The usual WIMP miracle is predicated on a simple 2$\to$2 interaction of the form $\overline\chi\chi\to\phi\to \overline\psi_{\rm SM}\psi_{\rm SM}$ where the dark-matter states $\chi_\ell$ are Dirac fermions, where $\psi_{\rm SM}$ denotes a visible-sector Dirac-fermion state, and where $\phi$ is an $s$-channel scalar mediator.  Moreover, the result underpinning the WIMP miracle implicitly assumes that $m_\chi \gg m_\phi, m_\psi$.   For a non-minimal dark sector, this process naturally generalizes to become $\overline\chi_\ell \chi_\ell \to\phi\to \overline\psi_{\rm SM}\psi_{\rm SM}$.   However, because we are now considering the $\chi_\ell$ states to be members of a DDM {\it ensemble}\/, the masses $m_\ell$ of the states in this ensemble can stretch across a large mass range for which the implicit WIMP assumption 
$m_\ell \gg m_\phi$ is no longer always valid.   Thus new features can emerge within the spectrum of abundances, even if we continue to take $m_\psi\ll m_{\phi},m_\ell$.     For example, when $m_\ell \gg m_\phi$, we find $\Omega_\ell\sim m_\ell^2$.  However, when $m_\ell\ll m_\phi$ we find $\Omega_\ell \sim m_\ell^{-2}$.   As discussed in Ref.~\cite{Dienes:2017zjq}, this implies that the abundance {\it falls}\/ as a function of mass for the lighter dark-matter constituents within the dark sector but then {\it rises}\/ as a function of mass for the heavier constituents.   Indeed, in this scenario the minimum abundance is obtained for the constituent with $m_\ell = m_\phi/2$, {\it implying that this portion of the ensemble may not be significantly populated through thermal freeze-out even though both the lighter and heavier parts of the ensemble are populated}\/!   It is especially noteworthy that thermal freeze-out can produce abundances which scale inversely with mass --- at least over a portion of the total ensemble --- since this property helps in creating the conditions for balancing lifetimes against abundances as described in Lesson~\#2.

%=============BEGIN FIGURE=================%
\begin{figure}[b!]
    \begin{center}
    \includegraphics[width=0.75 \textwidth, height=0.3975 \textwidth]{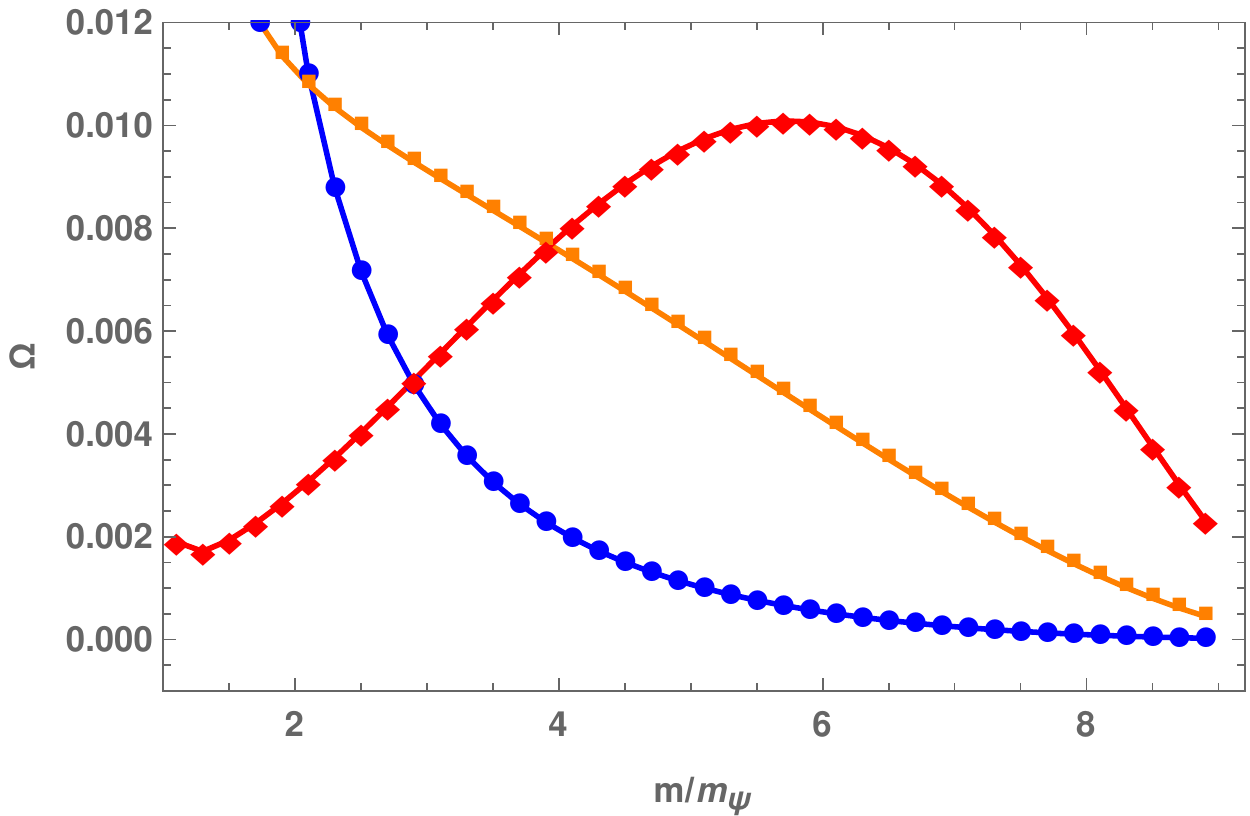}~
        \caption{
        The freeze-out abundances $\Omega_\ell$, plotted as a functions of the ensemble constituent dark-matter masses $m_\ell$
       over the kinematically allowed range $m_\psi \leq m_\ell \leq m_{\phi}/2$
       for the benchmark ratio $m_\phi/m_\psi=20$.   
       As discussed in Ref.~\cite{Dienes:2017zjq}, the different colors correspond to different spins and possible coupling structures for the fields that are involved in the dominant $2\to 2$ freeze-out process, and the abundances in each curve are normalized so that the entire dark sector in each case
          has total abundance $\sum_\ell \Omega_\ell= \Omega_{\rm CDM} \approx 0.26$.
        This figure is taken from Ref.~\cite{Dienes:2017zjq}.
            }
            \label{fig:freezeoutfigure}
    \end{center}
\end{figure}
%===============END FIGURE=================%

More complicated situations can also be imagined.   
For example, in addition to higher-point interactions involving more states, it is also possible to have {\it off}\/-diagonal $2\to 2$ interactions of the form $\chi_\ell \chi_m \to \phi\to \overline\psi_{\rm SM}\psi_{\rm SM}$. These may often (but not always) be suppressed relative to the diagonal interactions.  Likewise, there can also be intra-ensemble annihilation processes of the form $\overline\chi_\ell\chi_\ell \leftrightarrow \overline\chi_m \chi_m$, which we shall also assume to be suppressed.    However, even if we restrict our attention to processes of the form $\overline\chi_\ell \chi_\ell\to \phi\to \overline\psi_{\rm SM}\psi_{\rm SM}$,
it is possible to imagine choosing alternate spins for the $\chi_\ell$, $\phi$, and $\psi_{\rm SM}$ fields as well as alternate Lorentz coupling structures at each vertex.

In Fig.~\ref{fig:freezeoutfigure} we illustrate three of the possible sets of abundances $\lbrace \Omega_\ell\rbrace$ (shown in blue, red, and orange) that can result for different choices of spins and coupling structures.  
Given these results, we see that a wide set of possibilities are possible --- even including $\Omega_\ell$ functions which are not only non-monotonic but which also simultaneously exhibit local maxima as well as local minima.

This non-monotonic behavior in the ensemble abundances $\Omega_\ell$ as functions of the ensemble masses $m_\ell$
gives great flexibility to the DDM model-builder.
Indeed, through these mechanisms it is possible to build models of non-minimal dark sectors in which the individual constituents are endowed with cosmological abundances 
which either rise or fall with mass, or experience local maxima or minima.   
However, the presence of non-monotonicities in these cases also allows for other possibilities.
For example, through appropriate choices of constituent masses $m_\ell$ and the dominant interactions through which they are coupled to the visible sector, one can imagine situations
in which thermal freeze-out yields growing abundances for one dark-matter ensemble
(or one portion of a DDM ensemble)
and yet decreasing abundances for another.
Indeed, we see that it is even possible to preferentially populate only selected portions of an ensemble, and even have these portions be separated by significant energies.
These observations
thus significantly enrich the phenomenological possibilities for
model-building, all while continuing to assume only a single universal coupling between the dark and visible sectors.

\lesson{10}{Thermal freeze-out of  DDM-like non-minimal dark sectors can endow their constituents with richly varying cosmological abundances.  Depending on the properties of the dark sector in question, not only can these abundances rise or fall as a function of mass across a single ensemble, but they can also exhibit non-monotonic behaviors, with certain intermediate portions of the ensemble either becoming dominant or effectively invisible relative to their nearest ensemble neighbors either higher or lower in mass.  Indeed, all of this can be accomplished without varying the underlying interactions between the dark and visible sectors.   For further details, see Ref.~\cite{Dienes:2017zjq}.
}

%======================================================
\FloatBarrier
\subsection{\large Surprises in misalignment production:  ~Dark-matter resonances, re-overdamping, and a slingshot\hfill \label{misalignment}}

If freeze-out is the ``classic'' thermal mechanism for endowing dark matter with a cosmological abundance, then misalignment production is the classic {\it non-thermal}\/ mechanism for achieving the same goal for scalar fields. However, as we shall see, misalignment production is also capable of yielding a number of remarkable new features when a non-minimal dark sector is involved. 

Misalignment production ultimately works by endowing a massless scalar field with a non-zero energy density.
This happens as the result of a {\it mass-generating phase transition}\/ which gives this scalar a non-zero mass $\overline{m}$.   However, it is critical to realize that 
such cosmological phase transitions are generally not instantaneous.  Instead, they unfold across time intervals of non-zero duration.   Thus, under misalignment production, the scalar field accrues a mass according to a squared-mass function $m^2(t)$ which begins at $m^2(t)=0$ at early times prior to the phase transition, ends at some final value $m^2(t)=\overline{m}^2$ after the phase transition, but otherwise evolves smoothly and monotonically from $0$ to $\overline{m}^2$ during the time interval over which the phase transition unfolds.  

This much is standard and is familiar from the case in which $\phi$ denotes an axion.   Indeed, even in the case of the QCD axion, the non-perturbative instanton-induced mass-generating phase transition is not truly instantaneous, but 
instead unfolds smoothly over a non-zero time interval.

Given this, let us now consider how this picture generalizes if we have a non-minimal dark sector in which the single field $\phi$ is replaced by an entire {\it ensemble}\/ of states $\phi_\ell$ with masses $m_\ell$ prior to the phase transition.  For simplicity we shall assume that the $\phi_\ell$ fields share the same quantum numbers.   In this case, the time-evolving mass function $m^2(t)$ is replaced by a mass matrix $M^2_{\ell,\ell'}(t)$
which takes the general form
\beq 
         M^2_{\ell,\ell'}(t)~=~ m_\ell^2 \,\delta_{\ell,\ell'} + (\Delta m^2)_{\ell,\ell'}(t)~.
\label{massterm}
\eeq
Here the first term in Eq.~(\ref{massterm}) represents the masses of the ensemble constituents that existed prior to the phase transition while the second term represents the extra time-dependent mass contributions that accrue as the phase transition unfolds. Note that we assume that all components of the $\Delta m^2(t)$ matrix have the same time-dependence since they are all being produced through a common phase transition.  In general the magnitudes of the individual entries within the $\Delta m^2(t)$ matrix will depend on the properties of the specific phase transition, and self-consistency requires only that they lead to a Hermitian mass matrix $M^2_{\ell,\ell'}(t)$ with non-negative eigenvalues.   Note, however, that we have allowed the mass-generating phase transition to introduce off-diagonal mass terms.  We have done this because dynamical mass generation can generally give rise to mixing amongst
scalars which share the same quantum numbers.  This implies that the mass eigenstates of the theory need not be the same before and after the phase transition.

If all of the original ensemble masses $m_\ell$ had been vanishing, every entry in the total mass matrix $M^2_{\ell,\ell'}(t)$ would have carried the same time dependence.    The resulting mass eigenvalues $\lambda_\ell^2$ of  $M^2_{\ell,\ell'}(t)$ would thus all be monotonically increasing from zero, evolving together in a fixed ratios towards their final late-time values as the mass-generating phase transition unfolds.   However, because of the presence of non-zero ensemble masses $m_\ell$ prior to the phase transition, the different entries within $M^2_{\ell,\ell'}(t)$ do not carry the same time dependence.   It is therefore not required that the mass eigenvalues $\lambda_\ell(t)$ evolve monotonically as the phase transition unfolds.

We shall focus our attention on cases in which the lightest mass eigenvalue $\lambda_0^2$ evolves non-monotonically during the phase transition. While we have mathematically explained how such a non-monotonicity might arise, it is also easy to understand how such a non-monotonicity might arise in physical terms.   In general, the lightest eigenvalue $\lambda_0^2$ will begin to rise from zero as the mass-generating phase transition begins.   Indeed, if the mixing with the next-lightest mode is particularly strong, this eigenvalue will start to approach the second-lightest eigenvalue. However, as these two eigenvalues approach each other, level repulsion begins to set in.  This then causes the lightest eigenvalue to drop again, after which the mass-generating phase transition is complete.

%=============BEGIN FIGURE=================%
\begin{figure}[b!]
    \begin{center}
    \mbox{
       \includegraphics[width=0.98\textwidth, height=0.4\textwidth]{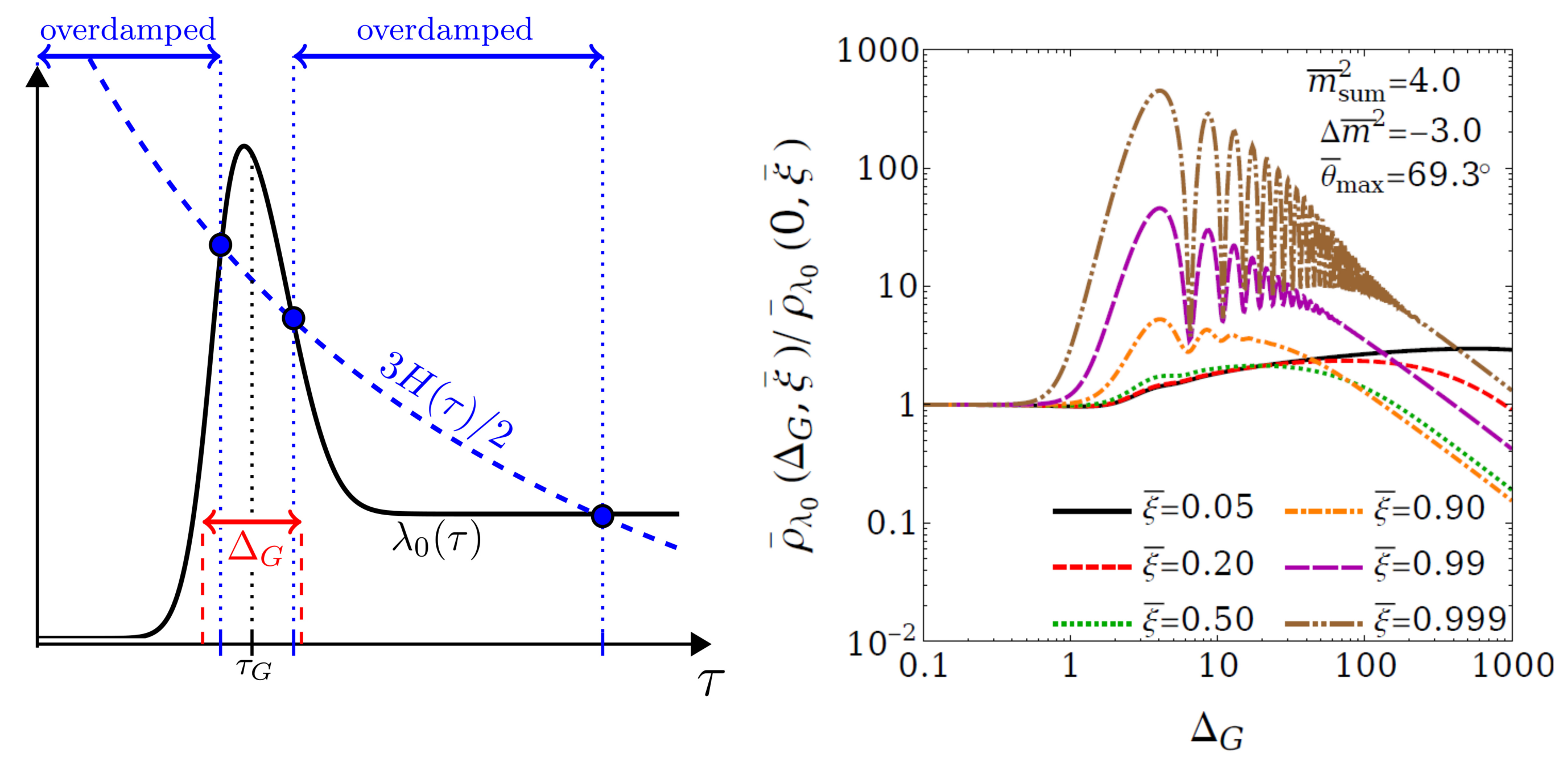}
        }
        \caption{
         {\it Left panel\/}:  In a non-minimal dark sector, the lightest state may experience a mixing with the next-lightest state.  During a mass-generating phase transition, the lightest state can thus experience an increase in its mass $\lambda_0$
           which is ultimately curtailed and reversed by level repulsion.   This process thereby effectively creates a ``pulse''
           in the mass of the lightest state, as shown.   The width of this pulse (and by extension its effective frequency) is correlated with the time interval $\Delta_G$ over which the mass-generating phase transition occurs.
           {\it Right panel}\/:
          This ``pulse'' can stimulate a parametric resonance which pumps huge amounts of energy into the lightest state.  Here we show the late-time energy density of the lightest state as a fraction of the value this quantity would have had for an instantaneous phase transition, plotted as a function of $\Delta_G$.  The different curves correspond to cases with different degrees of mixing with the next-highest state, with maximal mixing achieved as $\xi \to 1$.  
          We see that the parametric resonance becomes stronger as the mixing tends towards maximal, and occurs only for specific discrete values of $\Delta_G$. {\it  Likewise, we see that the parametric resonance induced by the pulse can enhance the resulting energy density of the lightest state by many orders of magnitude beyond traditional expectations.}
          These figures are taken from Ref.~\cite{Dienes:2015bka}.
          \label{fig:pulse}}
    \end{center}
\end{figure}
%===============END FIGURE=================%

The net result is that the lightest eigenvalue $\lambda_0(t)$ can experience a ``pulse'' during the mass-generating phase transition.
Such a ``pulse'' for $\lambda_0(t)$ is illustrated in left panel of  Fig.~\ref{fig:pulse}.  In this figure, $\tau$ represents a dimensionless time and the phase transition is presumed to occur at a time centered around $\tau_G$ unfolding over an interval of duration $\Delta_G$.    

The existence of such a pulse can then give rise to two dramatic effects.   The first is that this pulse can stimulate a parametric resonance which pumps huge amounts of energy into the lightest mode.  In general, a parametric resonance can be established for an oscillator if the mass term for the oscillator is varied sinusoidally at a frequency which is twice the natural frequency of the oscillator, or integer multiples thereof.    However, each $\phi_\ell$ mode behaves as an oscillator as long as it is underdamped;   likewise, even though the mass term for the lightest mode experiences only a single pulse, this pulse acts as if it were effectively a {\it portion}\/ of such a sinusoidal oscillation whose effective frequency is related to the magnitude $\Delta_G$ of the time interval through which the mass-generating phase transition occurs.  As it turns out, even such a portion is sufficient to induce a parametric resonance.  We therefore expect that a parametric resonance will be established for certain discrete ``resonant'' values $\Delta_G^{(n)}$ of the phase-transition time interval.

In the right panel of Fig.~\ref{fig:pulse}, we have adopted a specific two-state model for simplicity~\cite{Dienes:2015bka} and calculated the exact late-time energy density of the lightest state as a fraction of the value that this quantity would have had for an instantaneous phase transition.  This fraction is then plotted as a function of $\Delta_G$ for different degrees of mixing as quantified by a ``mixing saturation'' parameter $\xi$, with $\xi=0$ signifying the absence of mixing between the two states and $\xi=1$ signifying maximal mixing.  We see from the figure that significant enhancements occur for certain discrete ``resonant'' values of $\Delta_G^{(n)}$,
and that these enhancements are stronger when the mixing becomes stronger. Of course, in the $\xi\to 0$ limit the heavier state decouples;  this then corresponds to the case of single-component dark matter.  However we now see that as the mixing between the states becomes stronger, the parametric resonance induced by the pulse also strengthens {\it and can ultimately enhance the resulting energy density of the lightest state by many orders of magnitude beyond traditional expectations}\/!

The second remarkable feature induced by the pulse shown in the left panel of Fig.~\ref{fig:pulse} is that it allows a field which is already underdamped (with $3H<2m$) to revert back to an overdamped phase (with $3H>2m$).  This ``re-overdamping'' phenomenon is illustrated in the left panel of Fig.~\ref{fig:pulse}.  We stress that in single-component theories of dark matter, this cannot happen:   since the Hubble parameter is monotonically falling, the only transition that can ever occur for a scalar field of constant or monotonically increasing mass is one from an overdamped phase to an underdamped phase. However, as we have seen, mixing between different states within a DDM-like ensemble can induce a non-monotonic pulse for the lightest mass eigenvalue.   It is this monotonicity which then engenders the possibility of re-entering an overdamped phase.

Even though this process puts the field back into an overdamped phase, the properties of this re-overdamped phase need not resemble those of the original overdamped phase.  The reason for this is that the intervening underdamped phase will have generally endowed the field with a non-zero velocity.  It therefore follows that the field will generally enter the re-overdamped phase with a non-zero velocity  --- something it would never have had in the original overdamped phase.  {\it As a result, the field will generally not be fixed during this re-overdamped phase, but will instead find itself slowly ``skidding'' to a halt under the influence of Hubble friction.}

%=============BEGIN FIGURE=================%
\begin{figure}[h!]
    \begin{center}
    \mbox{
       \includegraphics[width=0.98\textwidth, height=0.4\textwidth]{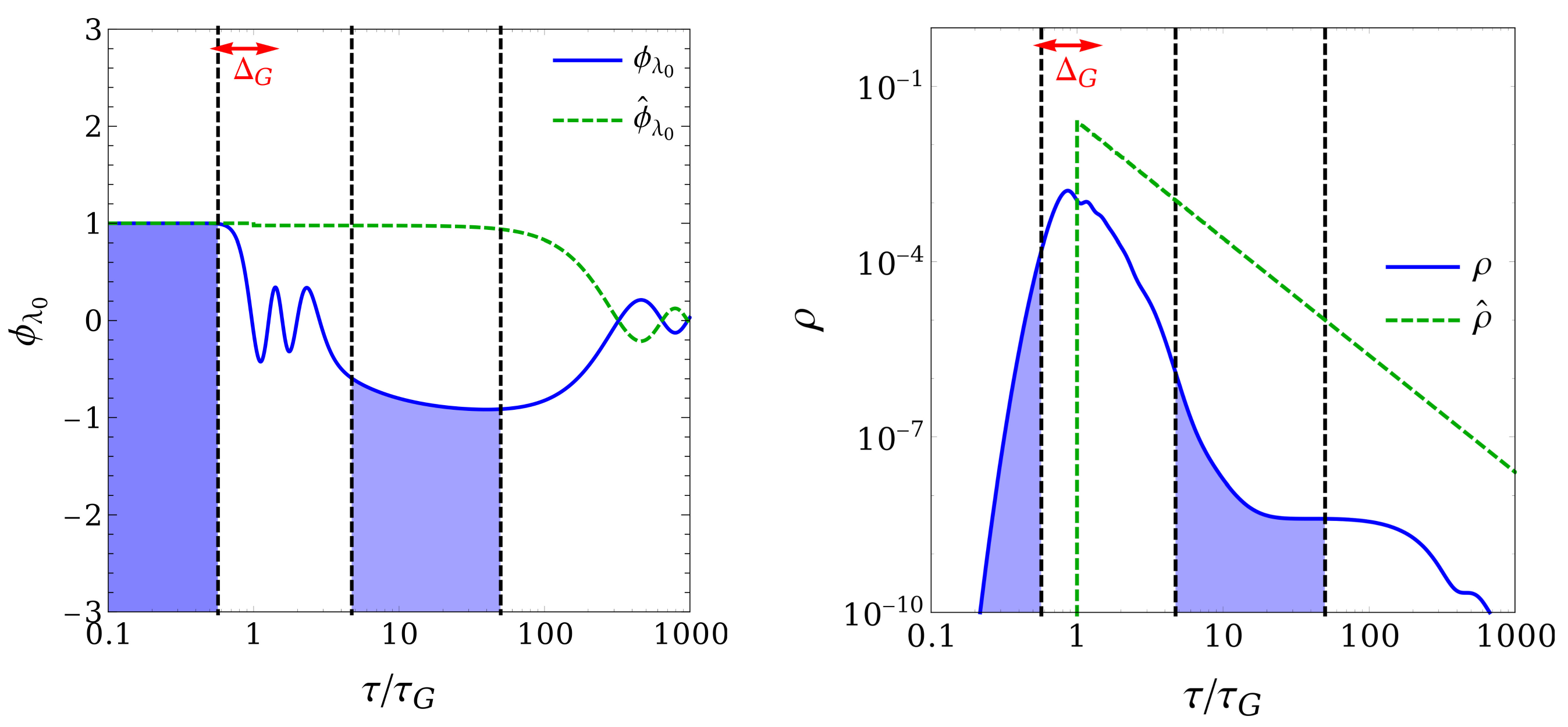}
        }
        \caption{
         The behavior of a scalar field (blue curve in left panel) and its associated energy density (blue curve in right panel) as the field transitions from an initial overdamped phase to an underdamped phase and then back to a re-overdamped phase, as illustrated in the left panel of Fig.~\protect\ref{fig:pulse}. Note that all quantities are plotted in Planck-scale units, and that intervals during which the field is overdamped are shaded in blue. In each panel, the green curves indicate the behavior that would have been expected with the same late-time mass eigenvalue but with an instantaneous phase transition (\ie, with no intervening pulse, or equivalently if there had been no level repulsion relative to other states in the system).   We see from the blue curves that because the field generally enters the re-overdamped phase with a non-zero velocity, it slowly ``skids'' to a halt.  During this skidding, we see that the behavior of the field and its energy density 
         are not like either vacuum energy or matter.  Instead, this represents a wholly new phase with a non-trivial, evolving equation of state.   This figure taken from Ref.~\cite{Dienes:2015bka}.
          \label{fig:vacter}}
    \end{center}
\end{figure}
%===============END FIGURE=================%

Because of the non-zero velocity that this field has while ``skidding'', this field does not generally have an equation of state $w= -1$ during the re-overdamped phase.  On the other hand, because this phase is overdamped, it experiences no tendency towards oscillation and we do not have $w=0$ either.  {\it Thus this phase is neither vacuum energy nor matter --- it is a wholly new phase which transcends the usual possibilities for scalar fields, one which gives rise to a non-trivial, evolving equation of state.}\/  The non-trivial behavior of the field and its energy density during such a re-overdamped phase are shown in Fig.~\ref{fig:vacter}.

%=============BEGIN FIGURE=================%
\begin{figure}[h!]
    \begin{center}
     \mbox{ \hskip -0.20 truein
      \includegraphics[width=0.75\textwidth,height=0.5\textwidth]{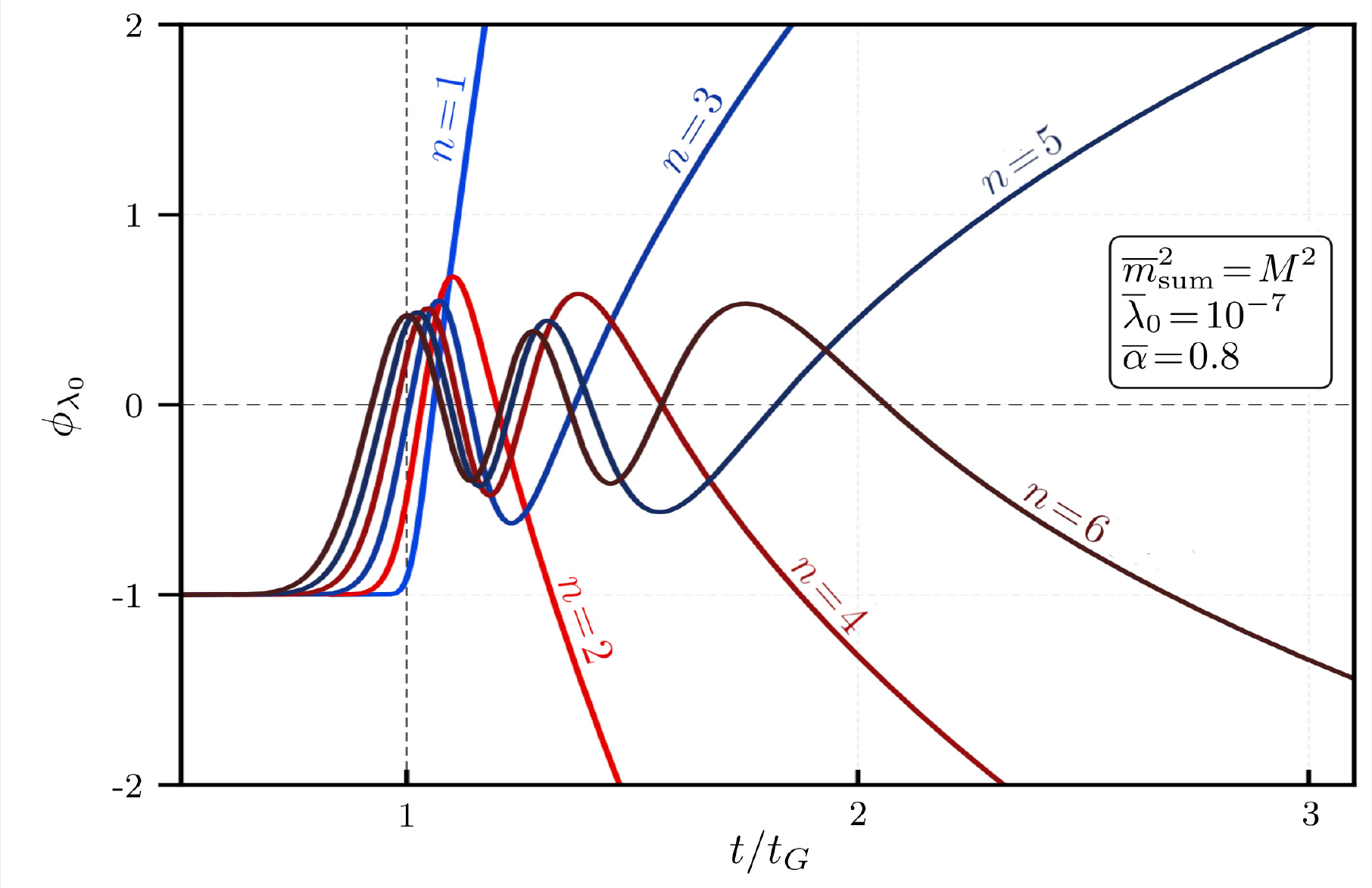}
        }
        \caption{A general ``slingshot'' mechanism through which a scalar field (here shown in Planck-scale units) can be dynamically propelled to extremely large field values.   This mechanism operates by having the field experience a parameteric resonance during which the field develops large velocities, followed by a carefully-timed ``release'' out of the oscillatory phase into a subsequent re-overdamped phase during which the field slowly skids to a halt.  In the special case in which the scalar field is an inflaton, this mechanism provides a dynamical method of generating the extremely large super-Planckian field values $\phi/M_P\sim {\cal O}(10)$ that are required for a subsequent period of inflation.  This figure is taken from Ref.~\cite{Dienes:2019chq}, where more details can be found.
          \label{fig:slingshot}}
    \end{center}
\end{figure}
%===============END FIGURE=================%

As we have seen, the pulse sketched in the left panel of Fig.~\ref{fig:pulse} has given rise to two distinct phenomena:  the parametric resonance and the entry into a re-overdamped phase.   However, under certain circumstances~\cite{Dienes:2019chq} these two features can be combined in order to ``slingshot'' the field to extremely large values.  This behavior is shown in Fig.~\ref{fig:slingshot}.  Indeed, the parametric resonance can be exploited in order to pump the field into a state with large oscillations and correspondingly large velocities, whereupon a carefully-timed transition out of the underdamped oscillatory phase ``releases'' the field at the time of its maximum velocity. Indeed it is shown in Ref.~\cite{Dienes:2019chq} that this can actually occur quite generically without fine-tuning.  Although the field will eventually skid to a halt during the subsequent re-overdamped phase, the field can nevertheless reach huge field values before coming to a halt. 

This mechanism for generating large field values thus functions as a ``slingshot''~\cite{Dienes:2019chq}.
Indeed, in ancient times, a
slingshot of the David/Goliath variety was fashioned by
attaching a projectile to a rope, twirling the rope around
overhead with increasing speed, and then releasing the
projectile at just the right moment so as to launch the
projectile in the desired direction. Our slingshot mechanism is essentially the same:  the field begins with an interval of periodic oscillations enhanced through a parametric resonance, followed by a ``release'' from the oscillatory
behavior at just the proper moment so as to propel the field towards large magnitudes with maximum velocity. Indeed, 
just as with an ancient slingshot, the time of release must occur at the moments during the oscillations when the field has its maximum velocity, with successive
higher orders of resonance alternating between forward or
backward motion of the projectile.

Although such a slingshot might have a plethora of uses, one natural idea~\cite{Dienes:2019chq} arises when $\phi$ represents the inflaton.  In such cases, the slingshot provides a dynamical way of propelling the inflaton to large super-Planckian field values $\phi/M_P\sim {\cal O}(10)$ --- exactly as often required as part of the initial conditions for a successful inflationary epoch.  This slingshot effect, operating during a pre-inflationary phase of the universe, can thereby  help to significantly enlarge the space of viable inflationary models.

\lesson{11}{Misalignment production of scalar non-minimal dark sectors can lead to a number of unique phenomena that arise when the different dark-sector components mix and which therefore would not have been possible for minimal dark sectors.  These include methods of enhancing the final abundances of dark-sector states by many orders of magnitudes;  the emergence of unusual scalar-field behaviors which correspond to non-trivial equations of state;  and new mechanisms for dynamically propelling scalar fields to large, super-Planckian values.  For more details, see Refs.~\cite{Dienes:2015bka,Dienes:2016zfr, Dienes:2019chq}.}

%======================================================
\FloatBarrier
\subsection{\large Non-traditional dark-matter phase-space distributions \hfill \label{phasespacedistributions}}

Dark-matter phase-space distributions often function as the central quantities in dark-matter physics, encoding whether the dark matter is hot or cold, thermal or non-thermal, and so forth.   These distributions also lead to direct predictions for the corresponding late-time matter power spectra which encode vital information about structure formation in the early universe, and eventually even lead to predictions for potentially observable late-time quantities such as the mass distributions of virialized dark-matter halos. 

As we shall now demonstrate, non-minimal dark sectors can lead to unexpected dark-matter momentum  distributions $f(p)$ --- even when the relevant dark-matter production mechanism is relatively simple (such as occurs for thermal freezeout), and even when all of the non-trivial dynamics associated with the non-minimal dark sector has concluded long before the current epoch.  Indeed, we have already seen in Lessons~\#10 and \#11 that relatively straightforward production mechanisms such as thermal freeze-out and misalignment production can lead to unexpected abundances within a non-minimal dark sector.  We now demonstrate that even the {\it phase-space distributions}\/ of the particles that comprise these abundances can be profoundly altered, taking shapes that are highly unorthodox from a single-particle perspective.

%---------------------BEGIN FIGURE -----------------------%
\begin{figure}[b!]
\centering
\mbox{
\hskip -0.2truein
\includegraphics[width=0.49\textwidth,height=0.49\textwidth]{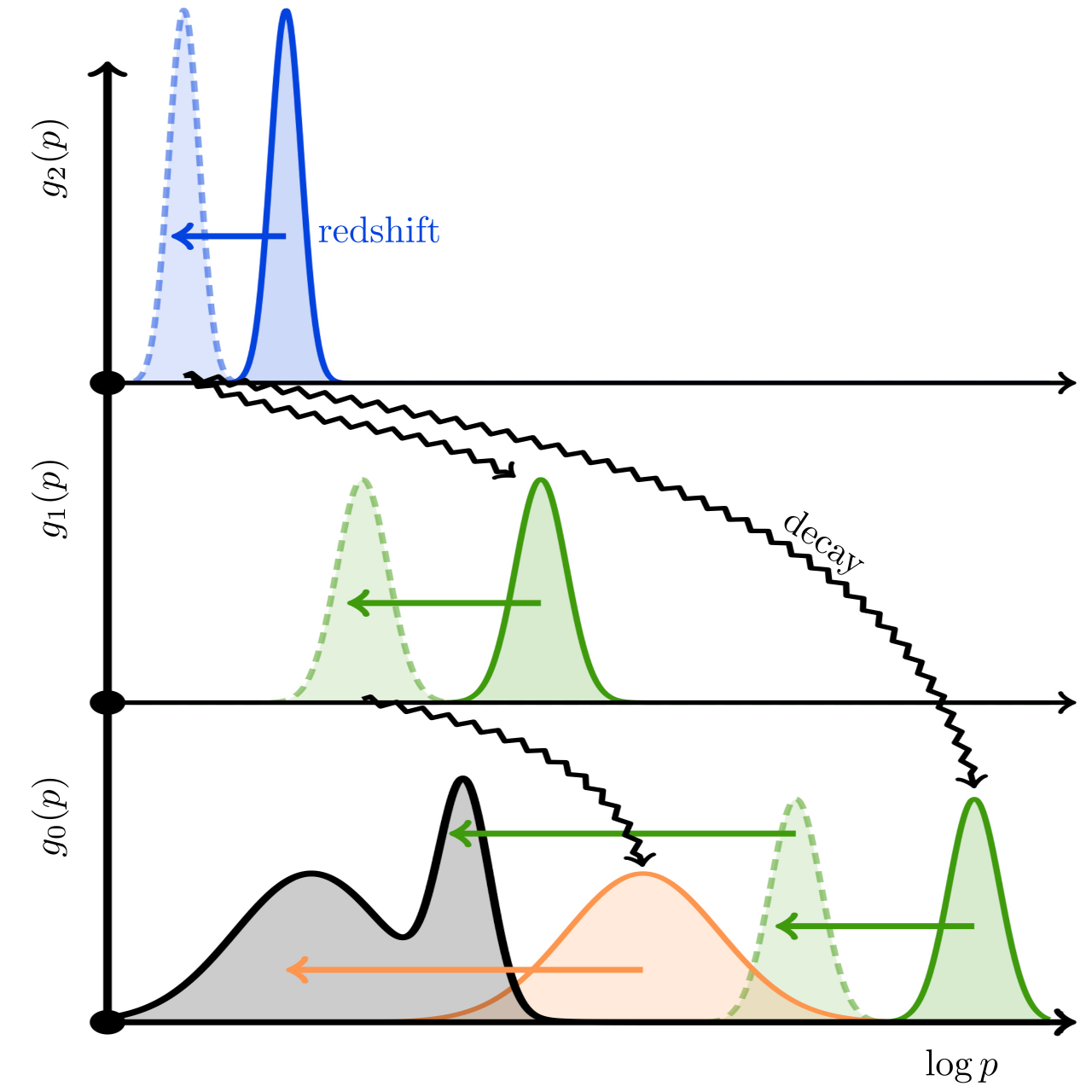}~~~
\includegraphics[width=.49\textwidth,height= 0.49 \textwidth]{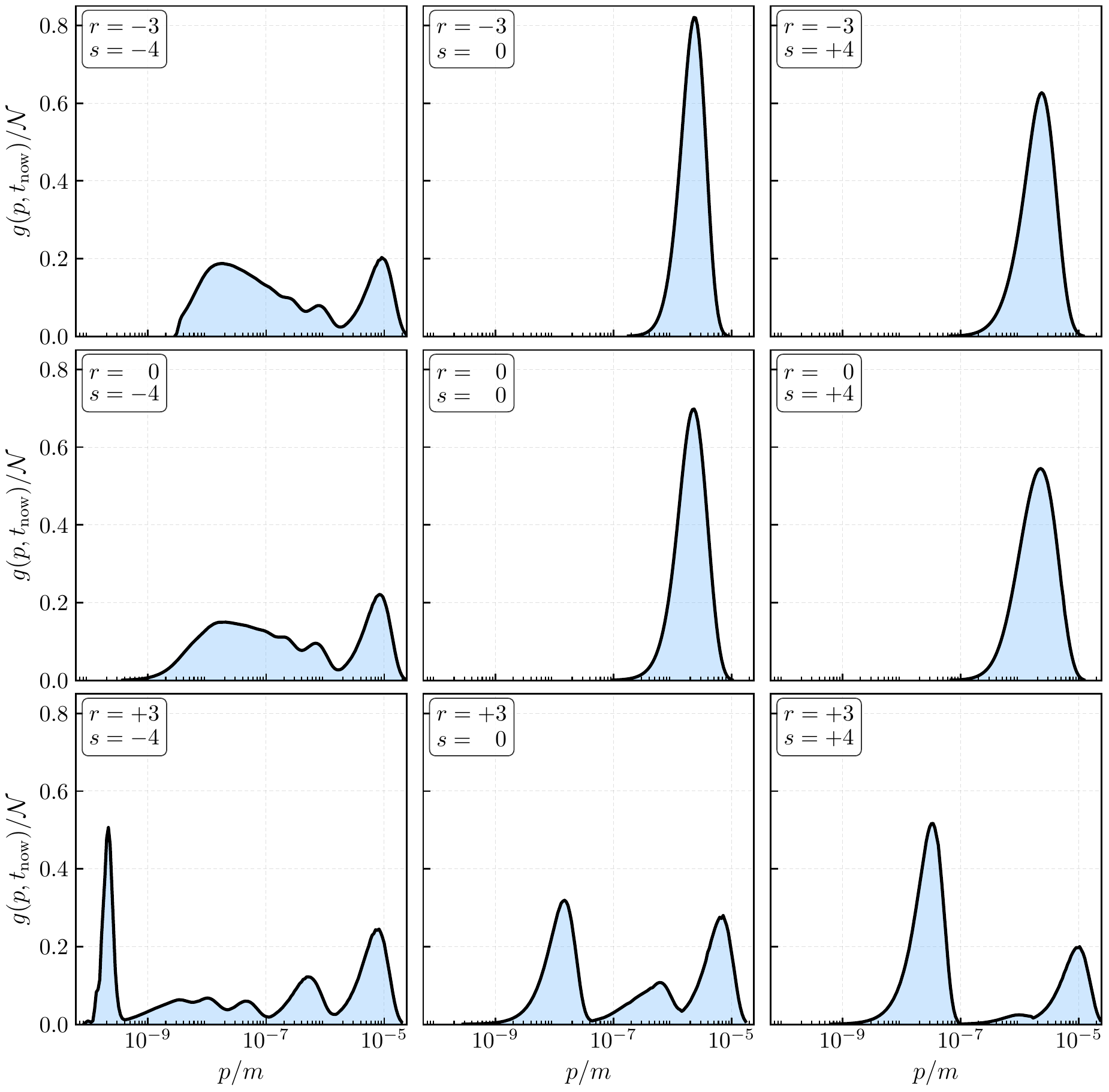}
}
\centering
\caption{ {\it Left panel}\/:  One possible scenario wherein the dynamics associated with a non-minimal dark sector makes it possible for a relatively straightforward production mechanism such as thermal freeze-out to give rise to a non-trivial (even bimodal!)\  dark-matter phase-space distribution at late times.  In this example, the results of two independent intra-ensemble decay chains superpose to produce the final complex dark-matter phase-space distribution.  {\it Right panel}\/:
The final phase-space distributions for nine different models of non-minimal dark sectors, each of which involves ten different levels. As discussed in the text, these models differ in their possible branching ratios for the decays at each stage, but in each model the number of possible decay chains quickly proliferates.   We see that a wide variety of final dark-matter distributions are possible, ranging from unimodal and bimodal distributions to multi-modal distributions exhibiting complex patterns of peaks and troughs as functions of momentum.  These figures are taken from Ref.~\cite{Dienes:2020bmn}.
}
\label{fig:conveyor2}
\end{figure}
%----------------------END FIGURE ------------------------%

One way in which this can happen in a non-minimal dark sector is through the sorts of intra-ensemble
decays already discussed, in which a heavier dark-sector component decays to lighter ones within the same ensemble.
To see this how this may lead to distortions in the final dark-matter phase-space distributions, let us consider  the three-state system~\cite{Dienes:2020bmn} 
illustrated in the left panel of Fig.~\ref{fig:conveyor2}.~
Here the three states are labeled $\ell=0,1,2$ in order of increasing mass,
and we shall assume that only the heaviest $\phi_2$ state is initially populated.
For simplicity, we shall even assume that this state has the simple thermal distribution function sketched in solid blue in the left panel of  Fig.~\ref{fig:conveyor2} when plotted versus $\log p$.  
Within an FRW universe, this ``packet'' then redshifts to the left (\ie, towards lower momenta, as indicated in dashed blue) until it decays.
For simplicity, let us assume that each decay is a two-body decay, with the $\ell=2$ state preferentially undergoing a decay of the form $2\to 1+0$.  This then produces 
two daughters (dark green), one for the $\ell=1$ state and the other for the $\ell=0$ state.
Each of these new daughter packets then redshifts until the $\ell=1$ daughter (now dashed green) undergoes its own decay
into two kinematically identical  $\ell=0$ granddaughters (orange).
As shown in the left panel of Fig.~\ref{fig:conveyor2}, these granddaughters may happen to have a non-negligible overlap
with the redshifted $\ell=0$ daughter (dashed green).
These distributions therefore superpose, and the resulting total distribution continues to redshift as a single unit.
The resulting dark-matter phase-space distribution (black) is thus highly complex and even bimodal --- even though our original distribution was thermal and hence unimodal!  

Although the left panel of Fig.~\ref{fig:conveyor2} is simply a sketch, its basic features are essentially accurate, as is its final lesson.
Indeed, each decay produces offspring packets
whose widths in momentum space are larger than those of the packet from the preceding generation;  likewise
these offspring packets also generally have higher momenta
than those of the corresponding parent packet.  This is all consistent with the underlying decay kinematics.  
Each of the offspring packets nevertheless has the same total area as the parent packet
from which it emerged, since each daughter is descended from a single parent.   Also note that in this sketch we are showing not the dark-matter phase-space distributions $f(p,t)$ but rather their cousins $g(p,t) \equiv (ap)^3 f(p,t)$, where $a$ is the FRW scale factor.   Since $f(p,t)$ is the quantity whose $d^3 p$ integral yields the physical number density $n(t)$, it then follows from isotropy that $g(p,t)$ is the quantity whose $d\log p$ integral yields the co-moving number density.  Under cosmological redshifting, the $g(p)$ distribution has the advantageous property that when plotted versus $\log p$ it simply slides rigidly to the left as if on a cosmological ``conveyor belt'' without any  distortion to its shape~\cite{Dienes:2020bmn}.

In sketching the left panel of Fig.~\ref{fig:conveyor2}, we have nevertheless neglected numerous effects arising from the full kinematics of particle decay, especially when the dark matter does not have a single momentum but instead has an actual phase-space {\it distribution}\/ stretching across many momenta.  We have also considered only a simple three-state system, whereas a larger non-minimal dark sector with a whole tower of independent states can give rise to a virtual proliferation of independent decay chains, each endowed with its own branching ratios.  One might therefore suspect that certain features shown in the left panel of Fig.~\ref{fig:conveyor2} --- such as the distinct bimodality of the final phase-space distribution $g(p)$ --- might ultimately be ``washed out'' when all of these effects are combined.  However, this is not the case.   In the right panel of Fig.~\ref{fig:conveyor2} we show the exact numerical results (obtained through direct numerical solution of appropriately coupled Boltzmann equations) for the final dark-matter phase-space distributions $g(p)$ that emerge within nine different models, each of which involves {\it ten}\/ distinct levels.  Each model has its own set of  possible two-body decay $\phi_\ell\to \phi_{\ell'} \phi_{\ell''}$ 
branching fractions, and we have varied these branching fractions across the different models so as to survey  decay phenomenologies that
involve different levels of asymmetry between the daughter masses in each decay as well as different levels of marginality (ranging from nearly marginal to highly exothermic).
As evident from the right panel of Fig.~\ref{fig:conveyor2}, the resulting phase-space distributions vary greatly across these models:   some are unimodal, some are bimodal, and others are multi-modal, exhibiting complex patterns of peaks and troughs as functions of $p$.  
The results shown in this figure are particularly remarkable since in each case we began with purely thermal phase-space distributions!   Thus, within a non-minimal dark sector, all of these different distributions with different internal structures can emerge from the same thermal freezeout.

The fact that the distributions $g(p)$ emerging from this analysis vary significantly across the different models considered implies that these distributions contain {\it imprints}\/ of this non-trivial early-universe dynamics.  As such, these imprints within $g(p)$ then translate to specific features in the predicted linear matter power spectrum $P(k)$ (where $k$ is a relevant wavenumber) and even to observable quantities such as the corresponding mass distribution $dn/d\log M$ of late-time virialized dark-matter halos.  Indeed, there exist sophisticated numerical codes which allow one to determine the $P(k)$ that corresponds to a given $g(p)$, and likewise there exist procedures (not only numerical techniques such as $N$-body simulations but also analytic methods such as the extended Press-Schechter formalism) which allow one to determine $dn/d\log M$ from $P(k)$.  Using this technology, one can then study how the unique features within phase-space distributions such as those shown in the right panel of Fig.~\ref{fig:conveyor2} translate to these latter quantities.   These issues were studied in some detail in Ref.~\cite{Dienes:2020bmn} (for the matter power spectrum) and Ref.~\cite{Dienes:2021itb} (for halo-mass distributions).  

Another potential consequence of dark-matter ensembles on the spatial distribution of matter is the modification of galactic morphologies.  Such modifications can arise as a consequence of dark-matter self-interactions and can involve an alteration of the dark-matter halo profile or even the development of a dark disk in addition to the halo.  Indeed, the latter possibility occurs even within two-component dark-matter scenarios~\cite{Fan:2013yva}, and additional degrees of freedom in the dark sector would almost certainly give rise to additional possibilities.

For simple umimodal $g(p)$ distributions of the sort that arise within single-component dark sectors, a standard procedure is to identify an ``average'' momentum $\langle p\rangle$ that characterizes the entire distribution.  This ``free-streaming scale'' then translates to a specific $k$-value within $P(k)$ beyond which a suppression of structure develops.  {\it However, for the complex $g(p)$ distributions that emerge within non-minimal dark sectors, the identification of an average momentum $\langle p\rangle$ no longer captures the underlying physics.}\/  This then translates to a more complex behavior for $P(k)$ and $dn/d\log M$.    

An important practical question is the  inverse ``archaeological'' problem:  {\it to what extent can we exploit the information encoded in a given $P(k)$ or $dn/d\log M$ in order to reconstruct the features of the dark-matter $g(p)$ distribution from which they must have arisen}\/?  As we have seen, this question becomes especially critical for non-miminal dark sectors, where such extremely complicated distributions can arise.  At first glance, answering this question would seem to be practically impossible, since the mapping from $g(p)$ to $P(k)$ and $dn/d\log M$ is highly complex and almost impossible to invert.   It is also highly non-local, in the sense that the dark matter contained within a specific momentum slice of $g(p)$ affects the values of $P(k)$ over an entire range of $k$-values.  However, in Refs.~\cite{Dienes:2020bmn} and \cite{Dienes:2021itb},  
remarkably simple empirical
analytic expressions were developed which were shown to permit the reconstruction of most of the salient features
of $g(p)$ directly from $P(k)$ and --- under certain conditions --- even $dn/d\log  M$. Even more interestingly, these reconstruction formulae are {\it local}\/, indicating that the individual features of $P(k)$ at a specific value of wavenumber $k$ and the features of $dn/d\log M$ at a specific halo mass $M$ give direct information about the behavior of $g(p)$ at a corresponding specific value of $p$.  Indeed, these reconstruction methods work for the simple phase-space distributions that arise in minimal dark sectors as well as the complicated distributions that emerge in non-minimal dark sectors. 
These results --- along with others in Refs.~\cite{Dienes:2020bmn} and \cite{Dienes:2021itb} --- therefore provide important tools  for learning about, and potentially constraining, the features of non-minimal dark sectors and their dynamics in the early universe.
   
\lesson{12}{Non-minimal dark sectors can give rise to
dark matter with extremely complex non-thermal phase-space distributions --- even if we begin with relatively simple dark-matter production mechanisms such as thermal freezeout.  Because of their complexity, the resulting phase-space distributions cannot be characterized by a single ``free-streaming scale'', as is typically assumed for minimal dark sectors.   Instead, such distributions contain important {\it imprints}\/ of this non-trivial early-universe dynamics.   This information can then be used to predict unexpected features in the corresponding matter power spectra $P(k)$ that govern structure formation, and even in the corresponding mass distributions $dn/d\log M$ of the eventual virialized dark-matter halos.
Even more interestingly, there exist relatively simple empirical ways of {\it inverting}\/ this process, starting with a given matter power spectrum and/or halo-mass distribution and reconstructing the dark-matter phase-space distribution from which it emerged.  These methods apply for the simple phase-space distributions that arise in minimal dark sectors as well as the complicated distributions that emerge in non-minimal dark sectors.  These issues are discussed further in Refs.~\cite{Dienes:2020bmn,Dienes:2021itb}.}

%======================================================
\FloatBarrier
\subsection{\large Stasis in an expanding universe \hfill \label{sec:stasis}}

We conclude our discussion by describing what is perhaps the most intriguing phenomenon that can arise within a non-minimal dark sector:   the theoretical possibility of
a new kind of {\it stable}\/ cosmological era in which no single kind of energy need dominate.    We are certainly used to the idea of having cosmological eras which are dominated by matter, radiation, or even vacuum energy, and indeed each of these eras is stable and would persist forever if the domination of matter, radiation, or vacuum energy during that epoch were exactly 100\%, with no admixture of the other components.   However, in an expanding universe that simultaneously contains matter, radiation, and vacuum energy, we know that the relative abundances of these different components cannot remain constant because these different kinds of energy scale differently as the universe expands.  We therefore find, for example, that a mixed-component  radiation-dominated epoch inevitably becomes a matter-dominated epoch --- simply as a result of cosmological expansion --- and similarly such a matter-dominated epoch inevitably gives way to one which is dominated by vacuum energy.   As a result, moments such as matter-radiation equality are fleeting.

For a non-minimal dark sector, however, this lore is incorrect.    Indeed, as we shall now discuss, {\it with non-minimal dark sectors it becomes possible to achieve a form of ``stasis'' for cosmologies consisting of both matter and radiation in which the total matter and radiation abundances each remain constant despite cosmological expansion}\/.  For example, it is possible to have a cosmological epoch during which the abundances of matter and radiation each remain stable at 50\% (or at 75\% versus 25\%, or any other desired values).  Indeed, as we shall demonstrate, such periods of ``stasis'' can persist for an arbitrarily large number of $e$-folds before coming to a natural end.
The existence of this kind of stasis epoch therefore gives
rise to a host of new theoretical possibilities across the entire cosmological timeline.

It may initially seem impossible to arrange such periods of stasis between matter and radiation. After all, as the universe expands,
the matter abundance will inevitably come to dominate the radiation abundance; this is so intrinsic a prediction of
the Friedmann equations that this conclusion seems unavoidable.  On the other hand, matter can decay back
into radiation. This can then provide a natural counterbalance to the effects of cosmological expansion.  Given this, it is natural to ask whether 
these two effects be balanced against each other in order to induce an extended
time interval of stasis during which the matter and radiation abundances each remain constant.

Of course, particle decay is a relatively short process,
localized in time. In order to have an extended period
of stasis we would therefore require an extended period
during which particle decays are continually occurring.
This could potentially be realized if we had a large tower of matter states $\phi_\ell$ ($\ell = 0, 1, ..., N -1$), with each state sequentially decaying directly into radiation.  However, as we have seen in Sect.~\ref{sec:dynDM}, precisely such towers of states emerge within the DDM framework (and within non-minimal dark sectors more generally), arising within many scenarios for physics beyond the SM and satisfying the exact or approximate scaling relations in Eq.~(\ref{scalings}).
The question is then whether the
sequential decays down such tower can be exploited in
order to counterbalance the effects of cosmological expansion and thereby sustain an extended period of cosmological stasis.

At first glance, such a balancing might seem to be impossible.  After all,
the abundance $\Omega_\ell(t)$ of each ensemble constituent initially grows according to
a common power-law as the result of cosmological expansion before ultimately experiencing exponential decay after its decay time $\tau_\ell\equiv 1/\Gamma_\ell$ is reached.  
Stasis would then require that all of this occurs for each ensemble constituent ---  each with its own independent abundance and lifetime according to Eq.~(\ref{scalings}) --- in such  a way that the total matter abundance $\Omega_M \equiv \sum_\ell \Omega_\ell(t)$ remains constant.

%======================== BEGIN FIGURE ======================================== 
\begin{figure*}[hbt]
\centering
\mbox{
\hskip -0.10 truein
\includegraphics[width=0.44\textwidth, height=0.38 \textwidth]{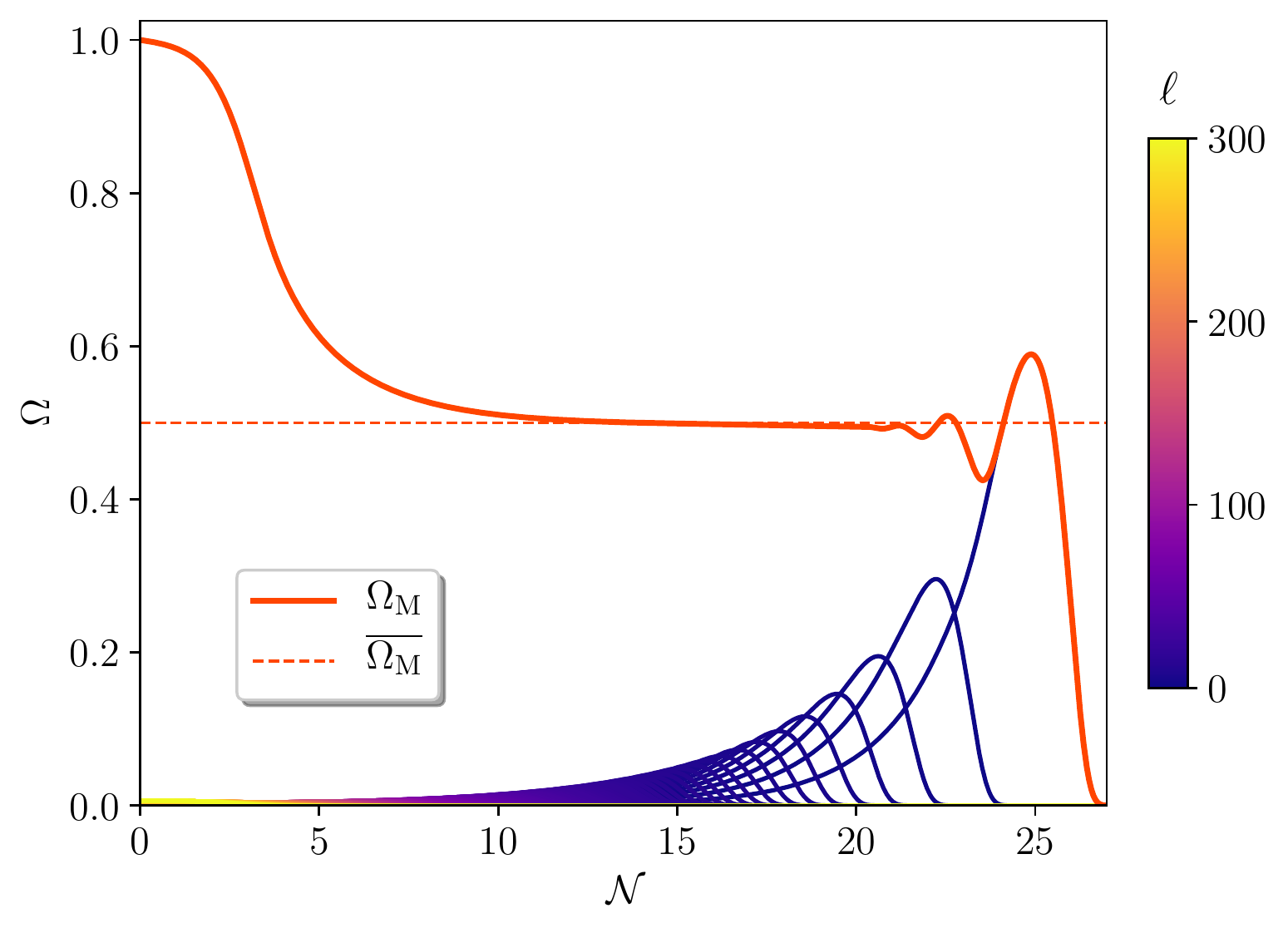}
~
\includegraphics[keepaspectratio, width=0.55 \textwidth]{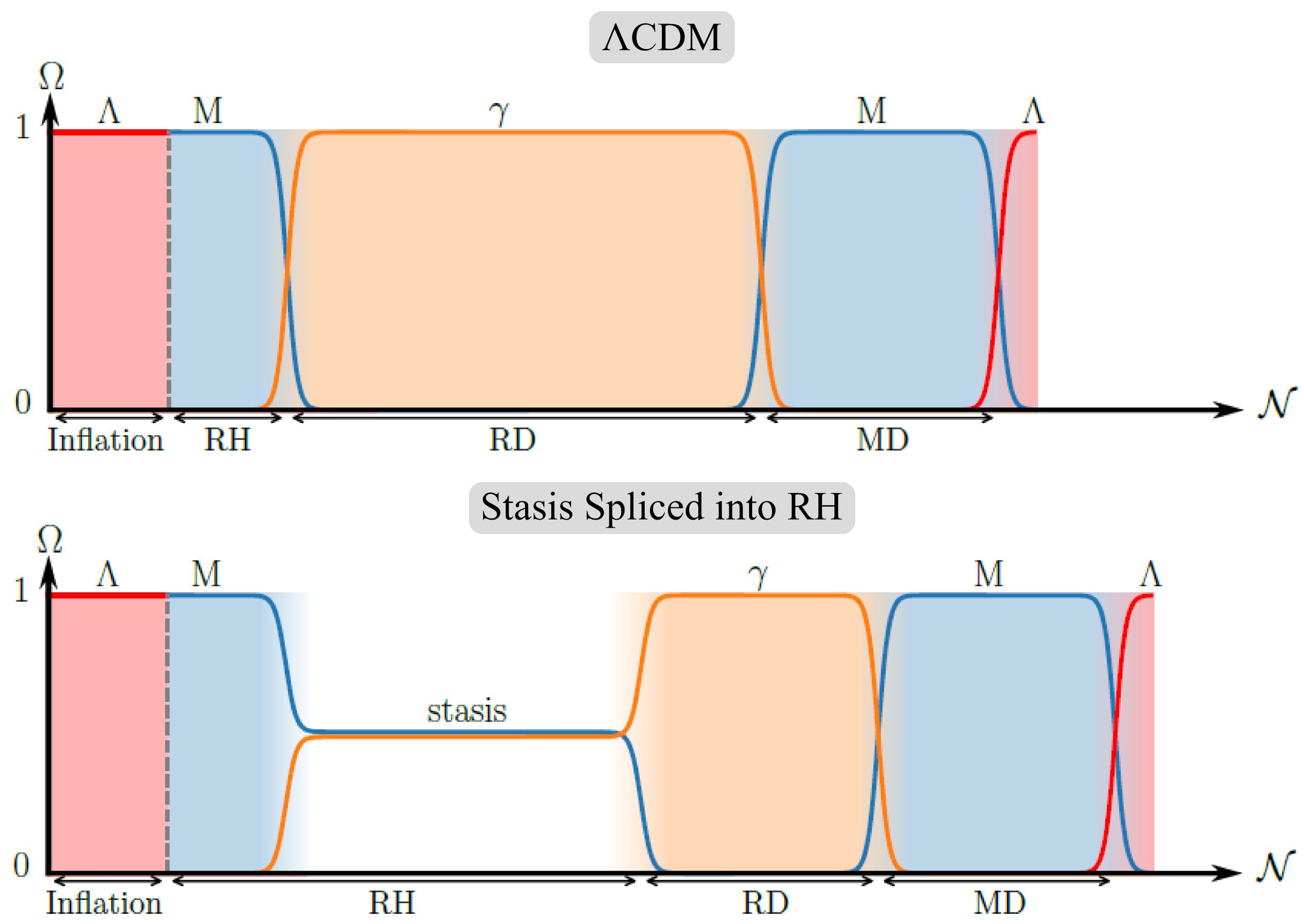}
}
\caption{ {\it Left panel}\/:  The individual matter abundances $\Omega_\ell(t)$ (orange/blue) and the corresponding total matter abundance $\Omega_M$ (red), plotted as functions of the number $\calN$ of $e$-folds since the initial $\phi_\ell$ production.   These curves were generated through a direct numerical solution of the relevant Boltzmann equations for our discrete tower of decaying states without invoking any approximations, and correspond to the parameter choices
 $(\alpha,\gamma,\delta)=(1,7,1)$, for which $\barOmega_M=1/2$.   
Although each individual $\Omega_\ell$ component exhibits a non-trivial behavior, first
growing as a power-law due to cosmological redshifting before ultimately decaying exponentially,
their sum $\Omega_M$ nevertheless evolves towards a stasis epoch in which $\Omega_M$ remains essentially constant
(in this case for approximately $15$ $e$-folds) before exiting stasis.   Even longer periods of stasis can be produced if $N$ is increased.  Eventually the stasis ends with the final decays of the lightest states. 
 {\it Right panel}\/:  
Sketch of the traditional $\Lambda$CDM cosmology (upper portion) as well
as a possible alternative cosmological history (lower portion) in which
a stasis epoch with $\barOmega_M=\barOmega_\gamma=1/2$
is ``spliced'' into the reheating portion of 
the cosmological timeline.
In each case we have sketched the abundances associated with vacuum energy (red), matter (blue), and radiation (orange) as time proceeds,
with corresponding background shadings indicating the dominant component in each epoch.
In the alternative cosmology shown, reheating occurs during the stasis epoch and results
from the decays of the $\phi_\ell$ states.
These figures are taken from Ref.~\cite{Dienes:2021woi}, where further details can be found.
}
\label{fig:stasis}
\end{figure*}
%======================== END FIGURE ======================================== 

However, it turns out that this is exactly what
occurs~\cite{Dienes:2021woi}.  This dynamics is plotted in the left panel of Fig.~\ref{fig:stasis}.   Before any decays have begun, no radiation has yet been generated;   we thus start with  $\Omega_M=1$ and $\Omega_\gamma=0$.  However, once the states with shortest lifetimes (here assumed to be the heaviest states at the top of the tower) begin to decay, $\Omega_M$ begins to fall while the radiation abundance $\Omega_\gamma$ begins to rise.  Ultimately, as the decays progress down the tower, we then evolve towards a ``stasis'' during which $\Omega_M$ and $\Omega_\gamma$ are constants.    This stasis state then persists throughout the remaining decay process until the lightest, longest-lived states in the tower finally decay.   This then signals the end of the stasis epoch, leaving us with $\Omega_M=0$ and $\Omega_\gamma=1$.

It is important to note that the emergence of such a stasis epoch is not fine-tuned.   Indeed, for {\it any}\/ tower of states parametrized by {\it any}\/ values of the scaling exponents $(\alpha,\gamma,\delta)$ in Eq.~(\ref{scalings}), an eventual period of stasis will emerge.  As such, the stasis state is 
actually a {\it global attractor}\/ within these systems, with the universe naturally evolving towards such long-lasting periods of
stasis for any values of $(\alpha,\gamma,\delta)$.   Indeed, the corresponding value of $\barOmega_M$ during this stasis epoch is then given by
\beq
       ~~\barOmega_M ~=~  \frac{  2 \gamma\delta - 4(1+\alpha\delta) }{2 \gamma\delta - (1+ \alpha \delta) }~.
\label{stasisOmegaM}
\eeq 
Interestingly, 
the special case with $(\alpha,\gamma,\delta)=(1,7,1)$ --- one of the well-motivated benchmarks discussed in Sect.~\ref{sec:dynDM} ---  
yields $\barOmega_M=1/2$.  This then corresponds
to a case of stasis exhibiting matter-radiation equality.   More details can be found in Ref.~\cite{Dienes:2021woi}, including the possibility of introducing into such matter-radiation scenarios additional energy components with other equations of state.

Needless to say, our results give rise to a host of new
theoretical possibilities for physics across the entire cosmological timeline. 
In particular, it is possible to imagine “splicing” an epoch of stasis into the standard $\Lambda$CDM timeline, either as an additional segment inserted into the timeline or as the replacement for
a segment which is removed.   In the right panel of Fig.~\ref{fig:stasis} we illustrate the standard $\Lambda$CDM cosmological timeline (upper portion) as well as a modified timeline (lower portion) in which an extended period of stasis with $\barOmega_M=1/2$ has been inserted during the reheating epoch.   This then yields an alternative cosmology in which reheating occurs during the stasis epoch and results from the decays of the $\phi_\ell$  states.   Other cosmological timelines involving stasis epochs are also possible.

\lesson{13}{
With non-minimal dark sectors it becomes possible to achieve a form of ``stasis'' for cosmologies consisting of both matter and radiation in which the total matter and radiation abundances each remain constant despite cosmological expansion.  Such periods of stasis can persist for an arbitrarily large number of $e$-folds.  During these periods, the effects of cosmological expansion 
are counterbalanced by the effects of intra-ensemble decays.
As discussed, the stasis state is not fine-tuned but actually represents a global attractor within the dynamics of such systems.
The existence of such stasis states therefore gives
rise to a host of new theoretical possibilities across the entire cosmological timeline, ranging from
potential implications for primordial density perturbations, dark-matter production, and structure
formation all the way to early reheating, early matter-dominated eras, and even the age of the
universe. For more details and discussions, see Ref.~\cite{Dienes:2021woi}.
}

%=========================================================================
\section{\large Conclusions}

In reviewing the salient features of the Dynamical Dark Matter edifice ---
and of non-minimal dark sectors more generally --- it is best to recall
that ``small [constructions]
may be finished by their first architects;
grand ones, true ones, ever leave the copestone to posterity''~\cite{Melville}.
Indeed, we have only begun to scratch the surface of the new dark-matter and  dark-sector phenomenologies which are possible for non-minimal dark sectors, choosing to highlight here only a handful which illustrate their tremendous range and scope.
Ultimately, of course, it is data which will confirm whether the ideas presented here
will have relevance for the natural world.  
However, as part of the Snowmass planning process, we believe that it is
of critical importance to bear these possibilities in mind, if only to avoid
the narrowness of perspective that comes from adhering to certain overly-constraining initial assumptions.
Indeed, the research community
has already been surprised by the relative difficulty in discovering
compelling evidence for either weak-scale SUSY or corresponding WIMP-like dark matter.
We therefore offer these ideas as examples of how far from those assumptions
the truth may lie, all while remaining within the space of well-motivated extensions
to the Standard Model.
Indeed, further observations along these lines can be
found in the papers cited.

\summarylesson{Multi-component dark sectors are well motivated, with many common extensions to the SM (extra spacetime dimensions, large hidden-sector gauge groups, string theories, {\it etc}\/.)\ 
involving large numbers of extra states.   Stability is not always 
guaranteed for such towers of states, implying that they can decay throughout the evolution of the universe and even today.  As realized through the Dynamical Dark Matter framework, these ideas lead to a number of unexpected possibilities for dark-matter phenomenology and model-building. 
It is therefore important that we shed our theoretical prejudices and embrace all the possibilities 
that dark-sector non-minimality and instability allow.
}

\bigskip
\bigskip
%====================================================================
% \begin{acknowledgments}
\section*{\large Acknowledgments}
We gratefully acknowledge our collaborators throughout the past decade
who have contributed to the ideas reviewed here
and who have served, in different combinations, as our co-authors
on the original papers cited herein.
In alphabetical order,
these include
Kim Boddy, Yusuf Buyukdag, David Curtin,
Aditi Desai, Emilian Dudas, Jacob Fennick, Max Fieg,
Jonathan Feng, Tony Gherghetta, Lucien Heurtier, Fei Huang, Doojin Kim, Jeff Kost,
Jason Kumar, Seung Lee, Tara Leininger, Kevin Manogue, Jong-Chul Park,
Huayang Song, Patrick Stengel, Shufang Su, Tim Tait, David Yaylali, and Hai-Bo Yu.
The research activities of KRD are supported in part by the U.S.\ Department of Energy
under Grant DE-FG02-13ER41976 / DE-SC0009913, and also by the U.S.\ National Science Foundation
through its employee IR/D program.
The research activities of BT are supported in part
by the U.S.\ National Science Foundation under Grant PHY-2014104.
The opinions and conclusions
expressed herein are those of the authors, and do not represent any funding agencies.

%\end{acknowledgments}

\bigskip
\bigskip

\bibliography{references}

\begin{thebibliography}{56}%
\makeatletter
\providecommand \@ifxundefined [1]{%
 \@ifx{#1\undefined}
}%
\providecommand \@ifnum [1]{%
 \ifnum #1\expandafter \@firstoftwo
 \else \expandafter \@secondoftwo
 \fi
}%
\providecommand \@ifx [1]{%
 \ifx #1\expandafter \@firstoftwo
 \else \expandafter \@secondoftwo
 \fi
}%
\providecommand \natexlab [1]{#1}%
\providecommand \enquote  [1]{``#1''}%
\providecommand \bibnamefont  [1]{#1}%
\providecommand \bibfnamefont [1]{#1}%
\providecommand \citenamefont [1]{#1}%
\providecommand \href@noop [0]{\@secondoftwo}%
\providecommand \href [0]{\begingroup \@sanitize@url \@href}%
\providecommand \@href[1]{\@@startlink{#1}\@@href}%
\providecommand \@@href[1]{\endgroup#1\@@endlink}%
\providecommand \@sanitize@url [0]{\catcode `\\12\catcode `\$12\catcode
  `\&12\catcode `\#12\catcode `\^12\catcode `\_12\catcode `\%12\relax}%
\providecommand \@@startlink[1]{}%
\providecommand \@@endlink[0]{}%
\providecommand \url  [0]{\begingroup\@sanitize@url \@url }%
\providecommand \@url [1]{\endgroup\@href {#1}{\urlprefix }}%
\providecommand \urlprefix  [0]{URL }%
\providecommand \Eprint [0]{\href }%
\providecommand \doibase [0]{http://dx.doi.org/}%
\providecommand \selectlanguage [0]{\@gobble}%
\providecommand \bibinfo  [0]{\@secondoftwo}%
\providecommand \bibfield  [0]{\@secondoftwo}%
\providecommand \translation [1]{[#1]}%
\providecommand \BibitemOpen [0]{}%
\providecommand \bibitemStop [0]{}%
\providecommand \bibitemNoStop [0]{.\EOS\space}%
\providecommand \EOS [0]{\spacefactor3000\relax}%
\providecommand \BibitemShut  [1]{\csname bibitem#1\endcsname}%
\let\auto@bib@innerbib\@empty
%</preamble>
\bibitem [{\citenamefont {Anderson}(1972)}]{anderson}%
  \BibitemOpen
  \bibfield  {author} {\bibinfo {author} {\bibfnamefont {P.~W.}\ \bibnamefont
  {Anderson}},\ }\href {\doibase 10.1126/science.177.4047.393} {\bibfield
  {journal} {\bibinfo  {journal} {Science}\ }\textbf {\bibinfo {volume}
  {177}},\ \bibinfo {pages} {393} (\bibinfo {year} {1972})}\BibitemShut
  {NoStop}%
\bibitem [{\citenamefont {Dienes}\ and\ \citenamefont
  {Thomas}(2012{\natexlab{a}})}]{Dienes:2011ja}%
  \BibitemOpen
  \bibfield  {author} {\bibinfo {author} {\bibfnamefont {K.~R.}\ \bibnamefont
  {Dienes}}\ and\ \bibinfo {author} {\bibfnamefont {B.}~\bibnamefont
  {Thomas}},\ }\href {\doibase 10.1103/PhysRevD.85.083523} {\bibfield
  {journal} {\bibinfo  {journal} {Phys. Rev. D}\ }\textbf {\bibinfo {volume}
  {85}},\ \bibinfo {pages} {083523} (\bibinfo {year} {2012}{\natexlab{a}})},\
  \Eprint {http://arxiv.org/abs/1106.4546} {arXiv:1106.4546 [hep-ph]}
  \BibitemShut {NoStop}%
\bibitem [{\citenamefont {Dienes}\ and\ \citenamefont
  {Thomas}(2012{\natexlab{b}})}]{Dienes:2011sa}%
  \BibitemOpen
  \bibfield  {author} {\bibinfo {author} {\bibfnamefont {K.~R.}\ \bibnamefont
  {Dienes}}\ and\ \bibinfo {author} {\bibfnamefont {B.}~\bibnamefont
  {Thomas}},\ }\href {\doibase 10.1103/PhysRevD.85.083524} {\bibfield
  {journal} {\bibinfo  {journal} {Phys. Rev. D}\ }\textbf {\bibinfo {volume}
  {85}},\ \bibinfo {pages} {083524} (\bibinfo {year} {2012}{\natexlab{b}})},\
  \Eprint {http://arxiv.org/abs/1107.0721} {arXiv:1107.0721 [hep-ph]}
  \BibitemShut {NoStop}%
\bibitem [{\citenamefont {Dienes}\ and\ \citenamefont
  {Thomas}(2012{\natexlab{c}})}]{Dienes:2012jb}%
  \BibitemOpen
  \bibfield  {author} {\bibinfo {author} {\bibfnamefont {K.~R.}\ \bibnamefont
  {Dienes}}\ and\ \bibinfo {author} {\bibfnamefont {B.}~\bibnamefont
  {Thomas}},\ }\href {\doibase 10.1103/PhysRevD.86.055013} {\bibfield
  {journal} {\bibinfo  {journal} {Phys. Rev. D}\ }\textbf {\bibinfo {volume}
  {86}},\ \bibinfo {pages} {055013} (\bibinfo {year} {2012}{\natexlab{c}})},\
  \Eprint {http://arxiv.org/abs/1203.1923} {arXiv:1203.1923 [hep-ph]}
  \BibitemShut {NoStop}%
\bibitem [{\citenamefont {Dienes}\ \emph
  {et~al.}(2012{\natexlab{a}})\citenamefont {Dienes}, \citenamefont {Kumar},\
  and\ \citenamefont {Thomas}}]{Dienes:2012cf}%
  \BibitemOpen
  \bibfield  {author} {\bibinfo {author} {\bibfnamefont {K.~R.}\ \bibnamefont
  {Dienes}}, \bibinfo {author} {\bibfnamefont {J.}~\bibnamefont {Kumar}}, \
  and\ \bibinfo {author} {\bibfnamefont {B.}~\bibnamefont {Thomas}},\ }\href
  {\doibase 10.1103/PhysRevD.86.055016} {\bibfield  {journal} {\bibinfo
  {journal} {Phys. Rev. D}\ }\textbf {\bibinfo {volume} {86}},\ \bibinfo
  {pages} {055016} (\bibinfo {year} {2012}{\natexlab{a}})},\ \Eprint
  {http://arxiv.org/abs/1208.0336} {arXiv:1208.0336 [hep-ph]} \BibitemShut
  {NoStop}%
\bibitem [{\citenamefont {Dienes}\ \emph {et~al.}(2013)\citenamefont {Dienes},
  \citenamefont {Kumar},\ and\ \citenamefont {Thomas}}]{Dienes:2013lxa}%
  \BibitemOpen
  \bibfield  {author} {\bibinfo {author} {\bibfnamefont {K.~R.}\ \bibnamefont
  {Dienes}}, \bibinfo {author} {\bibfnamefont {J.}~\bibnamefont {Kumar}}, \
  and\ \bibinfo {author} {\bibfnamefont {B.}~\bibnamefont {Thomas}},\ }\href
  {\doibase 10.1103/PhysRevD.88.103509} {\bibfield  {journal} {\bibinfo
  {journal} {Phys. Rev. D}\ }\textbf {\bibinfo {volume} {88}},\ \bibinfo
  {pages} {103509} (\bibinfo {year} {2013})},\ \Eprint
  {http://arxiv.org/abs/1306.2959} {arXiv:1306.2959 [hep-ph]} \BibitemShut
  {NoStop}%
\bibitem [{\citenamefont {Boddy}\ \emph {et~al.}(2016)\citenamefont {Boddy},
  \citenamefont {Dienes}, \citenamefont {Kim}, \citenamefont {Kumar},
  \citenamefont {Park},\ and\ \citenamefont {Thomas}}]{Boddy:2016fds}%
  \BibitemOpen
  \bibfield  {author} {\bibinfo {author} {\bibfnamefont {K.~K.}\ \bibnamefont
  {Boddy}}, \bibinfo {author} {\bibfnamefont {K.~R.}\ \bibnamefont {Dienes}},
  \bibinfo {author} {\bibfnamefont {D.}~\bibnamefont {Kim}}, \bibinfo {author}
  {\bibfnamefont {J.}~\bibnamefont {Kumar}}, \bibinfo {author} {\bibfnamefont
  {J.-C.}\ \bibnamefont {Park}}, \ and\ \bibinfo {author} {\bibfnamefont
  {B.}~\bibnamefont {Thomas}},\ }\href {\doibase 10.1103/PhysRevD.94.095027}
  {\bibfield  {journal} {\bibinfo  {journal} {Phys. Rev. D}\ }\textbf {\bibinfo
  {volume} {94}},\ \bibinfo {pages} {095027} (\bibinfo {year} {2016})},\
  \Eprint {http://arxiv.org/abs/1606.07440} {arXiv:1606.07440 [hep-ph]}
  \BibitemShut {NoStop}%
\bibitem [{\citenamefont {Boddy}\ \emph {et~al.}(2017)\citenamefont {Boddy},
  \citenamefont {Dienes}, \citenamefont {Kim}, \citenamefont {Kumar},
  \citenamefont {Park},\ and\ \citenamefont {Thomas}}]{Boddy:2016hbp}%
  \BibitemOpen
  \bibfield  {author} {\bibinfo {author} {\bibfnamefont {K.~K.}\ \bibnamefont
  {Boddy}}, \bibinfo {author} {\bibfnamefont {K.~R.}\ \bibnamefont {Dienes}},
  \bibinfo {author} {\bibfnamefont {D.}~\bibnamefont {Kim}}, \bibinfo {author}
  {\bibfnamefont {J.}~\bibnamefont {Kumar}}, \bibinfo {author} {\bibfnamefont
  {J.-C.}\ \bibnamefont {Park}}, \ and\ \bibinfo {author} {\bibfnamefont
  {B.}~\bibnamefont {Thomas}},\ }\href {\doibase 10.1103/PhysRevD.95.055024}
  {\bibfield  {journal} {\bibinfo  {journal} {Phys. Rev. D}\ }\textbf {\bibinfo
  {volume} {95}},\ \bibinfo {pages} {055024} (\bibinfo {year} {2017})},\
  \Eprint {http://arxiv.org/abs/1609.09104} {arXiv:1609.09104 [hep-ph]}
  \BibitemShut {NoStop}%
\bibitem [{\citenamefont {Dienes}\ \emph
  {et~al.}(2012{\natexlab{b}})\citenamefont {Dienes}, \citenamefont {Su},\ and\
  \citenamefont {Thomas}}]{Dienes:2012yz}%
  \BibitemOpen
  \bibfield  {author} {\bibinfo {author} {\bibfnamefont {K.~R.}\ \bibnamefont
  {Dienes}}, \bibinfo {author} {\bibfnamefont {S.}~\bibnamefont {Su}}, \ and\
  \bibinfo {author} {\bibfnamefont {B.}~\bibnamefont {Thomas}},\ }\href
  {\doibase 10.1103/PhysRevD.86.054008} {\bibfield  {journal} {\bibinfo
  {journal} {Phys. Rev. D}\ }\textbf {\bibinfo {volume} {86}},\ \bibinfo
  {pages} {054008} (\bibinfo {year} {2012}{\natexlab{b}})},\ \Eprint
  {http://arxiv.org/abs/1204.4183} {arXiv:1204.4183 [hep-ph]} \BibitemShut
  {NoStop}%
\bibitem [{\citenamefont {Dienes}\ \emph
  {et~al.}(2015{\natexlab{a}})\citenamefont {Dienes}, \citenamefont {Su},\ and\
  \citenamefont {Thomas}}]{Dienes:2014bka}%
  \BibitemOpen
  \bibfield  {author} {\bibinfo {author} {\bibfnamefont {K.~R.}\ \bibnamefont
  {Dienes}}, \bibinfo {author} {\bibfnamefont {S.}~\bibnamefont {Su}}, \ and\
  \bibinfo {author} {\bibfnamefont {B.}~\bibnamefont {Thomas}},\ }\href
  {\doibase 10.1103/PhysRevD.91.054002} {\bibfield  {journal} {\bibinfo
  {journal} {Phys. Rev. D}\ }\textbf {\bibinfo {volume} {91}},\ \bibinfo
  {pages} {054002} (\bibinfo {year} {2015}{\natexlab{a}})},\ \Eprint
  {http://arxiv.org/abs/1407.2606} {arXiv:1407.2606 [hep-ph]} \BibitemShut
  {NoStop}%
\bibitem [{\citenamefont {Curtin}\ \emph {et~al.}(2018)\citenamefont {Curtin},
  \citenamefont {Dienes},\ and\ \citenamefont {Thomas}}]{Curtin:2018ees}%
  \BibitemOpen
  \bibfield  {author} {\bibinfo {author} {\bibfnamefont {D.}~\bibnamefont
  {Curtin}}, \bibinfo {author} {\bibfnamefont {K.~R.}\ \bibnamefont {Dienes}},
  \ and\ \bibinfo {author} {\bibfnamefont {B.}~\bibnamefont {Thomas}},\ }\href
  {\doibase 10.1103/PhysRevD.98.115005} {\bibfield  {journal} {\bibinfo
  {journal} {Phys. Rev. D}\ }\textbf {\bibinfo {volume} {98}},\ \bibinfo
  {pages} {115005} (\bibinfo {year} {2018})},\ \Eprint
  {http://arxiv.org/abs/1809.11021} {arXiv:1809.11021 [hep-ph]} \BibitemShut
  {NoStop}%
\bibitem [{\citenamefont {Dienes}\ \emph
  {et~al.}(2020{\natexlab{a}})\citenamefont {Dienes}, \citenamefont {Kim},
  \citenamefont {Song}, \citenamefont {Su}, \citenamefont {Thomas},\ and\
  \citenamefont {Yaylali}}]{Dienes:2019krh}%
  \BibitemOpen
  \bibfield  {author} {\bibinfo {author} {\bibfnamefont {K.~R.}\ \bibnamefont
  {Dienes}}, \bibinfo {author} {\bibfnamefont {D.}~\bibnamefont {Kim}},
  \bibinfo {author} {\bibfnamefont {H.}~\bibnamefont {Song}}, \bibinfo {author}
  {\bibfnamefont {S.}~\bibnamefont {Su}}, \bibinfo {author} {\bibfnamefont
  {B.}~\bibnamefont {Thomas}}, \ and\ \bibinfo {author} {\bibfnamefont
  {D.}~\bibnamefont {Yaylali}},\ }\href {\doibase 10.1103/PhysRevD.101.075024}
  {\bibfield  {journal} {\bibinfo  {journal} {Phys. Rev. D}\ }\textbf {\bibinfo
  {volume} {101}},\ \bibinfo {pages} {075024} (\bibinfo {year}
  {2020}{\natexlab{a}})},\ \Eprint {http://arxiv.org/abs/1910.01129}
  {arXiv:1910.01129 [hep-ph]} \BibitemShut {NoStop}%
\bibitem [{\citenamefont {Dienes}\ \emph
  {et~al.}(2021{\natexlab{a}})\citenamefont {Dienes}, \citenamefont {Kim},
  \citenamefont {Leininger},\ and\ \citenamefont {Thomas}}]{Dienes:2021cxr}%
  \BibitemOpen
  \bibfield  {author} {\bibinfo {author} {\bibfnamefont {K.~R.}\ \bibnamefont
  {Dienes}}, \bibinfo {author} {\bibfnamefont {D.}~\bibnamefont {Kim}},
  \bibinfo {author} {\bibfnamefont {T.}~\bibnamefont {Leininger}}, \ and\
  \bibinfo {author} {\bibfnamefont {B.}~\bibnamefont {Thomas}},\ }\href@noop {}
  {\  (\bibinfo {year} {2021}{\natexlab{a}})},\ \Eprint
  {http://arxiv.org/abs/2108.02204} {arXiv:2108.02204 [hep-ph]} \BibitemShut
  {NoStop}%
\bibitem [{\citenamefont {Dienes}\ \emph
  {et~al.}(2020{\natexlab{b}})\citenamefont {Dienes}, \citenamefont {Huang},
  \citenamefont {Kost}, \citenamefont {Su},\ and\ \citenamefont
  {Thomas}}]{Dienes:2020bmn}%
  \BibitemOpen
  \bibfield  {author} {\bibinfo {author} {\bibfnamefont {K.~R.}\ \bibnamefont
  {Dienes}}, \bibinfo {author} {\bibfnamefont {F.}~\bibnamefont {Huang}},
  \bibinfo {author} {\bibfnamefont {J.}~\bibnamefont {Kost}}, \bibinfo {author}
  {\bibfnamefont {S.}~\bibnamefont {Su}}, \ and\ \bibinfo {author}
  {\bibfnamefont {B.}~\bibnamefont {Thomas}},\ }\href {\doibase
  10.1103/PhysRevD.101.123511} {\bibfield  {journal} {\bibinfo  {journal}
  {Phys. Rev. D}\ }\textbf {\bibinfo {volume} {101}},\ \bibinfo {pages}
  {123511} (\bibinfo {year} {2020}{\natexlab{b}})},\ \Eprint
  {http://arxiv.org/abs/2001.02193} {arXiv:2001.02193 [astro-ph.CO]}
  \BibitemShut {NoStop}%
\bibitem [{\citenamefont {Dienes}\ \emph
  {et~al.}(2021{\natexlab{b}})\citenamefont {Dienes}, \citenamefont {Huang},
  \citenamefont {Kost}, \citenamefont {Manogue},\ and\ \citenamefont
  {Thomas}}]{Dienes:2021itb}%
  \BibitemOpen
  \bibfield  {author} {\bibinfo {author} {\bibfnamefont {K.~R.}\ \bibnamefont
  {Dienes}}, \bibinfo {author} {\bibfnamefont {F.}~\bibnamefont {Huang}},
  \bibinfo {author} {\bibfnamefont {J.}~\bibnamefont {Kost}}, \bibinfo {author}
  {\bibfnamefont {K.}~\bibnamefont {Manogue}}, \ and\ \bibinfo {author}
  {\bibfnamefont {B.}~\bibnamefont {Thomas}},\ }\href@noop {} {\  (\bibinfo
  {year} {2021}{\natexlab{b}})},\ \Eprint {http://arxiv.org/abs/2101.10337}
  {arXiv:2101.10337 [astro-ph.CO]} \BibitemShut {NoStop}%
\bibitem [{\citenamefont {Desai}\ \emph {et~al.}(2020)\citenamefont {Desai},
  \citenamefont {Dienes},\ and\ \citenamefont {Thomas}}]{Desai:2019pvs}%
  \BibitemOpen
  \bibfield  {author} {\bibinfo {author} {\bibfnamefont {A.}~\bibnamefont
  {Desai}}, \bibinfo {author} {\bibfnamefont {K.~R.}\ \bibnamefont {Dienes}}, \
  and\ \bibinfo {author} {\bibfnamefont {B.}~\bibnamefont {Thomas}},\ }\href
  {\doibase 10.1103/PhysRevD.101.035031} {\bibfield  {journal} {\bibinfo
  {journal} {Phys. Rev. D}\ }\textbf {\bibinfo {volume} {101}},\ \bibinfo
  {pages} {035031} (\bibinfo {year} {2020})},\ \Eprint
  {http://arxiv.org/abs/1909.07981} {arXiv:1909.07981 [astro-ph.CO]}
  \BibitemShut {NoStop}%
\bibitem [{\citenamefont {Anchordoqui}\ \emph {et~al.}(2022)\citenamefont
  {Anchordoqui}, \citenamefont {Barger}, \citenamefont {Marfatia},\ and\
  \citenamefont {Soriano}}]{Anchordoqui:2022gmw}%
  \BibitemOpen
  \bibfield  {author} {\bibinfo {author} {\bibfnamefont {L.~A.}\ \bibnamefont
  {Anchordoqui}}, \bibinfo {author} {\bibfnamefont {V.}~\bibnamefont {Barger}},
  \bibinfo {author} {\bibfnamefont {D.}~\bibnamefont {Marfatia}}, \ and\
  \bibinfo {author} {\bibfnamefont {J.~F.}\ \bibnamefont {Soriano}},\
  }\href@noop {} {\  (\bibinfo {year} {2022})},\ \Eprint
  {http://arxiv.org/abs/2203.04818} {arXiv:2203.04818 [astro-ph.CO]}
  \BibitemShut {NoStop}%
\bibitem [{\citenamefont {Dienes}\ \emph
  {et~al.}(2021{\natexlab{c}})\citenamefont {Dienes}, \citenamefont {Huang},
  \citenamefont {Kost}, \citenamefont {Thomas},\ and\ \citenamefont
  {Yu}}]{Dienes:2021cxp}%
  \BibitemOpen
  \bibfield  {author} {\bibinfo {author} {\bibfnamefont {K.~R.}\ \bibnamefont
  {Dienes}}, \bibinfo {author} {\bibfnamefont {F.}~\bibnamefont {Huang}},
  \bibinfo {author} {\bibfnamefont {J.}~\bibnamefont {Kost}}, \bibinfo {author}
  {\bibfnamefont {B.}~\bibnamefont {Thomas}}, \ and\ \bibinfo {author}
  {\bibfnamefont {H.-B.}\ \bibnamefont {Yu}},\ }\href@noop {} {\  (\bibinfo
  {year} {2021}{\natexlab{c}})},\ \Eprint {http://arxiv.org/abs/2112.09105}
  {arXiv:2112.09105 [astro-ph.CO]} \BibitemShut {NoStop}%
\bibitem [{\citenamefont {Dienes}\ \emph
  {et~al.}(2015{\natexlab{b}})\citenamefont {Dienes}, \citenamefont {Kumar},
  \citenamefont {Thomas},\ and\ \citenamefont {Yaylali}}]{Dienes:2014via}%
  \BibitemOpen
  \bibfield  {author} {\bibinfo {author} {\bibfnamefont {K.~R.}\ \bibnamefont
  {Dienes}}, \bibinfo {author} {\bibfnamefont {J.}~\bibnamefont {Kumar}},
  \bibinfo {author} {\bibfnamefont {B.}~\bibnamefont {Thomas}}, \ and\ \bibinfo
  {author} {\bibfnamefont {D.}~\bibnamefont {Yaylali}},\ }\href {\doibase
  10.1103/PhysRevLett.114.051301} {\bibfield  {journal} {\bibinfo  {journal}
  {Phys. Rev. Lett.}\ }\textbf {\bibinfo {volume} {114}},\ \bibinfo {pages}
  {051301} (\bibinfo {year} {2015}{\natexlab{b}})},\ \Eprint
  {http://arxiv.org/abs/1406.4868} {arXiv:1406.4868 [hep-ph]} \BibitemShut
  {NoStop}%
\bibitem [{\citenamefont {Dienes}\ \emph
  {et~al.}(2017{\natexlab{a}})\citenamefont {Dienes}, \citenamefont {Kumar},
  \citenamefont {Thomas},\ and\ \citenamefont {Yaylali}}]{Dienes:2017ylr}%
  \BibitemOpen
  \bibfield  {author} {\bibinfo {author} {\bibfnamefont {K.~R.}\ \bibnamefont
  {Dienes}}, \bibinfo {author} {\bibfnamefont {J.}~\bibnamefont {Kumar}},
  \bibinfo {author} {\bibfnamefont {B.}~\bibnamefont {Thomas}}, \ and\ \bibinfo
  {author} {\bibfnamefont {D.}~\bibnamefont {Yaylali}},\ }\href {\doibase
  10.1103/PhysRevD.96.115009} {\bibfield  {journal} {\bibinfo  {journal} {Phys.
  Rev. D}\ }\textbf {\bibinfo {volume} {96}},\ \bibinfo {pages} {115009}
  (\bibinfo {year} {2017}{\natexlab{a}})},\ \Eprint
  {http://arxiv.org/abs/1708.09698} {arXiv:1708.09698 [hep-ph]} \BibitemShut
  {NoStop}%
\bibitem [{\citenamefont {Buyukdag}\ \emph {et~al.}(2020)\citenamefont
  {Buyukdag}, \citenamefont {Dienes}, \citenamefont {Gherghetta},\ and\
  \citenamefont {Thomas}}]{Buyukdag:2019lhh}%
  \BibitemOpen
  \bibfield  {author} {\bibinfo {author} {\bibfnamefont {Y.}~\bibnamefont
  {Buyukdag}}, \bibinfo {author} {\bibfnamefont {K.~R.}\ \bibnamefont
  {Dienes}}, \bibinfo {author} {\bibfnamefont {T.}~\bibnamefont {Gherghetta}},
  \ and\ \bibinfo {author} {\bibfnamefont {B.}~\bibnamefont {Thomas}},\ }\href
  {\doibase 10.1103/PhysRevD.101.075054} {\bibfield  {journal} {\bibinfo
  {journal} {Phys. Rev. D}\ }\textbf {\bibinfo {volume} {101}},\ \bibinfo
  {pages} {075054} (\bibinfo {year} {2020})},\ \Eprint
  {http://arxiv.org/abs/1912.10588} {arXiv:1912.10588 [hep-ph]} \BibitemShut
  {NoStop}%
\bibitem [{\citenamefont {Dienes}\ \emph
  {et~al.}(2017{\natexlab{b}})\citenamefont {Dienes}, \citenamefont {Huang},
  \citenamefont {Su},\ and\ \citenamefont {Thomas}}]{Dienes:2016vei}%
  \BibitemOpen
  \bibfield  {author} {\bibinfo {author} {\bibfnamefont {K.~R.}\ \bibnamefont
  {Dienes}}, \bibinfo {author} {\bibfnamefont {F.}~\bibnamefont {Huang}},
  \bibinfo {author} {\bibfnamefont {S.}~\bibnamefont {Su}}, \ and\ \bibinfo
  {author} {\bibfnamefont {B.}~\bibnamefont {Thomas}},\ }\href {\doibase
  10.1103/PhysRevD.95.043526} {\bibfield  {journal} {\bibinfo  {journal} {Phys.
  Rev. D}\ }\textbf {\bibinfo {volume} {95}},\ \bibinfo {pages} {043526}
  (\bibinfo {year} {2017}{\natexlab{b}})},\ \Eprint
  {http://arxiv.org/abs/1610.04112} {arXiv:1610.04112 [hep-ph]} \BibitemShut
  {NoStop}%
\bibitem [{\citenamefont {Dienes}\ \emph
  {et~al.}(2018{\natexlab{a}})\citenamefont {Dienes}, \citenamefont {Huang},
  \citenamefont {Su},\ and\ \citenamefont {Thomas}}]{Dienes:2018tux}%
  \BibitemOpen
  \bibfield  {author} {\bibinfo {author} {\bibfnamefont {K.~R.}\ \bibnamefont
  {Dienes}}, \bibinfo {author} {\bibfnamefont {F.}~\bibnamefont {Huang}},
  \bibinfo {author} {\bibfnamefont {S.}~\bibnamefont {Su}}, \ and\ \bibinfo
  {author} {\bibfnamefont {B.}~\bibnamefont {Thomas}},\ }\href {\doibase
  10.22323/1.336.0008} {\bibfield  {journal} {\bibinfo  {journal} {PoS}\
  }\textbf {\bibinfo {volume} {Confinement2018}},\ \bibinfo {pages} {008}
  (\bibinfo {year} {2018}{\natexlab{a}})}\BibitemShut {NoStop}%
\bibitem [{\citenamefont {Dienes}\ \emph
  {et~al.}(2016{\natexlab{a}})\citenamefont {Dienes}, \citenamefont {Fennick},
  \citenamefont {Kumar},\ and\ \citenamefont {Thomas}}]{Dienes:2016kgc}%
  \BibitemOpen
  \bibfield  {author} {\bibinfo {author} {\bibfnamefont {K.~R.}\ \bibnamefont
  {Dienes}}, \bibinfo {author} {\bibfnamefont {J.}~\bibnamefont {Fennick}},
  \bibinfo {author} {\bibfnamefont {J.}~\bibnamefont {Kumar}}, \ and\ \bibinfo
  {author} {\bibfnamefont {B.}~\bibnamefont {Thomas}},\ }\href {\doibase
  10.1103/PhysRevD.93.083506} {\bibfield  {journal} {\bibinfo  {journal} {Phys.
  Rev. D}\ }\textbf {\bibinfo {volume} {93}},\ \bibinfo {pages} {083506}
  (\bibinfo {year} {2016}{\natexlab{a}})},\ \Eprint
  {http://arxiv.org/abs/1601.05094} {arXiv:1601.05094 [hep-ph]} \BibitemShut
  {NoStop}%
\bibitem [{\citenamefont {Cs\'aki}\ \emph
  {et~al.}(2021{\natexlab{a}})\citenamefont {Cs\'aki}, \citenamefont {Hong},
  \citenamefont {Kurup}, \citenamefont {Lee}, \citenamefont {Perelstein},\ and\
  \citenamefont {Xue}}]{Csaki:2021gfm}%
  \BibitemOpen
  \bibfield  {author} {\bibinfo {author} {\bibfnamefont {C.}~\bibnamefont
  {Cs\'aki}}, \bibinfo {author} {\bibfnamefont {S.}~\bibnamefont {Hong}},
  \bibinfo {author} {\bibfnamefont {G.}~\bibnamefont {Kurup}}, \bibinfo
  {author} {\bibfnamefont {S.~J.}\ \bibnamefont {Lee}}, \bibinfo {author}
  {\bibfnamefont {M.}~\bibnamefont {Perelstein}}, \ and\ \bibinfo {author}
  {\bibfnamefont {W.}~\bibnamefont {Xue}},\ }\href@noop {} {\  (\bibinfo {year}
  {2021}{\natexlab{a}})},\ \Eprint {http://arxiv.org/abs/2105.07035}
  {arXiv:2105.07035 [hep-ph]} \BibitemShut {NoStop}%
\bibitem [{\citenamefont {Cs\'aki}\ \emph
  {et~al.}(2021{\natexlab{b}})\citenamefont {Cs\'aki}, \citenamefont {Hong},
  \citenamefont {Kurup}, \citenamefont {Lee}, \citenamefont {Perelstein},\ and\
  \citenamefont {Xue}}]{Csaki:2021xpy}%
  \BibitemOpen
  \bibfield  {author} {\bibinfo {author} {\bibfnamefont {C.}~\bibnamefont
  {Cs\'aki}}, \bibinfo {author} {\bibfnamefont {S.}~\bibnamefont {Hong}},
  \bibinfo {author} {\bibfnamefont {G.}~\bibnamefont {Kurup}}, \bibinfo
  {author} {\bibfnamefont {S.~J.}\ \bibnamefont {Lee}}, \bibinfo {author}
  {\bibfnamefont {M.}~\bibnamefont {Perelstein}}, \ and\ \bibinfo {author}
  {\bibfnamefont {W.}~\bibnamefont {Xue}},\ }\href@noop {} {\  (\bibinfo {year}
  {2021}{\natexlab{b}})},\ \Eprint {http://arxiv.org/abs/2105.14023}
  {arXiv:2105.14023 [hep-ph]} \BibitemShut {NoStop}%
\bibitem [{\citenamefont {Cabrer}\ \emph {et~al.}(2010)\citenamefont {Cabrer},
  \citenamefont {von Gersdorff},\ and\ \citenamefont {Quiros}}]{Cabrer:2009we}%
  \BibitemOpen
  \bibfield  {author} {\bibinfo {author} {\bibfnamefont {J.~A.}\ \bibnamefont
  {Cabrer}}, \bibinfo {author} {\bibfnamefont {G.}~\bibnamefont {von
  Gersdorff}}, \ and\ \bibinfo {author} {\bibfnamefont {M.}~\bibnamefont
  {Quiros}},\ }\href {\doibase 10.1088/1367-2630/12/7/075012} {\bibfield
  {journal} {\bibinfo  {journal} {New J. Phys.}\ }\textbf {\bibinfo {volume}
  {12}},\ \bibinfo {pages} {075012} (\bibinfo {year} {2010})},\ \Eprint
  {http://arxiv.org/abs/0907.5361} {arXiv:0907.5361 [hep-ph]} \BibitemShut
  {NoStop}%
\bibitem [{\citenamefont {Dienes}\ \emph
  {et~al.}(2016{\natexlab{b}})\citenamefont {Dienes}, \citenamefont {Kost},\
  and\ \citenamefont {Thomas}}]{Dienes:2015bka}%
  \BibitemOpen
  \bibfield  {author} {\bibinfo {author} {\bibfnamefont {K.~R.}\ \bibnamefont
  {Dienes}}, \bibinfo {author} {\bibfnamefont {J.}~\bibnamefont {Kost}}, \ and\
  \bibinfo {author} {\bibfnamefont {B.}~\bibnamefont {Thomas}},\ }\href
  {\doibase 10.1103/PhysRevD.93.043540} {\bibfield  {journal} {\bibinfo
  {journal} {Phys. Rev. D}\ }\textbf {\bibinfo {volume} {93}},\ \bibinfo
  {pages} {043540} (\bibinfo {year} {2016}{\natexlab{b}})},\ \Eprint
  {http://arxiv.org/abs/1509.00470} {arXiv:1509.00470 [hep-ph]} \BibitemShut
  {NoStop}%
\bibitem [{\citenamefont {Dienes}\ \emph
  {et~al.}(2017{\natexlab{c}})\citenamefont {Dienes}, \citenamefont {Kost},\
  and\ \citenamefont {Thomas}}]{Dienes:2016zfr}%
  \BibitemOpen
  \bibfield  {author} {\bibinfo {author} {\bibfnamefont {K.~R.}\ \bibnamefont
  {Dienes}}, \bibinfo {author} {\bibfnamefont {J.}~\bibnamefont {Kost}}, \ and\
  \bibinfo {author} {\bibfnamefont {B.}~\bibnamefont {Thomas}},\ }\href
  {\doibase 10.1103/PhysRevD.95.123539} {\bibfield  {journal} {\bibinfo
  {journal} {Phys. Rev. D}\ }\textbf {\bibinfo {volume} {95}},\ \bibinfo
  {pages} {123539} (\bibinfo {year} {2017}{\natexlab{c}})},\ \Eprint
  {http://arxiv.org/abs/1612.08950} {arXiv:1612.08950 [hep-ph]} \BibitemShut
  {NoStop}%
\bibitem [{\citenamefont {Dienes}\ \emph {et~al.}(2019)\citenamefont {Dienes},
  \citenamefont {Kost},\ and\ \citenamefont {Thomas}}]{Dienes:2019chq}%
  \BibitemOpen
  \bibfield  {author} {\bibinfo {author} {\bibfnamefont {K.~R.}\ \bibnamefont
  {Dienes}}, \bibinfo {author} {\bibfnamefont {J.}~\bibnamefont {Kost}}, \ and\
  \bibinfo {author} {\bibfnamefont {B.}~\bibnamefont {Thomas}},\ }\href
  {\doibase 10.1103/PhysRevD.100.083516} {\bibfield  {journal} {\bibinfo
  {journal} {Phys. Rev. D}\ }\textbf {\bibinfo {volume} {100}},\ \bibinfo
  {pages} {083516} (\bibinfo {year} {2019})},\ \Eprint
  {http://arxiv.org/abs/1907.10074} {arXiv:1907.10074 [hep-th]} \BibitemShut
  {NoStop}%
\bibitem [{\citenamefont {Dienes}\ \emph
  {et~al.}(2018{\natexlab{b}})\citenamefont {Dienes}, \citenamefont {Fennick},
  \citenamefont {Kumar},\ and\ \citenamefont {Thomas}}]{Dienes:2017zjq}%
  \BibitemOpen
  \bibfield  {author} {\bibinfo {author} {\bibfnamefont {K.~R.}\ \bibnamefont
  {Dienes}}, \bibinfo {author} {\bibfnamefont {J.}~\bibnamefont {Fennick}},
  \bibinfo {author} {\bibfnamefont {J.}~\bibnamefont {Kumar}}, \ and\ \bibinfo
  {author} {\bibfnamefont {B.}~\bibnamefont {Thomas}},\ }\href {\doibase
  10.1103/PhysRevD.97.063522} {\bibfield  {journal} {\bibinfo  {journal} {Phys.
  Rev. D}\ }\textbf {\bibinfo {volume} {97}},\ \bibinfo {pages} {063522}
  (\bibinfo {year} {2018}{\natexlab{b}})},\ \Eprint
  {http://arxiv.org/abs/1712.09919} {arXiv:1712.09919 [hep-ph]} \BibitemShut
  {NoStop}%
\bibitem [{\citenamefont {Dienes}\ \emph {et~al.}(2022)\citenamefont {Dienes},
  \citenamefont {Heurtier}, \citenamefont {Huang}, \citenamefont {Kim},
  \citenamefont {Tait},\ and\ \citenamefont {Thomas}}]{Dienes:2021woi}%
  \BibitemOpen
  \bibfield  {author} {\bibinfo {author} {\bibfnamefont {K.~R.}\ \bibnamefont
  {Dienes}}, \bibinfo {author} {\bibfnamefont {L.}~\bibnamefont {Heurtier}},
  \bibinfo {author} {\bibfnamefont {F.}~\bibnamefont {Huang}}, \bibinfo
  {author} {\bibfnamefont {D.}~\bibnamefont {Kim}}, \bibinfo {author}
  {\bibfnamefont {T.~M.~P.}\ \bibnamefont {Tait}}, \ and\ \bibinfo {author}
  {\bibfnamefont {B.}~\bibnamefont {Thomas}},\ }\href {\doibase
  10.1103/PhysRevD.105.023530} {\bibfield  {journal} {\bibinfo  {journal}
  {Phys. Rev. D}\ }\textbf {\bibinfo {volume} {105}},\ \bibinfo {pages}
  {023530} (\bibinfo {year} {2022})},\ \Eprint
  {http://arxiv.org/abs/2111.04753} {arXiv:2111.04753 [astro-ph.CO]}
  \BibitemShut {NoStop}%
\bibitem [{\citenamefont {Strassler}\ and\ \citenamefont
  {Zurek}(2007)}]{Strassler:2006im}%
  \BibitemOpen
  \bibfield  {author} {\bibinfo {author} {\bibfnamefont {M.~J.}\ \bibnamefont
  {Strassler}}\ and\ \bibinfo {author} {\bibfnamefont {K.~M.}\ \bibnamefont
  {Zurek}},\ }\href {\doibase 10.1016/j.physletb.2007.06.055} {\bibfield
  {journal} {\bibinfo  {journal} {Phys. Lett. B}\ }\textbf {\bibinfo {volume}
  {651}},\ \bibinfo {pages} {374} (\bibinfo {year} {2007})},\ \Eprint
  {http://arxiv.org/abs/hep-ph/0604261} {arXiv:hep-ph/0604261} \BibitemShut
  {NoStop}%
\bibitem [{\citenamefont {Strassler}(2006)}]{Strassler:2006qa}%
  \BibitemOpen
  \bibfield  {author} {\bibinfo {author} {\bibfnamefont {M.~J.}\ \bibnamefont
  {Strassler}},\ }\href@noop {} {\  (\bibinfo {year} {2006})},\ \Eprint
  {http://arxiv.org/abs/hep-ph/0607160} {arXiv:hep-ph/0607160} \BibitemShut
  {NoStop}%
\bibitem [{\citenamefont {Martin}(2007)}]{Martin:2007gf}%
  \BibitemOpen
  \bibfield  {author} {\bibinfo {author} {\bibfnamefont {S.~P.}\ \bibnamefont
  {Martin}},\ }\href {\doibase 10.1103/PhysRevD.75.115005} {\bibfield
  {journal} {\bibinfo  {journal} {Phys. Rev. D}\ }\textbf {\bibinfo {volume}
  {75}},\ \bibinfo {pages} {115005} (\bibinfo {year} {2007})},\ \Eprint
  {http://arxiv.org/abs/hep-ph/0703097} {arXiv:hep-ph/0703097} \BibitemShut
  {NoStop}%
\bibitem [{\citenamefont {Giromini}\ \emph {et~al.}(2008)\citenamefont
  {Giromini}, \citenamefont {Happacher}, \citenamefont {Kim}, \citenamefont
  {Kruse}, \citenamefont {Pitts}, \citenamefont {Ptohos},\ and\ \citenamefont
  {Torre}}]{Giromini:2008xh}%
  \BibitemOpen
  \bibfield  {author} {\bibinfo {author} {\bibfnamefont {P.}~\bibnamefont
  {Giromini}}, \bibinfo {author} {\bibfnamefont {F.}~\bibnamefont {Happacher}},
  \bibinfo {author} {\bibfnamefont {M.~J.}\ \bibnamefont {Kim}}, \bibinfo
  {author} {\bibfnamefont {M.}~\bibnamefont {Kruse}}, \bibinfo {author}
  {\bibfnamefont {K.}~\bibnamefont {Pitts}}, \bibinfo {author} {\bibfnamefont
  {F.}~\bibnamefont {Ptohos}}, \ and\ \bibinfo {author} {\bibfnamefont
  {S.}~\bibnamefont {Torre}},\ }\href@noop {} {\  (\bibinfo {year} {2008})},\
  \Eprint {http://arxiv.org/abs/0810.5730} {arXiv:0810.5730 [hep-ph]}
  \BibitemShut {NoStop}%
\bibitem [{\citenamefont {Strassler}(2008)}]{Strassler:2008jq}%
  \BibitemOpen
  \bibfield  {author} {\bibinfo {author} {\bibfnamefont {M.~J.}\ \bibnamefont
  {Strassler}},\ }\href@noop {} {\  (\bibinfo {year} {2008})},\ \Eprint
  {http://arxiv.org/abs/0811.1560} {arXiv:0811.1560 [hep-ph]} \BibitemShut
  {NoStop}%
\bibitem [{\citenamefont {Juknevich}\ \emph {et~al.}(2009)\citenamefont
  {Juknevich}, \citenamefont {Melnikov},\ and\ \citenamefont
  {Strassler}}]{Juknevich:2009ji}%
  \BibitemOpen
  \bibfield  {author} {\bibinfo {author} {\bibfnamefont {J.~E.}\ \bibnamefont
  {Juknevich}}, \bibinfo {author} {\bibfnamefont {D.}~\bibnamefont {Melnikov}},
  \ and\ \bibinfo {author} {\bibfnamefont {M.~J.}\ \bibnamefont {Strassler}},\
  }\href {\doibase 10.1088/1126-6708/2009/07/055} {\bibfield  {journal}
  {\bibinfo  {journal} {JHEP}\ }\textbf {\bibinfo {volume} {07}},\ \bibinfo
  {pages} {055} (\bibinfo {year} {2009})},\ \Eprint
  {http://arxiv.org/abs/0903.0883} {arXiv:0903.0883 [hep-ph]} \BibitemShut
  {NoStop}%
\bibitem [{\citenamefont {Juknevich}(2010)}]{Juknevich:2009gg}%
  \BibitemOpen
  \bibfield  {author} {\bibinfo {author} {\bibfnamefont {J.~E.}\ \bibnamefont
  {Juknevich}},\ }\href {\doibase 10.1007/JHEP08(2010)121} {\bibfield
  {journal} {\bibinfo  {journal} {JHEP}\ }\textbf {\bibinfo {volume} {08}},\
  \bibinfo {pages} {121} (\bibinfo {year} {2010})},\ \Eprint
  {http://arxiv.org/abs/0911.5616} {arXiv:0911.5616 [hep-ph]} \BibitemShut
  {NoStop}%
\bibitem [{\citenamefont {Craig}\ \emph {et~al.}(2015)\citenamefont {Craig},
  \citenamefont {Katz}, \citenamefont {Strassler},\ and\ \citenamefont
  {Sundrum}}]{Craig:2015pha}%
  \BibitemOpen
  \bibfield  {author} {\bibinfo {author} {\bibfnamefont {N.}~\bibnamefont
  {Craig}}, \bibinfo {author} {\bibfnamefont {A.}~\bibnamefont {Katz}},
  \bibinfo {author} {\bibfnamefont {M.}~\bibnamefont {Strassler}}, \ and\
  \bibinfo {author} {\bibfnamefont {R.}~\bibnamefont {Sundrum}},\ }\href
  {\doibase 10.1007/JHEP07(2015)105} {\bibfield  {journal} {\bibinfo  {journal}
  {JHEP}\ }\textbf {\bibinfo {volume} {07}},\ \bibinfo {pages} {105} (\bibinfo
  {year} {2015})},\ \Eprint {http://arxiv.org/abs/1501.05310} {arXiv:1501.05310
  [hep-ph]} \BibitemShut {NoStop}%
\bibitem [{\citenamefont {Schwaller}\ \emph {et~al.}(2015)\citenamefont
  {Schwaller}, \citenamefont {Stolarski},\ and\ \citenamefont
  {Weiler}}]{Schwaller:2015gea}%
  \BibitemOpen
  \bibfield  {author} {\bibinfo {author} {\bibfnamefont {P.}~\bibnamefont
  {Schwaller}}, \bibinfo {author} {\bibfnamefont {D.}~\bibnamefont
  {Stolarski}}, \ and\ \bibinfo {author} {\bibfnamefont {A.}~\bibnamefont
  {Weiler}},\ }\href {\doibase 10.1007/JHEP05(2015)059} {\bibfield  {journal}
  {\bibinfo  {journal} {JHEP}\ }\textbf {\bibinfo {volume} {05}},\ \bibinfo
  {pages} {059} (\bibinfo {year} {2015})},\ \Eprint
  {http://arxiv.org/abs/1502.05409} {arXiv:1502.05409 [hep-ph]} \BibitemShut
  {NoStop}%
\bibitem [{\citenamefont {Cohen}\ \emph {et~al.}(2015)\citenamefont {Cohen},
  \citenamefont {Lisanti},\ and\ \citenamefont {Lou}}]{Cohen:2015toa}%
  \BibitemOpen
  \bibfield  {author} {\bibinfo {author} {\bibfnamefont {T.}~\bibnamefont
  {Cohen}}, \bibinfo {author} {\bibfnamefont {M.}~\bibnamefont {Lisanti}}, \
  and\ \bibinfo {author} {\bibfnamefont {H.~K.}\ \bibnamefont {Lou}},\ }\href
  {\doibase 10.1103/PhysRevLett.115.171804} {\bibfield  {journal} {\bibinfo
  {journal} {Phys. Rev. Lett.}\ }\textbf {\bibinfo {volume} {115}},\ \bibinfo
  {pages} {171804} (\bibinfo {year} {2015})},\ \Eprint
  {http://arxiv.org/abs/1503.00009} {arXiv:1503.00009 [hep-ph]} \BibitemShut
  {NoStop}%
\bibitem [{\citenamefont {Knapen}\ \emph {et~al.}(2017)\citenamefont {Knapen},
  \citenamefont {Pagan~Griso}, \citenamefont {Papucci},\ and\ \citenamefont
  {Robinson}}]{Knapen:2016hky}%
  \BibitemOpen
  \bibfield  {author} {\bibinfo {author} {\bibfnamefont {S.}~\bibnamefont
  {Knapen}}, \bibinfo {author} {\bibfnamefont {S.}~\bibnamefont {Pagan~Griso}},
  \bibinfo {author} {\bibfnamefont {M.}~\bibnamefont {Papucci}}, \ and\
  \bibinfo {author} {\bibfnamefont {D.~J.}\ \bibnamefont {Robinson}},\ }\href
  {\doibase 10.1007/JHEP08(2017)076} {\bibfield  {journal} {\bibinfo  {journal}
  {JHEP}\ }\textbf {\bibinfo {volume} {08}},\ \bibinfo {pages} {076} (\bibinfo
  {year} {2017})},\ \Eprint {http://arxiv.org/abs/1612.00850} {arXiv:1612.00850
  [hep-ph]} \BibitemShut {NoStop}%
\bibitem [{\citenamefont {Park}\ and\ \citenamefont
  {Zhang}(2019)}]{Park:2017rfb}%
  \BibitemOpen
  \bibfield  {author} {\bibinfo {author} {\bibfnamefont {M.}~\bibnamefont
  {Park}}\ and\ \bibinfo {author} {\bibfnamefont {M.}~\bibnamefont {Zhang}},\
  }\href {\doibase 10.1103/PhysRevD.100.115009} {\bibfield  {journal} {\bibinfo
   {journal} {Phys. Rev. D}\ }\textbf {\bibinfo {volume} {100}},\ \bibinfo
  {pages} {115009} (\bibinfo {year} {2019})},\ \Eprint
  {http://arxiv.org/abs/1712.09279} {arXiv:1712.09279 [hep-ph]} \BibitemShut
  {NoStop}%
\bibitem [{\citenamefont {Cohen}\ \emph {et~al.}(2019)\citenamefont {Cohen},
  \citenamefont {D'Agnolo},\ and\ \citenamefont {Low}}]{Cohen:2018cnq}%
  \BibitemOpen
  \bibfield  {author} {\bibinfo {author} {\bibfnamefont {T.}~\bibnamefont
  {Cohen}}, \bibinfo {author} {\bibfnamefont {R.~T.}\ \bibnamefont {D'Agnolo}},
  \ and\ \bibinfo {author} {\bibfnamefont {M.}~\bibnamefont {Low}},\ }\href
  {\doibase 10.1103/PhysRevD.99.031702} {\bibfield  {journal} {\bibinfo
  {journal} {Phys. Rev. D}\ }\textbf {\bibinfo {volume} {99}},\ \bibinfo
  {pages} {031702} (\bibinfo {year} {2019})},\ \Eprint
  {http://arxiv.org/abs/1808.02031} {arXiv:1808.02031 [hep-ph]} \BibitemShut
  {NoStop}%
\bibitem [{\citenamefont {D'Agnolo}\ and\ \citenamefont
  {Low}(2019)}]{DAgnolo:2019cio}%
  \BibitemOpen
  \bibfield  {author} {\bibinfo {author} {\bibfnamefont {R.~T.}\ \bibnamefont
  {D'Agnolo}}\ and\ \bibinfo {author} {\bibfnamefont {M.}~\bibnamefont {Low}},\
  }\href {\doibase 10.1007/JHEP08(2019)163} {\bibfield  {journal} {\bibinfo
  {journal} {JHEP}\ }\textbf {\bibinfo {volume} {08}},\ \bibinfo {pages} {163}
  (\bibinfo {year} {2019})},\ \Eprint {http://arxiv.org/abs/1902.05535}
  {arXiv:1902.05535 [hep-ph]} \BibitemShut {NoStop}%
\bibitem [{\citenamefont {Chou}\ \emph {et~al.}(2017)\citenamefont {Chou},
  \citenamefont {Curtin},\ and\ \citenamefont {Lubatti}}]{Chou:2016lxi}%
  \BibitemOpen
  \bibfield  {author} {\bibinfo {author} {\bibfnamefont {J.~P.}\ \bibnamefont
  {Chou}}, \bibinfo {author} {\bibfnamefont {D.}~\bibnamefont {Curtin}}, \ and\
  \bibinfo {author} {\bibfnamefont {H.~J.}\ \bibnamefont {Lubatti}},\ }\href
  {\doibase 10.1016/j.physletb.2017.01.043} {\bibfield  {journal} {\bibinfo
  {journal} {Phys. Lett. B}\ }\textbf {\bibinfo {volume} {767}},\ \bibinfo
  {pages} {29} (\bibinfo {year} {2017})},\ \Eprint
  {http://arxiv.org/abs/1606.06298} {arXiv:1606.06298 [hep-ph]} \BibitemShut
  {NoStop}%
\bibitem [{\citenamefont {Curtin}\ \emph {et~al.}(2019)\citenamefont {Curtin}
  \emph {et~al.}}]{Curtin:2018mvb}%
  \BibitemOpen
  \bibfield  {author} {\bibinfo {author} {\bibfnamefont {D.}~\bibnamefont
  {Curtin}} \emph {et~al.},\ }\href {\doibase 10.1088/1361-6633/ab28d6}
  {\bibfield  {journal} {\bibinfo  {journal} {Rept. Prog. Phys.}\ }\textbf
  {\bibinfo {volume} {82}},\ \bibinfo {pages} {116201} (\bibinfo {year}
  {2019})},\ \Eprint {http://arxiv.org/abs/1806.07396} {arXiv:1806.07396
  [hep-ph]} \BibitemShut {NoStop}%
\bibitem [{\citenamefont {Feng}\ \emph {et~al.}(2018)\citenamefont {Feng},
  \citenamefont {Galon}, \citenamefont {Kling},\ and\ \citenamefont
  {Trojanowski}}]{Feng:2017uoz}%
  \BibitemOpen
  \bibfield  {author} {\bibinfo {author} {\bibfnamefont {J.~L.}\ \bibnamefont
  {Feng}}, \bibinfo {author} {\bibfnamefont {I.}~\bibnamefont {Galon}},
  \bibinfo {author} {\bibfnamefont {F.}~\bibnamefont {Kling}}, \ and\ \bibinfo
  {author} {\bibfnamefont {S.}~\bibnamefont {Trojanowski}},\ }\href {\doibase
  10.1103/PhysRevD.97.035001} {\bibfield  {journal} {\bibinfo  {journal} {Phys.
  Rev. D}\ }\textbf {\bibinfo {volume} {97}},\ \bibinfo {pages} {035001}
  (\bibinfo {year} {2018})},\ \Eprint {http://arxiv.org/abs/1708.09389}
  {arXiv:1708.09389 [hep-ph]} \BibitemShut {NoStop}%
\bibitem [{\citenamefont {Gligorov}\ \emph {et~al.}(2018)\citenamefont
  {Gligorov}, \citenamefont {Knapen}, \citenamefont {Papucci},\ and\
  \citenamefont {Robinson}}]{Gligorov:2017nwh}%
  \BibitemOpen
  \bibfield  {author} {\bibinfo {author} {\bibfnamefont {V.~V.}\ \bibnamefont
  {Gligorov}}, \bibinfo {author} {\bibfnamefont {S.}~\bibnamefont {Knapen}},
  \bibinfo {author} {\bibfnamefont {M.}~\bibnamefont {Papucci}}, \ and\
  \bibinfo {author} {\bibfnamefont {D.~J.}\ \bibnamefont {Robinson}},\ }\href
  {\doibase 10.1103/PhysRevD.97.015023} {\bibfield  {journal} {\bibinfo
  {journal} {Phys. Rev. D}\ }\textbf {\bibinfo {volume} {97}},\ \bibinfo
  {pages} {015023} (\bibinfo {year} {2018})},\ \Eprint
  {http://arxiv.org/abs/1708.09395} {arXiv:1708.09395 [hep-ph]} \BibitemShut
  {NoStop}%
\bibitem [{\citenamefont {Aielli}\ \emph {et~al.}(2020)\citenamefont {Aielli}
  \emph {et~al.}}]{Aielli:2019ivi}%
  \BibitemOpen
  \bibfield  {author} {\bibinfo {author} {\bibfnamefont {G.}~\bibnamefont
  {Aielli}} \emph {et~al.},\ }\href {\doibase 10.1140/epjc/s10052-020-08711-3}
  {\bibfield  {journal} {\bibinfo  {journal} {Eur. Phys. J. C}\ }\textbf
  {\bibinfo {volume} {80}},\ \bibinfo {pages} {1177} (\bibinfo {year}
  {2020})},\ \Eprint {http://arxiv.org/abs/1911.00481} {arXiv:1911.00481
  [hep-ex]} \BibitemShut {NoStop}%
\bibitem [{\citenamefont {Feng}\ \emph {et~al.}(2022)\citenamefont {Feng} \emph
  {et~al.}}]{Feng:2022inv}%
  \BibitemOpen
  \bibfield  {author} {\bibinfo {author} {\bibfnamefont {J.~L.}\ \bibnamefont
  {Feng}} \emph {et~al.},\ }in\ \href@noop {} {\emph {\bibinfo {booktitle}
  {{2022 Snowmass Summer Study}}}}\ (\bibinfo {year} {2022})\ \Eprint
  {http://arxiv.org/abs/2203.05090} {arXiv:2203.05090 [hep-ex]} \BibitemShut
  {NoStop}%
\bibitem [{\citenamefont {Dienes}\ \emph {et~al.}(2000)\citenamefont {Dienes},
  \citenamefont {Dudas},\ and\ \citenamefont {Gherghetta}}]{Dienes:1999gw}%
  \BibitemOpen
  \bibfield  {author} {\bibinfo {author} {\bibfnamefont {K.~R.}\ \bibnamefont
  {Dienes}}, \bibinfo {author} {\bibfnamefont {E.}~\bibnamefont {Dudas}}, \
  and\ \bibinfo {author} {\bibfnamefont {T.}~\bibnamefont {Gherghetta}},\
  }\href {\doibase 10.1103/PhysRevD.62.105023} {\bibfield  {journal} {\bibinfo
  {journal} {Phys. Rev. D}\ }\textbf {\bibinfo {volume} {62}},\ \bibinfo
  {pages} {105023} (\bibinfo {year} {2000})},\ \Eprint
  {http://arxiv.org/abs/hep-ph/9912455} {arXiv:hep-ph/9912455} \BibitemShut
  {NoStop}%
\bibitem [{\citenamefont {Dienes}\ \emph {et~al.}()\citenamefont {Dienes},
  \citenamefont {Dudas}, \citenamefont {Gherghetta},\ and\ \citenamefont
  {Thomas}}]{decohtoappear}%
  \BibitemOpen
  \bibfield  {author} {\bibinfo {author} {\bibfnamefont {K.~R.}\ \bibnamefont
  {Dienes}}, \bibinfo {author} {\bibfnamefont {E.}~\bibnamefont {Dudas}},
  \bibinfo {author} {\bibfnamefont {T.}~\bibnamefont {Gherghetta}}, \ and\
  \bibinfo {author} {\bibfnamefont {B.}~\bibnamefont {Thomas}},\ }\href@noop {}
  {\ }\Eprint {http://arxiv.org/abs/to appear} {to appear} \BibitemShut
  {NoStop}%
\bibitem [{\citenamefont {Fan}\ \emph {et~al.}(2013)\citenamefont {Fan},
  \citenamefont {Katz}, \citenamefont {Randall},\ and\ \citenamefont
  {Reece}}]{Fan:2013yva}%
  \BibitemOpen
  \bibfield  {author} {\bibinfo {author} {\bibfnamefont {J.}~\bibnamefont
  {Fan}}, \bibinfo {author} {\bibfnamefont {A.}~\bibnamefont {Katz}}, \bibinfo
  {author} {\bibfnamefont {L.}~\bibnamefont {Randall}}, \ and\ \bibinfo
  {author} {\bibfnamefont {M.}~\bibnamefont {Reece}},\ }\href {\doibase
  10.1016/j.dark.2013.07.001} {\bibfield  {journal} {\bibinfo  {journal} {Phys.
  Dark Univ.}\ }\textbf {\bibinfo {volume} {2}},\ \bibinfo {pages} {139}
  (\bibinfo {year} {2013})},\ \Eprint {http://arxiv.org/abs/1303.1521}
  {arXiv:1303.1521 [astro-ph.CO]} \BibitemShut {NoStop}%
\bibitem [{\citenamefont {Melville}(1851)}]{Melville}%
  \BibitemOpen
  \bibfield  {author} {\bibinfo {author} {\bibfnamefont {H.}~\bibnamefont
  {Melville}},\ }\href@noop {} {\emph {\bibinfo {title} {{Moby Dick, or The
  Whale}}}}\ (\bibinfo  {publisher} {Richard Bentley Publishers},\ \bibinfo
  {address} {London},\ \bibinfo {year} {1851})\BibitemShut {NoStop}%
\end{thebibliography}%

\end{document}